\documentclass[aps,onecolumn,nofootinbib,amsmath,amssymb,floatfix]{revtex4} 
\usepackage{graphicx}  
\usepackage{dcolumn}   
\usepackage{bm,color}        
\usepackage{amssymb}   
\usepackage{adjustbox}
\usepackage{hyperref}

\definecolor{MyBlue}{rgb}{0.15,0.15,0.70}

\hypersetup{
colorlinks=true,
citecolor=red,
linkcolor=MyBlue,
urlcolor=MyBlue
}

\newcommand{\be}{\begin{equation}}
\newcommand{\ee}{\end{equation}}
\newcommand{\beq}{\begin{equation}}
\newcommand{\eeq}{\end{equation}}
\newcommand{\bea}{\begin{eqnarray}}
\newcommand{\eea}{\end{eqnarray}}

\newcommand{\iu}{{i\mkern1mu}}

\def\k{{\bf k}}
\def\q{{\bf q}}

\def\x{{\bf x}}

\def\O{\mathcal{O}}

\newcommand\ees{\end{eqnarray}}
\newcommand\bees{\begin{eqnarray}}

\def\fnl{f_{\rm NL}}

\def\pa{\partial}

\def\M{\mathcal{M}}

\begin{document}

\pagenumbering{arabic}

\title{Primordial physics from large-scale structure beyond the power spectrum}
\author{Roland de Putter}
\address{California Institute of Technology, Pasadena, CA}

\begin{abstract}
We study
constraints on primordial mode-coupling from the power spectrum, squeezed-limit bispectrum and collapsed trispectrum of matter and halos.
We describe these statistics in terms of long-wavelength
$2$-point functions
involving the matter/halo density and position-dependent power spectrum.
This allows us to derive simple, analytic expression for the information content,
treating constraints from scale-dependent bias in the halo power spectrum on the same footing as those from higher order statistics.
In particular, we include non-Gaussian covariance due to long-short mode-coupling from non-linear evolution,
which manifests itself as long-mode cosmic variance in the position-dependent power spectrum.
We find that bispectrum forecasts that ignore this cosmic variance
may underestimate $\sigma(\fnl)$ by up to a factor $\sim 3$ for the matter density (at $z=1$) and commonly a factor $\sim 2$ for the halo bispectrum.
Constraints from the bispectrum can be improved by combining it with the power spectrum and trispectrum.
The reason is that, in the position-dependent power spectrum picture,
the bispectrum and trispectrum intrinsically incorporate multitracer
cosmic variance cancellation, which is optimized in a joint analysis.
For halo statistics, we discuss the roles of scale-dependent bias, matter mode-coupling, and non-linear, non-Gaussian biasing ($b_{11}^{(h)}$).
While scale-dependent bias in the halo power spectrum is already very constraining,
higher order halo statistics are competitive
in the regime where stochastic noise in the position-dependent halo power spectrum is low enough for cosmic variance cancellation to be effective,
i.e.~for large halo number density and large $k_{\rm max}$.
This motivates exploring this regime observationally.

\end{abstract}

\maketitle

\section{Introduction}

A key goal of observational cosmology is to elucidate the nature of inflation and more generally the primordial physics describing the initial conditions of the Universe.
A powerful and general method for doing this is offered by the single-field consistency conditions \cite{Maldacena:2002vr, Creminelli:2004yq}, which describe relations between $n$-point functions of cosmological perturbations at long and short wavelengths.
The physics behind these relations is that, in single-field inflation, perturbations are purely adiabatic\footnote{We assume standard attractor solution single-field inflation. Another implicit assumption in the consistency conditions is the absence of local-type mode-coupling in the initial quantum state, which is satisfied if the perturbations start in the Bunch-Davies vacuum. See e.g.~\cite{Namjoo:2012aa,Assassi:2012et,Chen:2013aj,mooijpalma15} and \cite{chenetal07,holtol08,agupar11,ganc11}, respectively, for modifications of these two assumptions.} (i.e.~described by a single ``clock'') ,
which means that, modulo gradients, a long-wavelength potential perturbation $\phi_L$ can locally be removed by a coordinate transformation.
Therefore, to zeroth order in gradients, $\phi_L$ has no effect
on local physics (e.g.~$n$-point functions of short modes) in a region small compared to the long mode.

In practice, the consistency conditions and deviations thereof are
tested by constraining the mode-coupling described by {\it local primordial non-Gaussianity} (see \cite{RdPetal17a} for a discussion of the exact relation between the local ansatz and the consistency relations).
To leading order, local non-Gaussianity describes a modulation of the {\it amplitude} of short modes by $\phi_L$,
with an amplitude given by the parameter $\fnl$ \cite{komsper01}. Schematically, $\delta_S \supset 2 \fnl \, \phi_L \, \tilde{\delta}_S$, where $\delta_S$ is the short-mode matter density perturbation and $\tilde{\delta}_S$ the short-mode in the absence of $\phi_L$.
The single-field consistency conditions predict a long-short mode-coupling corresponding to\footnote{The consistency conditions predict zero modulation of local, physical quantities by the long mode $\phi_L$ (modulo gradient).
The non-zero value is a projection effect, arising from the conversion between local physical coordinates and the global coordinate system.} $\fnl = -5/12 \, (n_s - 1)$ \cite{Maldacena:2002vr, Creminelli:2004yq}, where $n_s$ is the scalar spectral index.
On the other hand, the mode-coupling is {\it a priori} unconstrained in multifield inflation, with a particularly interesting class of multifield models typically predicting $\fnl$ of order unity \cite{Alvarez:2014vva,RdPetal2017b}.

It is the long-short mode-coupling described above, as quantified by the non-Gaussianity parameter $\fnl$, that
is the ``signal'' of interest in this paper.
Probing this signal thus serves both as a method for distinguishing between single-field and multifield models of inflation and as a general test of the single-field inflation paradigm and the assumptions therein.

\vskip 7pt

The tightest current observational constraints on $\fnl$ come from the bispectra of cosmic microwave background (CMB) fluctuations and are fully consistent with single-field inflation,
$\fnl = 0.8 \pm 5.0$ \cite{Ade:2015ava}.
However, in the future, large-scale structure measurements have the potential to improve on this
and may in principle reach a precision of $\sigma(\fnl) < 1$ \cite{Alvarez:2014vva,giannanetal12,dePutter:2014lna,ferrsmith15,Yamauchi:2014ioa,baldaufetal11a,baldaufetal16,scoccetal04,Sefusatti:2007ih}, which is an interesting target set by multifield models.

Local-type primordial mode-coupling manifests itself in large-scale structure in multiple ways and we here consider three types of probes.
First, the mode-coupling leads to a characteristic signature in the squeezed-limit bispectrum and collapsed trispectrum of matter density perturbations. We will refer to this signal as ``higher order matter statistics'' (even though we may include the matter power spectrum).
The mode-coupling also leads to a modulation by a long mode $\phi_L$ of the number density of halos, otherwise known as scale-dependent halo bias \cite{dalaletal08,matver08,slosaretal08,desjseljak10,schmidtkam10}. Our second probe will therefore be the scale-dependent bias information obtained from the power spectrum (and cross-spectra) of halos
and we will refer to this simply as ``scale-dependent bias''.
For the third set of probes, we additionally include the halo squeezed bispectrum and trispectrum, i.e.~``higher order halo statistics''.
This final set of statistics probes the primordial signal both
through the long-short mode-coupling of halos (due to the mode-coupling in the underlying matter density and due to non-Gaussian, non-linear biasing)
and through scale-dependent halo bias.
Previous studies of higher order statistics as a probe of primordial mode-coupling focus mostly on the (halo) bispectrum and include \cite{Bernardeau:2001qr,scocc00,sefu09,jeongkom09,giannporc10,tasetal14,tellarinietal15,baldaufetal11a,baldaufetal16,scoccetal04,Sefusatti:2007ih,sefuetal06,gilmaretal14,wellingetal16}.

\vskip 7pt

The goal of this paper is to present a comprehensive and systematic study of the information content on primordial mode-coupling ($\fnl$) contained in the above large-scale structure measurements,
paying specific attention to the comparison between scale-dependent bias and higher order statistics, and the complementarity between probes.

\vskip 7pt

An invaluable realization for such a study is that the squeezed-limit higher order statistics of interest can be captured in terms of the {\it long-wavelength} modulation of a {\it position-dependent power spectrum},
where the latter measures the {\it short-wavelength} clustering amplitude \cite{chiangetal14,chiangetal15,chiang17,chiangetal17}, see also \cite{schmittfulletal15} (note this idea can naturally be extended to position-dependent higher order statistics \cite{adhietal16}).
This allows us to describe the otherwise complicated higher order correlations purely in terms of power and cross-spectra of long-mode fluctuations, i.e.~in terms of simple $2$-point statistics.
The information content is then captured by the ``multitracer'' machinery commonly used
for describing multiple biased tracers of the matter density \cite{Seljak:2008xr,mcdseljak09,hamausetal12}.
This approach allows for fast calculations and often helps gain physical insight into the results.
Concretely, in the above picture the matter/halo bispectrum is equivalent to the correlation between a long-mode matter/halo overdensity and a long-mode fluctuation in the position-dependent matter/halo power spectrum,
whereas the squeezed or collapsed trispectrum is equivalent to the cross-spectrum between two instances of the
position-dependent power spectrum.

A major advantage of the position-dependent power spectrum approach is that it enables us to include non-Gaussian contributions to the bispectrum covariance (see e.g.~\cite{sefuetal06}) analytically.
In conventional bispectrum forecasts, one often
neglects such contributions
so that
different bispectrum configurations are independent and the information content
(or likelihood) can be computed as a single sum over triangles. However,
there are important non-Gaussian terms in the bispectrum covariance that correlate different configurations sharing a common long mode, and that therefore
may significantly lower the bispectrum information content.
These non-Gaussian covariance contributions are tedious to include in the conventional approach, requiring numerical inversion of a non-diagonal bispectrum covariance matrix and performing a double sum over triangles, but they are naturally included in our forecasts.

We will see below that, in the position-dependent power spectrum picture, the non-Gaussian covariance is associated with cosmic variance in the long modes. Essentially the same cosmic variance terms are by default included in any forecast for the halo power spectrum (and/or cross-spectra). Thus, since we want to perform a consistent and fair comparison between information in scale-dependent bias from the halo power spectrum on the one hand, and that in the bispectrum (or higher order statistics in general) on the other hand, we have no choice but to include the non-Gaussian/cosmic variance terms in the latter.

Another important benefit of the position-dependent power spectrum approach is that it will allow us to easily calculate not just the information content in the bispectrum and trispectrum, but also the {\it joint} information content of e.g.~the combination of the halo power spectrum, bispectrum and trispectrum, in a way that includes the covariance between the different statistics.

\vskip 7pt

The outline of this paper is as follows. In Section \ref{sec:form}, we will introduce local non-Gaussianity and the mode-coupling signal in the matter perturbations, introduce our conventions and definitions, and we will briefly review the multitracer Fisher matrix formalism.
In Section \ref{sec:matter}, we will discuss the information content in higher order matter statistics, with the matter bispectrum as the main focus. It is here, specifically in Section \ref{subsec:posdeppk}, that we will introduce the position-dependent power spectrum approach in detail.
In Section \ref{sec:mat2 vs halo}, we consider the information content in scale-dependent bias
and in Section \ref{sec:halos}, we discuss the constraining power of higher order halo statistics.
Finally, we discuss our results and conclude in Section \ref{sec:disc}.

\begin{figure*}[]
\centering
\includegraphics[width=0.74\textwidth]{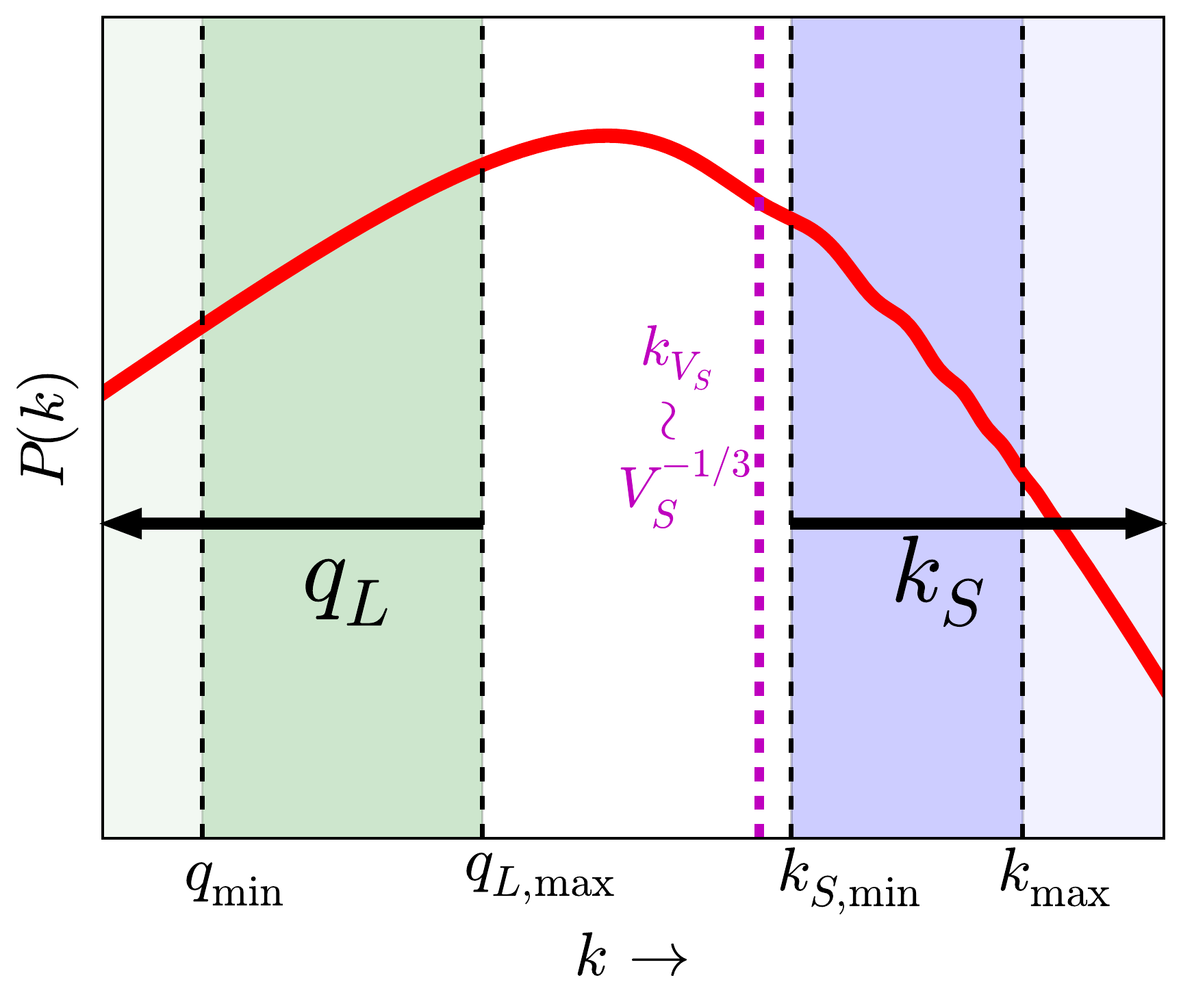} 
\caption{Illustration of the scales of interest in our analysis (with the linear matter power spectrum in arbitrary units shown in red for comparison). We consider the primordial non-Gaussianity information contained in the modulation of short modes, with wave number $k_S$, by long-wavelength perturbations with wave number $q_L$. We study long modes in the range $q_{\rm min} < q_L < q_{L,{\rm max}}$ and short modes in the range $k_{S,{\rm min}} < k_S < k_{\rm max}$. The squeezed-limit is enforced by imposing a hierarchy $q_{L,{\rm max}} \ll k_{S,{\rm min}}$. In this limit, the position-dependent power spectrum approach introduced in Section \ref{subsec:posdeppk} gives an extremely useful description of the squeezed-limit bispectrum and collapsed trispectrum information content. The dashed magenta line indicates the wave number corresponding to the ``local volume'' $V_S$ over which the position-dependent power spectrum is estimated (see Section \ref{subsec:posdeppk} for details).}
\label{fig:sepscales}
\end{figure*}

\section{Formalism}
\label{sec:form}

\subsection{Local primordial non-Gaussianity}

\begin{figure*}[]
\centering
\includegraphics[width=0.74\textwidth]{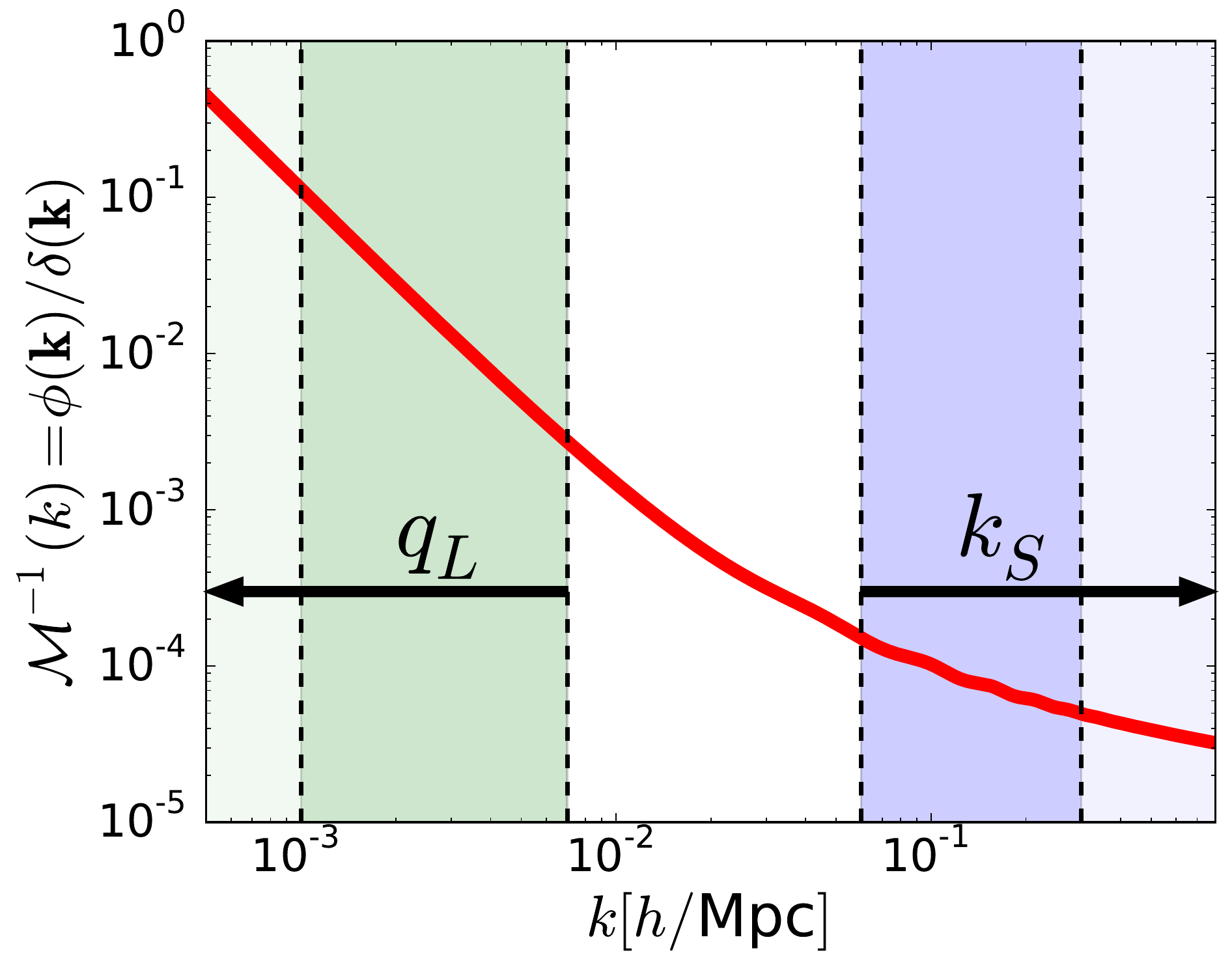} 
\caption{The quantity $\M^{-1}(k)$, which gives the ratio of a primordial potential fluctuation $\phi$ with wave number $k$ to the corresponding matter density perturbation $\delta$ (at $z = 1$). This quantity determines the scale-dependence (relative to $\delta$) of the primordial modulation with long-wavelength perturbations of small-scale power and halo number density.
}
\label{fig:Mk}
\end{figure*}

We consider primordial non-Gaussianity given by the local ansatz \cite{komsper01},
\beq
\phi(\x) = \tilde{\phi}(\x) + \fnl \, \left( \tilde{\phi}^2(\x) - \langle \tilde{\phi}^2 \rangle \right),
\eeq
where $\phi$ is the primordial Bardeen potential
and $\tilde{\phi}$ is a Gaussian auxiliary field.
In Fourier space\footnote{We use the Fourier convention $\delta(\k) = \int d^3 \x \, e^{\iu \k \cdot \x} \, \delta(\x)$.}, the matter density perturbation at redshift $z$ is to linear order related to $\phi$ by,
\bea
\label{eq:matphi}
\delta(\k) &=& \M(k) \, \phi(\k), \quad {\rm with} \quad \M(k) = \frac{2 \, k^2 \, T(k) \, D(z)}{3 \, \Omega_m \, H_0^2},
\eea
where $\Omega_m$ is the matter density relative to the critical density
and $H_0$ is the Hubble parameter, both at $z=0$.
The factor $T(q)$ is the transfer function of matter perturbations, normalized to unity at low wave number $q$, and $D(z)$
is the linear growth function, normalized such that $D(z)=1/(1 + z)$ during matter domination.
We show $\M^{-1}(k)$, the ratio between the primordial potential and the matter density perturbation as a function of scale, at redshift $z=1$, in Figure \ref{fig:Mk}.

\subsection{Long-short mode coupling and primordial non-Gaussianity}
\label{subsec:modecoupling}

Throughout this article, we focus on the squeezed-limit signal induced by $\fnl$, i.e.~the mode-coupling between long and short wavelength perturbations. We thus formally introduce a hierarchy of scales by defining long modes to have wave number $q_L$ and short modes wave number $k_S$ with,
\beq
q_L < q_{L,{\rm max}} \ll k_{S,{\rm min}} < k_S.
\eeq
As illustrated schematically in Figure \ref{fig:sepscales},
we focus mainly on long wavelength modes larger than the matter-radiation equality scale,
i.e.~we take $q_{L,{\rm max}} \lesssim k_{\rm eq} \sim 0.02 h/$Mpc.
To be fully in the squeezed limit, one wants $k_{S,{\rm min}}$ to be significantly larger than
$q_{L,{\rm max}}$, as illustrated in the Figure, although in practice we will relax this requirement a little in our forecasts.
We will also define a {\it longest long mode}, $q_{\rm min}$, effectively determined by the survey volume, and a ``shortest short mode'', $k_{\rm max}$, set by how deep into the non-linear regime we are able to probe.
We discuss in more detail the motivation behind the choice of these scales, and in particular the choice of the range of long modes, at the end of this subsection,
and we will discuss the exact numerical choices of $q_{L,{\rm max}}$, $k_{S,{\rm min}}$, $k_{\rm max}$, $q_{\rm min}$ when we first introduce quantitative results in Section \ref{subsec:matbk res}.

We can now express the squeezed-limit mode-coupling in Fourier space by writing the response of the short modes to the long modes,
\beq
\delta(\k_S) = \tilde{\delta}(\k_S) + 2 \fnl \, \int_{L} \, \frac{d^3 \q_L}{(2 \pi)^3} \, \phi(\q_L) \, \tilde{\delta}(\k_S - \q_L)
= \tilde{\delta}(\k_S) + 2 \fnl \, \int_{L} \, \frac{d^3 \q_L}{(2 \pi)^3} \, \M^{-1}(q_L) \, \delta(\q_L) \, \tilde{\delta}(\k_S - \q_L),
\eeq
where integrals with subscript $L$ are over $q_L < q_{L,{\rm max}}$ (i.e.~we have explicitly only written the squeezed-limit mode-coupling).
Here, $\tilde{\delta}(\k_S)$ is the short mode in the absence of long mode perturbations.
Throughout this paper, we will ignore the mode-coupling {\it between} short modes (at fixed realization of the long modes), which means we will treat $\tilde{\delta}(\k_S)$ as Gaussian. This approximation of course breaks down if we consider short modes deep in the non-linear regime. In principle, the short-short mode-coupling due to non-linear evolution can also be incorporated in our formalism (through higher-order stochastic noise terms), but this is beyond the scope of this paper.
We treat the long mode itself to linear order, i.e.~we will not consider the non-linear effect of pairs of short modes feeding into the long mode.
Schematically, we may write the squeezed-limit mode-coupling as
\beq
\label{eq:modecoupling}
\delta_S = \tilde{\delta}_S + 2 \fnl \, \phi_L \, \tilde{\delta}_S.
\eeq
It is this modulation of small-scale physics by a long-wavelength primordial potential fluctuation $\phi_L$ that we will treat as our signal throughout this paper.

\begin{figure*}[]
\centering
\includegraphics[width=0.47\textwidth]{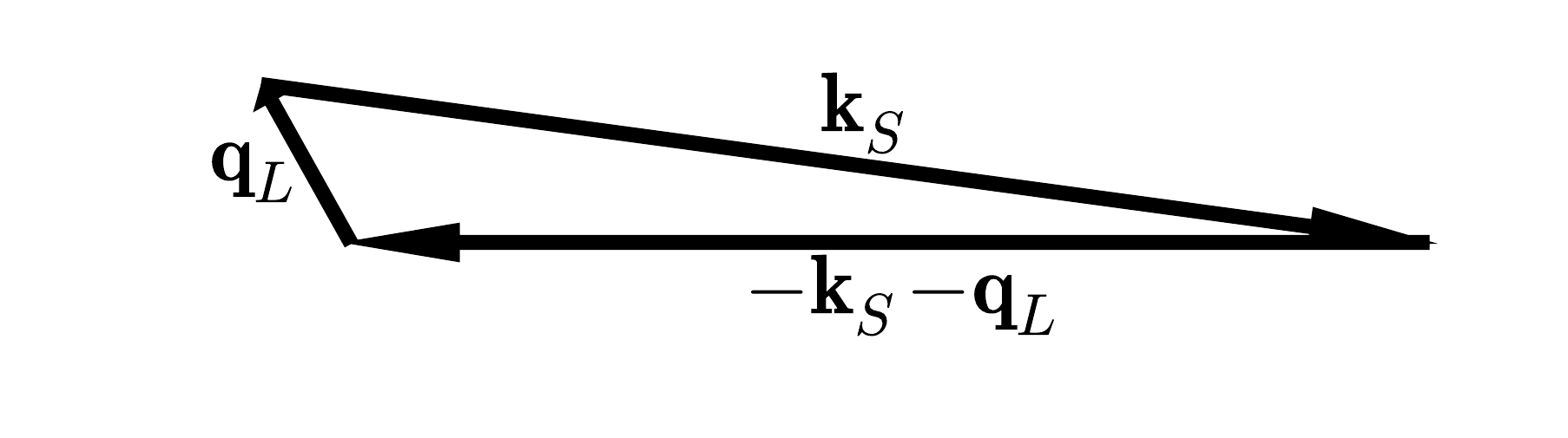} 
\includegraphics[width=0.47\textwidth]{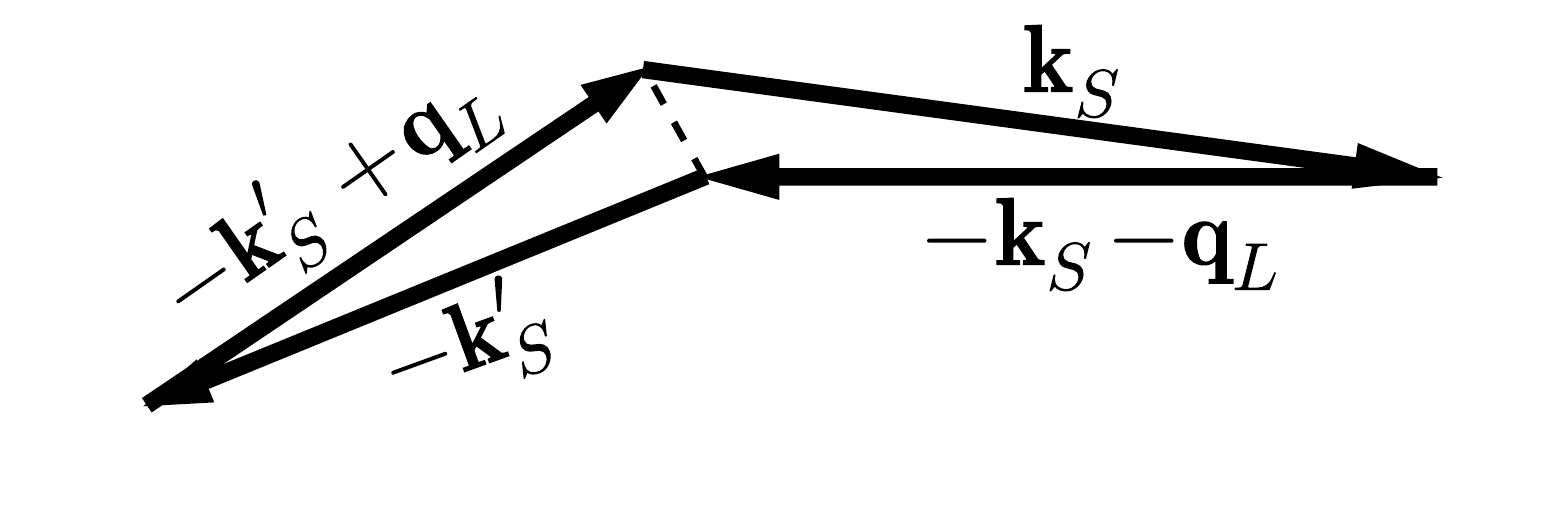} 
\caption{{\it Left:} The squeezed-limit bispectrum is the correlation between one long and two short modes. The pair of short modes is contained in the long-wavelength modulation of the position-dependent power spectrum, $\delta({\bf k}_s) \, \delta(-\k_S - \q_L) \subset \delta \hat{P}(\k_i; -\q_L)$,
where $\k_i$ indicates a bin of short modes containing $\k_S$ (see Section \ref{sec:matter}).
The squeezed-limit matter bispectrum is thus equivalent to the cross-spectrum between a long-mode matter perturbation and the long-mode position-dependent power spectrum perturbation.
{\it Right:}
The collapsed trispectrum correlates two pairs of short modes. It can be seen as the (cross-)power spectrum of
two instances of the position-dependent power spectrum, $\langle \delta \hat{P}(\k_i; \q_L) \, \delta \hat{P}(\k_j; -\q_L) \rangle$.}
\label{fig:bi-tri config}
\end{figure*}

The matter mode-coupling signal is most directly probed by the squeezed matter bispectrum (Figure \ref{fig:bi-tri config}, left panel) and the collapsed matter trispectrum (right panel).
The former describes the three-way correlation of one long mode (wave vector $\q_L$) and one pair of short modes (wave vectors $\k_S$ and $-\k_S - \q_L$, adding up to $-\q_L$).
The collapsed trispectrum  describes the four-way correlation of two pairs of short modes (one pair adding up to $\q_L$ and the other to $-\q_L$).
For convenience of notation, we will not always explicitly write ``squeezed-limit''/``collapsed'', but when we discuss the bi- or trispectrum, we always have in mind specifically those configurations.
We will describe in detail in the following sections how the bi- and trispectrum can be thought of in terms of long-wavelength fluctuations in a position-dependent small-scale power spectrum.
While we first consider modulation of the amplitude of small-scale {\it matter density perturbations} $\delta_S$, as in Eq.~(\ref{eq:modecoupling}), in Sections \ref{sec:mat2 vs halo} and \ref{sec:halos} we will extend the analysis to the modulation of the {\it halo number density} $n_h$ and of the amplitude of {\it small-scale halo number density perturbations} $\delta_{h,S}$ respectively. This means that in the course of this paper, we will study signals from power spectra, cross-power spectra, bispectra and trispectra, of both matter and halos.

Returning to the matter density perturbations for now, non-linear evolution also induces a mode-coupling in the matter overdensity of its own. Adding the standard perturbation theory (SPT, \cite{Bernardeau:2001qr}) result gives to leading order,
\beq
\delta(\k_S) = \tilde{\delta}(\k_S) + 2 \fnl \, \int_{L} \, \frac{d^3 q_L}{(2 \pi)^3} \, \phi(\q_L) \, \tilde{\delta}(\k_S - \q_L)
+ 2 \int_{L} \frac{d^3 q_L}{(2 \pi)^3} \, F_2(\q, \k_S - \q) \delta(\q_L) \, \tilde{\delta}(\k_S - q_L),
\eeq
where $F_2$ is the SPT kernel,
\beq
F_2(\k_1, \k_2) = \frac{5}{7} + \frac{1}{2} \, \mu  \, \left( \frac{k_1}{k_2} + \frac{k_2}{k_1}\right) + \frac{2}{7} \, \mu^2, \quad  \quad \mu \equiv \hat{\k}_1 \cdot \hat{\k}_2.
\eeq

Schematically, we thus have the modulation,
\beq
\delta_S = \tilde{\delta}_S + 2 \fnl \, \phi_L \, \tilde{\delta}_S + 2 F_2 \, \delta_L \, \tilde{\delta}_S.
\eeq
The primordial and non-primordial mode-couplings have a manifestly different dependence on the scale (or gradients) of the long mode. On large scales, $q_L \ll k_{\rm eq} \approx 0.02 h/$Mpc, we have $\M(q_L) \sim q_L^2 \sim \nabla_L^2$, so that primordial non-Gaussianity leads to a modulation $\propto q_L^{-2} \, \delta(q_L)$.
This is a very characteristic scale-dependence, robust against degeneracies from non-primordial contributions.
To give a sense of the amplitude of the primordial signal, for $|\fnl | \sim 1$,
the primordial modulation will be of the same order as non-primordial modulation due to non-linear evolution at approximately the Hubble scale, $q_L \sim c^{-1} \, H$, while it is suppressed at smaller scales.

\vskip 7pt

As illustrated in Figure \ref{fig:sepscales} and Figure \ref{fig:bi-tri config}, we focus our analysis on squeezed configurations, $k_S \gg q_L$, and we restrict the long modes to be larger than the matter-radiation equality scale, $q_L \lesssim k_{\rm eq} \approx 0.02 h/$Mpc).
The former choice is made because the information from the local-type primordial mode-coupling
is dominated by squeezed configurations, but the latter choice deserves some more explanation.
Indeed, it is not in general true that the information on $\fnl$ comes exclusively from the very largest scales.
In particular, considering the information content in the primordial matter density field, Appendix \ref{app:scaledep} shows that the information content is proportional to the variance in $\phi$
generated by the full range of long modes included in the analysis. Concretely, as long as $q_L \ll k_{\rm max}$, the $q_L$ dependence of the signal-to-noise squared scales like,
\beq
\left(\frac{S}{N}\right)^2 \propto \int d\ln q_L \, \Delta_\phi^2(q_L),
\eeq
where $\Delta_\phi^2(q_L) \propto q_L^{n_s - 1}$ is the dimensionless power spectrum of primordial potential fluctuations. This means that the amount of information per decade in $q_L$ is approximately scale-independent.
Thus, in principle there is additional information to be gained, although in practice it is a very modest amount, by extending the range of long modes to smaller scales.
We nonetheless stick to very long modes only, for the following reasons.

First, it is only at $q_L \ll k_{\rm eq}$ that the primordial mode-coupling has the characteristic $\propto 1/q_L^2$ scale-dependence relative to non-primordial mode-coupling, $\phi_L \sim q_L^{-2} \, \delta_L$. As shown in Figure \ref{fig:Mk}, at larger $q_L$, the scale-dependence becomes flatter. In principle, this is still a distinct signal from the non-primordial modulation, but in practice it may become much harder to distinguish the two and the primordial signal is likely to be much more degenerate with cosmological parameters describing (non-linear) evolution \cite{baldaufetal16,gleyzesetal17}.
Related to this point, because of the scale-dependence, for large $q_L$, the primordial mode-coupling becomes extremely small compared to the non-primordial one.
The only reason the information content is not similarly suppressed is that the number of independent modes (per $d \ln q_L$) is much larger at small scales.
The information contained at large $q_L$, while in principle there, may therefore be difficult to extract as this is effectively a very ``foreground dominated'' regime.

A second reason for focusing on very small $q_L$ is that we want the long modes to be safely inside the linear regime. This way, we are justified in treating the {\it long} modes as Gaussian so that in particular all information is contained in just the $2$-point functions of long modes.
Finally, the restriction to small $q_L$ allows us to keep $k_{S,{\rm min}}$ fixed, whereas if we wanted to include a larger range of $q_L$ values, we would have to adjust $k_{S,{\rm min}}$ so that the condition $k_{S,{\rm min}} > q_L$ is always satisfied (technically, $k_{S,{\rm min}} \gg q_L$, to be in the squeezed limit).

\subsection{Information content multiple biased tracers}
\label{subsec:infocontent}

In the next sections, we will use that not only the matter and halo power spectrum, but also the squeezed-limit bispectrum and collapsed trispectrum can be treated in terms of (cross-)power spectra of multiple biased tracers of the long-wavelength matter perturbations.
We here briefly review the general formalism
for computing
the Fisher information \cite{fisher,tegetal97,heavens09} from multiple tracers.

Consider a set of biased tracers (since we will later specifically study tracers of the {\it long mode} matter perturbations, we here use the wave vector $\q_L$),
\beq
\delta_a(\q_L) = b_a(\q_L) \, \delta(\q_L) + \epsilon_a(\q_L),
\eeq
where $\delta(\q_L)$ is the matter overdensity with power spectrum defined by,
\beq
\langle \delta(\k) \, \delta(\k')\rangle = (2 \pi)^3 \, \delta^{(D)}(\k + \k') \, P(k),
\eeq
$b_a(\q_L)$ is the tracer bias, and $\epsilon_a(\q_L)$ is a stochastic noise contribution, which is uncorrelated with $\delta(\q_L)$,
and has power spectrum,
\beq
\langle \epsilon_a(\k) \, \epsilon_b(\k') \rangle = (2 \pi)^3 \, \delta^{(D)}(\k + \k') \, {\bf N}_{ab}
= (2 \pi)^3 \, \delta^{(D)}(\k + \k') \, N_a \, \delta^{(K)}_{ab}.
\eeq
The last equality captures the assumption, which we will apply in the following, that the shot noise is uncorrelated between different tracers.
Then, the tracer (cross-)power spectra are given by,
\beq
P_{ab}(\q_L) = b_a(\q_L) \, b_b(\q_L) \, P(q_L) + N_a \, \delta^{(K)}_{ab}.
\eeq

Let us now consider the information contained in some (sub)set of such spectra, i.e.~our observables are the (cross-)power spectrum estimators,
\beq
\hat{\O}_{A}(\q_L) \equiv \hat{P}_{ab}(\q_L),
\eeq
for some set of tracer pairs $\{A = (ab)\}$.
The Fisher information in a parameter $p$ is then,
\beq
F = V \, \int \frac{d^3 \q_L}{(2 \pi)^3} \, F(\q_L),
\eeq
where $V$ is the survey volume, and $F(\q_L)$ is the Fisher information per mode $\q_L$,
\beq
\label{eq:fisher1}
F(\q_L) = \sum_{A B} \frac{\pa \O_A(\q_L)}{\pa p} \, {\bf C}^{-1}_{AB}(\q_L) \, \frac{\pa \O_B(\q_L)}{\pa p}.
\eeq
Here, $\O_A(\q_L)$ is the expectation value $\langle \hat{\O}_A(\q_L)\rangle$ and
${\bf C}$ is the covariance matrix of the tracer spectra,
\beq
{\bf C}_{AB}(\q_L) = P_{ac}(\q_L) \, P_{bd}(\q_L) + P_{ad}(\q_L) \, P_{bc}(\q_L),
\quad \text{with} \quad  A=(ab), \, B=(cd).
\eeq
The above covariance matrix makes the approximation that the long-wavelength perturbations are Gaussian.

We will quantify the information available on the non-Gaussianity parameter $\fnl$ in terms of the Fisher information $F_{\fnl \fnl} = F$ or, equivalently, the {\it unmarginalized} uncertainty,
\beq
\sigma(\fnl) = F^{-1/2}.
\eeq
The true expected uncertainty is generally larger than this due to parameter degeneracies. However, the local-type modulation by long wavelength primordial potential modes is only weakly degenerate with other parameters \cite{dePutter:2014lna,baldaufetal16,gleyzesetal17} and based on forecasts of scale-dependent halo bias, one expects marginalization to have only a modest effect ($10 - 30 \%$).

In the case of the parameter $\fnl$, the information will be contained fully in the $\fnl$-dependence of the biases. To make this explicit, the bias parameters above should really be written as $b_a(\q_L) \to b_a(\q_L; \fnl)$. The derivative of the signal with respect to $\fnl$ is then,
\beq
\frac{\pa \O_A(\q_L)}{\pa \fnl} = \left( b_a(\q_L) \, b_b'(\q_L)  + b_a'(\q_L) \, b_b(\q_L) \right) \, P(\q_L),
\eeq
where now $b_a(\q_L) \equiv b_a(\q_L; f_{\rm NL, fid})$ is to be interpreted as the fiducial bias
and $b_a'(\q_L) \equiv \pa b_a(\q_L; f_{\rm NL, fid})/\pa \fnl$ the derivative with respect to $\fnl$ evaluated at the fiducial value of $\fnl$.
Our default fiducial model is $f_{\rm NL, fid} = 0$.

In the special case where, for a given set of tracers $\{ a \}$, {\it all} possible auto- and cross-spectra are used, a useful alternative form for the Fisher information per mode is,
\beq
\label{eq:Fisheralt}
F(\q_L) = \frac{1}{2} \, {\rm Tr}\left[ {\bf P}^{-1}(\q_L) \, {\bf P}'(\q_L) \, {\bf P}^{-1}(\q_L) \, {\bf P}'(\q_L) \right],
\eeq
where ${\bf P}(\q_L)$ is the matrix of auto- and cross-spectra, with components,
\beq
{\bf P}_{ab}(\q_L) = P_{ab}(\q_L),
\eeq
and primes again denote derivatives with respect to $\fnl$.
This form can be very convenient as, for $n$ tracers, it only involves inversion of matrices of size $n \times n$,
while the form Eq.~(\ref{eq:fisher1}) requires inversion of a matrix of size $\tfrac{1}{2} n (n + 1) \times \tfrac{1}{2} n (n + 1)$.

Finally, even when the matrix ${\bf C}$ or ${\bf P}$ is not diagonal, it turns out that it can often still be analytically inverted in a simple way. The reason is that it can often be written in the form,
\beq
\label{eq:shermmorform}
{\bf A} + {\bf u} \, {\bf u}^T,
\eeq
with ${\bf A}$ a diagonal matrix and ${\bf u}$ a vector. Thus, the Sherman-Morrison formula \cite{Sherman-Morrison-1950,Bartlett-Sherm_Morr-1951} allows for its
inversion,
\beq
\label{eq:shermmor}
\left({\bf A} + {\bf u} \, {\bf u}^T \right)^{-1} = {\bf A}^{-1} - \frac{{\bf A}^{-1} \, {\bf u} \, {\bf u}^T \, {\bf A}^{-1}}{1 + {\bf u}^T \, {\bf A}^{-1} \, {\bf u}},
\eeq
which leads to simple analytic expressions because ${\bf A}$ is diagonal (see e.g.~\cite{hamausetal12}).
We will make extensive use of this throughout this work.

\subsection{Higher order statistics definitions}

The matter bispectrum and trispectrum are defined by,
\bea
\langle \delta(\k_1) \, \delta(\k_2) \, \delta(\k_3) \rangle &=& B(\k_1, \k_2, \k_3)   \,  (2\pi)^3 \, \delta^{(D)}(\k_1 + \k_2 + \k_3) \nonumber \\
\langle \delta(\k_1) \, \delta(\k_2) \, \delta(\k_3) \, \delta(\k_4) \rangle_c &=& T(\k_1, \k_2, \k_3, \k_4)  \, (2\pi)^3 \, \delta^{(D)}(\k_1 + \k_2 + \k_3 +  \k_4), \nonumber
\eea
where the subscript $c$ means we consider the ``connected'' four-point function, subtracting out contributions that are products of two-point functions.
The definitions for halo bi- and trispectra follow analogously.
In the following, we will always treat these higher order statistics as two-point functions of long-mode tracers.

\section{$\fnl$ from the matter bispectrum and trispectrum}
\label{sec:matter}

We first consider
the information contained in higher order statistics of the (low-redshift) matter overdensity, specifically the squeezed-limit matter bispectrum and collapsed trispectrum.
The key insight is that these quantities can be treated in terms of a {\it position-dependent power spectrum}, which, just like the number density of halos, is a biased tracer of long-wavelength matter perturbations.
The squeezed-limit bispectrum then is the cross-power spectrum of the position-dependent small-scale power spectrum in Fourier space with a long-mode density perturbation.
We consider ``measurements'' of the matter density (and later the halo density) in real space and do not include redshift space distortions.

\subsection{The position-dependent power spectrum}
\label{subsec:posdeppk}

Let us define $\hat{P}(\k_S; \x)$ as the power spectrum estimator for a single mode $\k_S$, estimated over some volume $V_S$ centered on the point $\x$,
and let
\beq
\hat{P}(\k_{i}; \x) \equiv \int_{\k_i} \frac{d^3 \k_S}{V_{\k,i}} \, \hat{P}(\k_S; \x)
\eeq
denote that same position-dependent power spectrum estimator, but averaged over a bin of short modes centered on $\k_i$, with Fourier-space volume $V_{\k,i}$.
We identify the modes $\k_S$ (and $\k_i$) measured inside $V_S$ as the {\it short} modes so that we choose $V_S \equiv k_{V_S}^{-3} \gtrsim k_{S,{\rm min}}^{-3}$ (cf.~Section \ref{subsec:modecoupling} and Figure \ref{fig:sepscales}).
We are now interested in the long-wavelength modulation of this position-dependent power spectrum. Transforming $\x$ to Fourier space, gives the quantity $\hat{P}(\k_{i}; \q_L) = \delta \hat{P}(\k_{i}; \q_L)$ (the equality follows because for $\q_L \ne 0$, the estimator $\hat{P}$ has zero expectation value). Since we have imposed the separation of scales,
$q_L \ll k_{S,{\rm min}}$, this means that we are by definition considering fluctuations on scales $\lambda_L \sim q_L^{-1} \gg V_S$.
In this limit, it is straightforward to show that (see Appendix \ref{app:app1}),
\beq
\label{eq:dPk est}
\delta \hat{P}(\k_{i}; \q_L) \approx \int_{\k_{i}} \frac{d^3 \k_S}{V_{\k,i}} \, \delta(\k_S) \, \delta(-\k_S + \q_L),
\eeq
where the integral is over the bin of short modes, and $V_{\k,i}$ is the Fourier-space volume of that bin.

The squeezed-limit bispectrum estimator for a long mode $\q_L$ and averaged over the bin of short modes centered on a mode $\k_i$ can be written as the cross-power spectrum of $\delta \hat{P}(\k_i; \q_L)$ with the matter overdensity $\delta(\q_L)$,
\beq
\hat{B}(\q_L, \k_i, - \k_i - \q_L)
\equiv
\int_{\k_{i}} \frac{d^3 \k_S}{V_{\k,i}} \, \hat{B}(\q_L, \k_S, - \k_S - \q_L)
= \hat{P}_{\delta, \delta \hat{P}(\k_i)}(\q_L)
\propto \delta(\q_L) \, \delta \hat{P}(\k_i; - \q_L) + c.c.
\eeq
where $c.c.$ indicates the complex conjugate to ensure the estimator is real.
Recall that in our notation, the use of $\k_i$ on the left hand side on the first line of the above expression implies averaging over a bin of short modes.

\vspace{0.2cm}

Let us now consider the long wavelength behavior of the position-dependent power spectrum.
Assuming the size of the short-mode bins is small compared to the mean wave number in the bin, i.e.~$\Delta k_S \ll k_i$, we get,
\beq
\label{eq:dPk qL}
\delta \hat{P}(\k_{i}; \q_L) = P(k_i) \, \left[ 4 \, \bar{F}_2(\hat{\k}_i \cdot \hat{\q}_L, k_i) \, \delta(\q_L) \\
+ 4 \, \fnl \, \M^{-1}(q_L) \, \delta(\q_L) + \epsilon_{\k_{i}}(\q_L) \right]
\eeq
where we have defined
\beq
\label{eq:meankernelmat}
\bar{F}_2(\mu, k) \equiv \frac{F_2(\q,-\k) \, P(k) + F_2(\q,\k-\q) \, P(|\k - \q|)}{2 P(k)}
= \frac{13}{28} + \left( \frac{2}{7} - \frac{1}{4} \, \frac{d\ln P(k_{i})}{d\ln k} \right) \, \mu^2
+ \mathcal{O}\left( \frac{q}{k} \right),   \quad  \mu \equiv \hat{k} \cdot \hat{q}.
\eeq
The first two terms in the parentheses on the right hand side of Eq.~(\ref{eq:dPk qL}) represent the mode-coupling due to non-linear evolution and PNG respectively.
This response to the long mode $\delta(\q_L)$ is obtained by taking the expectation value of Eq.~(\ref{eq:dPk est}) for fixed realization of the long mode, using the mode-coupling expressions in Section \ref{subsec:modecoupling}.
A powerful alternative method (closely connected to the position-dependent power spectrum picture)
is the ``separate Universe'' or ``power spectrum response'' approach, where the modulation
by a long-mode perturbation is realized as the response of the small-scale power spectrum to a rescaling of background Universe properties such as the spatial curvature and initial amplitude of perturbations (see e.g.~\cite{baldaufetal11,baldaufetal16a,Lietal16,BarSchmidt17,chiang17}).

\renewcommand{\arraystretch}{1.4}
\begin{table*}[t]
\small
\begin{center}
\begin{adjustbox}{max width=\textwidth}
\begin{tabular}{|l||l|l|l|}
\hline
Perturbation &  Description & Expression & Properties  \\
\hline \hline
$\delta_1(\q_L)$ & Matter overdensity & $\delta \rho_m(\q_L)/\bar{\rho}_m$ & Eq.~(\ref{eq:d1})  \\
$\delta_{2(i)}(\q_L)$  & Pos.-dep.~matter power spectrum  & $\delta \hat{P}(\k_i; \q_L)/P(k_i)$ & Eq.~(\ref{eq:bias dP}) \\
$\delta_h(\q_L)$  & Halo overdensity & $\delta n_h(\q_L)/\bar{n}_h$ & Eq.~(\ref{eq:haloprops}) \\
$\delta_{2h(i)}(\q_L)$ &  Pos.-dep.~halo power spectrum  & $\delta \hat{P}_{hh}(\k_i; \q_L)/\left(b_{10}^{(h) \, 2}P(k_i)\right)$ & Eq.~(\ref{eq:d2h}) \\
\hline
\end{tabular}
\end{adjustbox}
\end{center}
\caption{Summary of the long-wavelength tracers of which we study the information content. More details are given in the text: the matter overdensity $\delta_1$ and the (relative fluctuation in the) position-dependent matter power spectrum $\delta_{2(i)}$ are introduced in Section \ref{sec:matter}, the halo overdensity $\delta_h$ in Section \ref{sec:mat2 vs halo}, and the position-dependent halo power spectrum $\delta_{2h(i)}$ in Section \ref{sec:halos}.}
\label{table:t1}
\end{table*}

We have also included a stochastic noise contribution to the small-scale power spectrum measurement, $P(k_{i}) \, \epsilon_{\k_{i}}(\q_L)$, in analogy with the shot noise in the halo number density measurement.
This term appears due to the variance in the small-scale power spectrum measurement for fixed realization of the long mode, which is uncorrelated on scales much larger than $V_S^{1/3}$.
The covariance in the local matter power spectrum measurement (divided by $P(k_i)$) in the volume $V_S$ centered on $\x$ is,
\beq
\frac{\langle \delta \hat{P}(\k_{i}; \x) \, \delta \hat{P}(\k_{j}; \x) \rangle}{P(k_{i}) \, P(k_{j})} = \frac{2}{N_{\k,i}} \, \delta^{(K)}_{ij} = \frac{2 (2 \pi)^3}{V_{\k,i} \, V_S} \, \delta^{(K)}_{ij},
\eeq
where $N_{\k,i}$ is the number of independent modes in the bin $\k_i$ as measured in the local volume $V_S$
and we have assumed small, non-overlapping bins $\{ \k_i \}$.
For points $\x, \x'$ with separations much larger than $V_S^{1/3}$, the local power spectrum estimators
are uncorrelated, meaning that on these large scales, the stochastic noise $\epsilon_{\k_i}(\q)$ is described by a white noise power spectrum,
\beq
\label{eq:mat sn}
\langle \epsilon_{\k_i}(\q) \, \epsilon_{\k_j}(\q') \rangle
=  \left[ V_S \,\frac{\langle \delta \hat{P}(\k_{i}; \x) \, \delta \hat{P}(\k_{j}; \x) \rangle}{P(k_{i}) \, P(k_{j})}  \right]
\, (2 \pi)^3 \, \delta^{(D)}(\q + \q')
=  \left[ \frac{2 (2 \pi)^3}{V_{\k,i}} \, \delta^{(K)}_{ij} \right] \, (2 \pi)^3 \, \delta^{(D)}(\q + \q').
\eeq
We could have obtained the same result directly using Eq.~(\ref{eq:dPk est}).
We note that the above expression for the stochastic noise, and in particular the lack of correlation between position-dependent power spectra in different short-mode bins, uses our previously stated approximation that the short modes at fixed long mode realization are Gaussian. For an analysis incorporating short modes well into the non-linear regime, short-short mode-coupling (corresponding to trispectrum configurations $T(\k_S, -\k_S, \k_S', -\k_S')$) will become important and this assumption will break down. This will lower the information content relative to our calculation.

\vskip 10pt

{\bf Summary:} In the notation established in Section \ref{subsec:infocontent}, the position-dependent matter power spectrum is a set of biased tracers (with labels $2(i)$ indexing the bin of short modes $\k_i$) of the underlying large-scale matter density,
with relative perturbations given to leading order by,
\beq
\label{eq:dPk}
\delta_{2(i)}(\q_L) \equiv \delta \ln \hat{P}(\k_i; \q_L) = \frac{\delta \hat{P}(\k_i; \q_L)}{P(k_i)},
\eeq
and,
\beq
\label{eq:bias dP}
b_{2(i)}(\q_L) \equiv 4 \bar{F}_2(\hat{\k}_i \cdot \hat{\q}_L, k_i), \quad b_{2(i)}'(\q_L) \equiv 4 \M^{-1}(q_L),
\quad N_{2(i)} = \frac{2 (2 \pi)^3}{V_{\k,i}}.
\eeq
Note that $b_{2(i)}$ in general also contains a term proportional to the fiducial value of $\fnl$ which we have set to zero.

For the sake of notation, let us also write the long-mode {\it matter} overdensity itself as an unbiased tracer with index $1$,
\beq
\label{eq:d1}
\delta_1(\q_L) \equiv \delta(\q_L), \quad \text{with} \quad b_1 = 1, \quad b_1' = 0, \quad N_1 = 0.
\eeq
We summarize the tracers of which we consider the information content throughout this paper in Table \ref{table:t1}

\vskip 10pt

Before we continue to actual forecasts, we note that the position-dependent power spectrum method corresponds to a different perturbative expansion than in the standard approach to the bispectrum and higher order statistics.
In a conventional perturbation theory expansion, all modes of the matter overdensity $\delta$, long and short, are of the same order.
In our approach, however, the expansion is fundamentally in terms of {\it long} mode perturbations, with in particular the long-mode stochastic noise $\epsilon_{2(i),L}$ treated at the same order as the linear response to $\delta_L$ (and to $\phi_L$).
This is different because
in a conventional perturbative expansion, what we call $\epsilon_{2(i),L}$ is of zeroth order in perturbations, while $b_{2(i)} \delta_L$ (and $\fnl \, b_{2(i)}' \, \phi_L$) are of first order.

In our approach,
the leading order matter statistics are all $2$-point functions of $\delta_{1,L}$ and $\delta_{2(i),L}$, or in other words the matter power spectrum, squeezed bispectrum and trispectrum.
These $2$-point functions contain {\it all} information on $\fnl$ in the limit that (1) the long modes can be treated as Gaussian and (2) the short modes for fixed realization of long modes, i.e.~the $\tilde{\delta}_S$, are Gaussian  (in other words, when neglecting short-short mode-coupling).
As we shall see, it is because of the difference in expansion parameters that the position-dependent power spectrum approach
includes important non-Gaussian covariance of the bispectrum at the same order as the usual Gaussian contributions.
Note also that
our expansion in long modes is exactly what one applies in the standard treatment of halo clustering (see Section \ref{sec:mat2 vs halo}). There, the direct analogue of the position-dependent power spectrum is the halo number density and we indeed treat shot noise in the halo density (fundamentally also due to small-scale stochasticity) at the same order as the linear response to the matter overdensity even though they are of different order in $\delta$.

\renewcommand{\arraystretch}{1.4}
\begin{table*}[t]
\small
\begin{center}
\begin{adjustbox}{max width=\textwidth}
\begin{tabular}{|l||l|l|}
\hline
Spectrum &  Description & Equivalent expression   \\
\hline \hline
$\hat{P}_{11}(\q_L)$ & Matter power spectrum (PK or $mm$) & $\hat{P}(\q_L)$  \\
$\hat{P}_{12(i)}(\q_L)$ & Sq.-lim.~matter bispectrum (BK)  & $\hat{B}(\q_L, \k_i, -\k_i - \q_L)/P(k_i)$  \\
$\hat{P}_{2(i)2(j)}(\q_L)$ & Collapsed matter trispectrum (TK) & $\hat{T}(\k_i, -\k_i + \q_L, \k_j, -\k_j - \q_L)/P(k_i)/P(k_j)$  \\
\hline
\end{tabular}
\end{adjustbox}
\end{center}
\caption{The power and cross-spectrum estimators of long-mode perturbations considered in Section \ref{sec:matter}, and their equivalent expressions in terms of higher order matter statistics. The bispectrum and trispectrum in the third column are averages over bins of short modes $\k_{S}$ labeled by the index $i$ (with central wave vector $\k_i$). See text for more details.}
\label{table:t2}
\end{table*}

\subsection{Matter bispectrum information content - Formalism}

In the notation introduced above, the bispectrum is equivalent to the set of cross-spectra\\
$P_{12(i)}(\q_L)$ ($ = B(\q_L, \k_i, -\k_i - \q_L)/P(k_i)$), see Table \ref{table:t2}.
We can now use Eq.~(\ref{eq:fisher1}) to quantify the Fisher information per long mode on local non-Gaussianity.
Following Section \ref{subsec:infocontent}, the covariance matrix of the data vector is given by,
\beq
\label{eq:covmatbispec}
{\bf C}_{12(i),12(j)}(\q_L) = 2 b_{2(i)}(\q_L) \, b_{2(j)}(\q_L) \, P^2(q_L) + \delta_{ij}^{(K)} \, N_{2(i)} \, P(q_L).
\eeq
The common approximation of keeping only the Gaussian contributions to the covariance matrix (from here on ``Gaussian covariance approximation'' or GCA) corresponds
to setting the biases $b_{2(i)} = 0$ in the above equation. Thus, in the language of this paper, the Gaussian covariance approximation is equivalent to {\it ignoring the cosmic variance with respect to the long mode in the position-dependent power spectrum $\delta_{2(i)}(\q_L)$}.
In the conventional bispectrum language, it corresponds schematically to keeping only the terms of the form,
\beq
{\rm Cov}\left(\hat{B}(\q_L, \k_S, -\k_S - \q_L), \hat{B}(\q_L, \k_S', -\k_S' - \q_L)\right)
\sim P(q_L) \, P(k_S) \, P(k_S').\nonumber
\eeq
However,
this is not a well justified approximation\footnote{Note also that for the halo power spectrum (see Section \ref{sec:mat2 vs halo}), one typically does not and should not make the approximation of ignoring cosmic variance in the halo overdensity.} for short modes $\k_S$ in the (quasi-)non-linear regime.
Fortunately, the position-dependent power spectrum approach neatly allows us to go beyond the GCA by keeping the cosmic variance terms in Eq.~(\ref{eq:covmatbispec}).

In general (i.e.~even for $b_{2(i)} \neq 0$), since the matrix in Eq.~(\ref{eq:covmatbispec}) can be written in the ``Sherman-Morrison form'' (\ref{eq:shermmorform}), it can be analytically inverted using Eq.~(\ref{eq:shermmor}), giving,
\beq
{\bf C}^{-1}_{12(i),12(j)} = P^{-1}(q_L) \, N_{2(i)}^{-1} \, \delta^{(K)}_{ij}
- \frac{2 \, N^{-1}_{2(i)} b_{2(i)} \, N^{-1}_{2(j)} b_{2(j)}}{1 + 2 P(q_L) \sum_a N_{2(a)}^{-1} \, b_{2(a)}^2}.
\eeq
This leads to the Fisher information per mode,
\beq
\label{eq:fishermatbispec}
F(q_L) = \Sigma_2''(q_L) - \frac{2 \left( \Sigma_2'(q_L) \right)^2}{1 + 2 \Sigma_2(q_L)}
\quad \text{{\bf (matter bispectrum, BK)}},
\eeq
with
\bea
\Sigma_2(q) &\equiv& P(q) \, \sum_i N^{-1}_{2(i)} \, b_{2(i)}^2(\q) \to P(q) \, \int_S \frac{d^3 \k}{2 (2 \pi)^3} \, \left(  4 \bar{F}_2(\hat{\k} \cdot \hat{\q}, k) \right)^2 \nonumber \\
\Sigma_2'(q) &\equiv& P(q) \, \sum_i N^{-1}_{2(i)} \, b_{2(i)}(\q) \, b_{2(i)}'(\q)
\to P(q) \, \left( 4 \M^{-1}(q)\right)\, \int_S \frac{d^3 \k}{2 (2 \pi)^3} \, \left(  4 \bar{F}_2(\hat{\k} \cdot \hat{\q}, k) \right)
\nonumber \\
\Sigma_2''(q) &\equiv& P(q) \, \sum_i N^{-1}_{2(i)} \, \left( b_{2(i)}'(\q) \right)^2
\to P(q) \, \left( 4 \M^{-1}(q)\right)^2 \, \int_S \frac{d^3 \k}{2 (2 \pi)^3}.
\eea
On the right hand side of the arrows, we have inserted the specific bias and stochastic noise parameters. The integrals are over short modes.
The above quantities, and especially $\Sigma_2(q)$ should be seen as analogous to ``$\bar{n} P$'', the cosmic variance to shot noise ratio in galaxy clustering. Large values of these parameters correspond to the signal or cosmic variance dominated regime, whereas small values correspond to the (effective) shot noise dominated regime.

On the other hand, when the Gaussian covariance approximation is made, ignoring the cosmic variance in $\delta_{2(i)}$, the Fisher information simply becomes,
\beq
\label{eq:matbispecGCA}
F(q_L) = \Sigma_2''(q_L) \quad \text{{\bf (matter BK Gauss.~covariance approx. (GCA))}}.
\eeq
We have checked explicitly (see Appendix \ref{app:F GCA}) that the above expression reproduces (the squeezed limit of) the standard expression for the bispectrum Fisher information in the GCA.
Comparing the Fisher information in the GCA to the full result including non-Gaussian covariance shows, as expected, that including the cosmic variance term, always decreases the bispectrum information.

\vskip 7pt

In the conventional bispectrum language, the cosmic variance contribution to the covariance matrix in Eq.~(\ref{eq:covmatbispec})
describes non-Gaussian covariance between triangle configurations that share a long mode $\q_L$ but involve different short modes $\k_S$.
Schematically,
it captures terms of the type
\beq
{\rm Cov}\left(\hat{B}(\q_L, \k_S, -\k_S - \q_L), \hat{B}(\q_L, \k_S', -\k_S' - \q_L)\right)
\sim B(\q_L, \k_S', -\k_S' - \q_L) \, B(\q_L, \k_S, -\k_S - \q_L), \nonumber
\eeq
and
\beq
{\rm Cov}\left(\hat{B}(\q_L, \k_S, -\k_S - \q_L), \hat{B}(\q_L, \k_S', -\k_S' - \q_L)\right) \sim P(q_L) \, T(\k_S, -\k_S - \q_L, -\k_S', \k_S' + \q_L), \nonumber
\eeq
where $T$ is the trispectrum.
These contributions are generated by non-linear evolution encoded by the $b_{2(i)}$ parameters. Primordial mode-coupling would also contribute to the non-Gaussian covariance (see e.g.~\cite{cremetal07}), but we assume a fiducial $\fnl = 0$ throughout this work, so that the primordial contributions vanish.

Interestingly, the cosmic variance in Eq.~(\ref{eq:covmatbispec}) corresponds to ``super-sample covariance'' \cite{takadahu13,taksper14,schaanetal14} or ``beat coupling'' \cite{rimesham06,hamrimscoc06,RdPetal12} (which is the manifestation of super-sample covariance in the context of perturbation theory) in the small-scale power spectrum, with respect to the the local subvolume ($V_S$), generated by modes $\q_L$ on scales larger than $V_S$.
Note however that we do not include super-sample covariance/beat coupling due to modes larger than the {\it full} survey size $V$. This is justified because we focus on long modes $q_L$ well inside the linear regime.

\begin{figure*}[]
\centering
\includegraphics[width=0.74\textwidth]{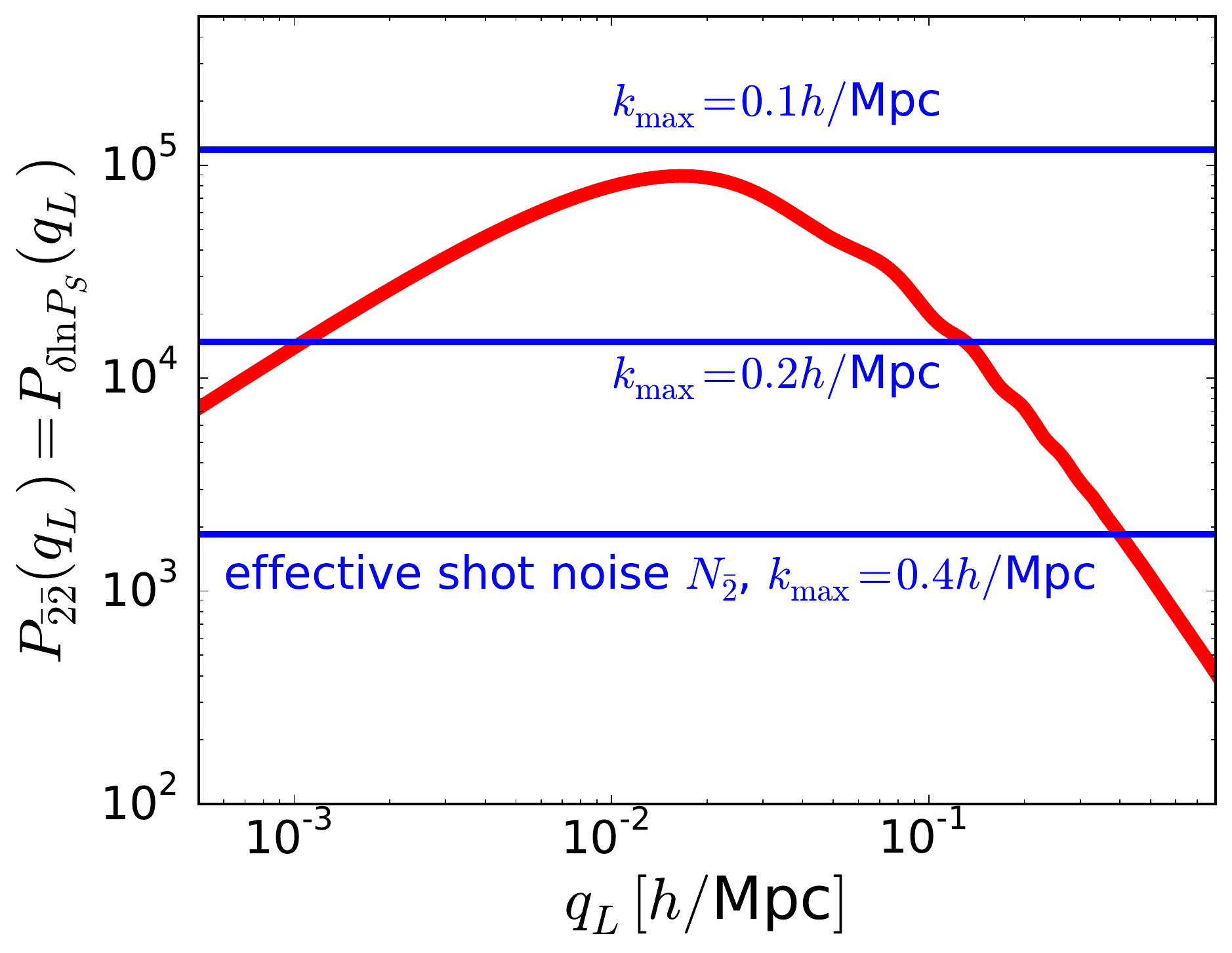} 
\caption{Power spectrum (red) and stochastic noise spectrum (blue) of the position-dependent small-scale power spectrum averaged over short modes, $\delta_{\bar{2}}(\q_L)$. We assume a fiducial redshift $z = 1$ and $\langle n^{\delta}_s(k_S) \rangle = -2.5$ (see Appendix \ref{app:d2bar} for details). The effective shot noise, $N_{\bar{2}}$, is due to the variance in realizations of the short modes and is thus smaller for larger $k_{\rm max}$ where one averages over more modes. The power spectrum gives rise to the cosmic variance contributions to the matter bispectrum covariance (Eq.~(\ref{eq:covmatbispec})), which are not included in the Gaussian covariance approximation. The figure shows that for $k_{\rm max} \gtrsim 0.1 h/$Mpc, ignoring the cosmic variance contribution is not a good approximation.}
\label{fig:CV vs SN}
\end{figure*}

\subsection{Matter bispectrum information content - Results}
\label{subsec:matbk res}

Let us now consider quantitatively the difference between the information content using the full covariance and the GCA. As a default choice, we will throughout this paper assume a survey
with volume $V = 100 \, (h^{-1} $Gpc$)^3$ at an effective redshift $z = 1$.
This volume approximately corresponds to a full-sky sample in the redshift range $z = 0.5 - 1.5$.
We note that $V = 100 \, (h^{-1} $Gpc$)^3$ is less than the volume covered by galaxy samples expected from some planned/proposed surveys and, accordingly, it will in principle be possible to reach lower $\sigma(\fnl)$ than the values plotted throughout this paper.

We study bispectrum configurations defined by the ``shortest long mode'' wave number, $q_{\rm max} = 0.02 h/$Mpc, and the "longest short mode" wave number, $k_{\rm min} = 0.02 h/$Mpc. Technically this means we include configurations where $q_L$ is not much smaller than $k_S$, which are not squeezed at all, and we are pushing the range of scales included a bit beyond the schematic illustration in Figure \ref{fig:sepscales}. However, in practice, the information content is dominated by the squeezed configurations and is only weakly dependent on the choices of $q_{\rm max}$ and $k_{\rm min}$ (see Appendix \ref{app:scaledep}).
Our default choice for the longest mode included is $q_{\rm min} = 10^{-3} \, h/$Mpc.
This is an appropriate (slightly conservative) scale given the survey volume $V = 100 \, (h^{-1}$Gpc$)^3$ and the rule of thumb,
$q_{\rm min} \approx \pi/V^{1/3}$.
We note that the absolute values we obtain for $\sigma(\fnl)$ are strongly dependent on fiducial survey volume and other parameters and that our main focus throughout this paper will be on the relative {\it differences} in $\sigma(\fnl)$ between different probes and approaches.

Before turning to the consequences for the constraining power on $\fnl$, it is instructive to directly compare the cosmic variance and stochastic noise contributions to the fluctuations in the position-dependent power spectrum.
For convenience, let us consider the modulation of the small-scale power spectrum {\it averaged} over all short modes, see Appendix \ref{app:d2bar}. This corresponds to
averaging the multiple tracers $\delta_{2(i)}(\q_L)$ together into a single mean tracer, $\delta_{\bar{2}}(\q_L)$. Figure \ref{fig:CV vs SN} shows the fiducial power spectrum of position-dependent power spectrum fluctuations, $P_{\bar{2}\bar{2}}(q_L) = b_{\bar{2}}^2 \, P(q_L)$ (red) and the effective shot noise $N_{\bar{2}}$ (blue) for three choices of the maximum wave number (see Appendix \ref{app:d2bar} for details).
It is clear from the Figure that for the long modes of interest, the cosmic variance contribution
is {\it not} negligible compared to the effective shot noise, especially for $k_{\rm max} = 0.2 h/$Mpc and
$0.4 h/$Mpc. It is therefore important to consider the information content calculation that includes the cosmic variance, Eq.~(\ref{eq:fishermatbispec}), and one expects significant differences between the full calculation and the Gaussian covariance approximation, which ignores the cosmic variance.

\begin{figure*}[]
\centering
\includegraphics[width=0.48\textwidth]{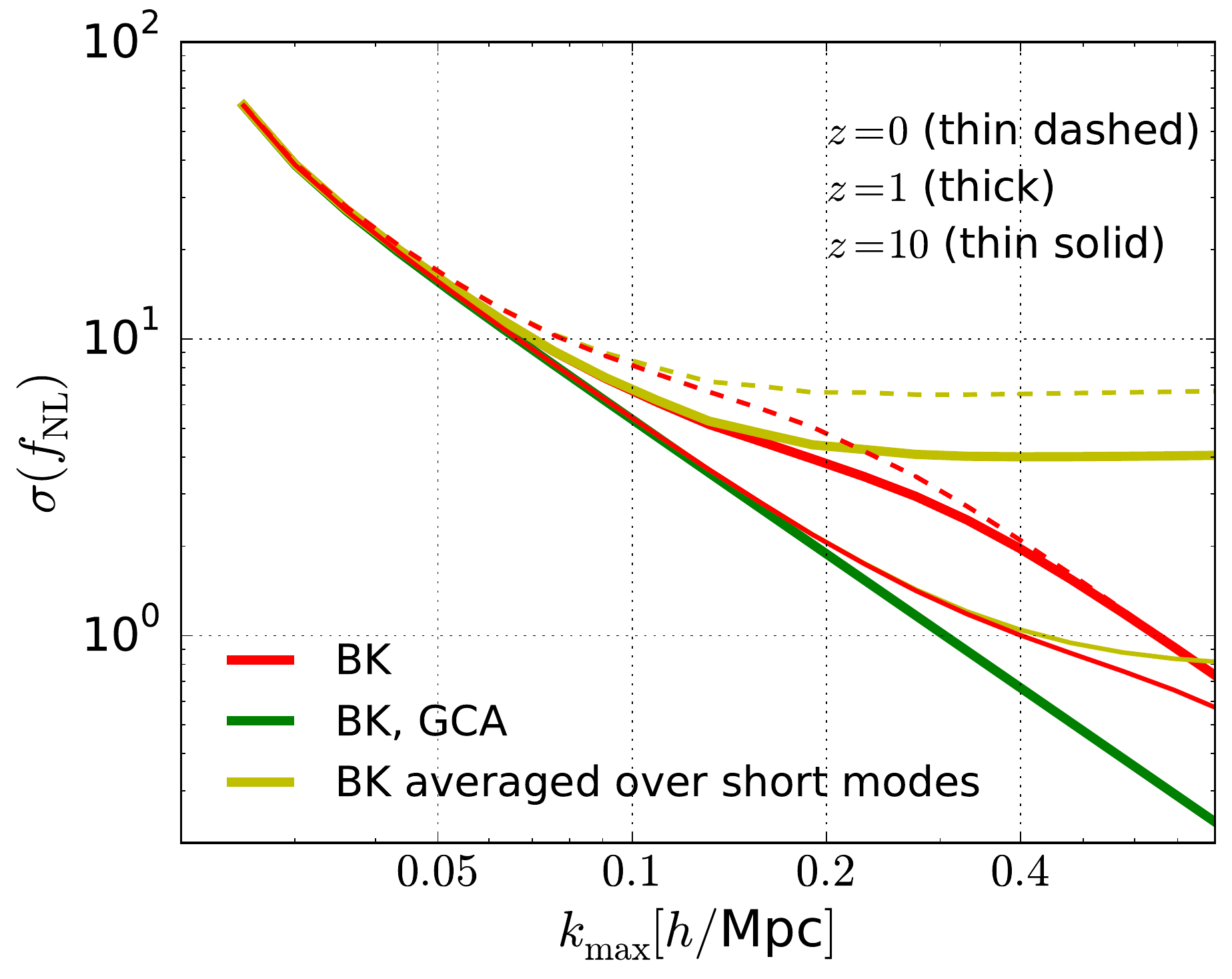} 
\includegraphics[width=0.48\textwidth]{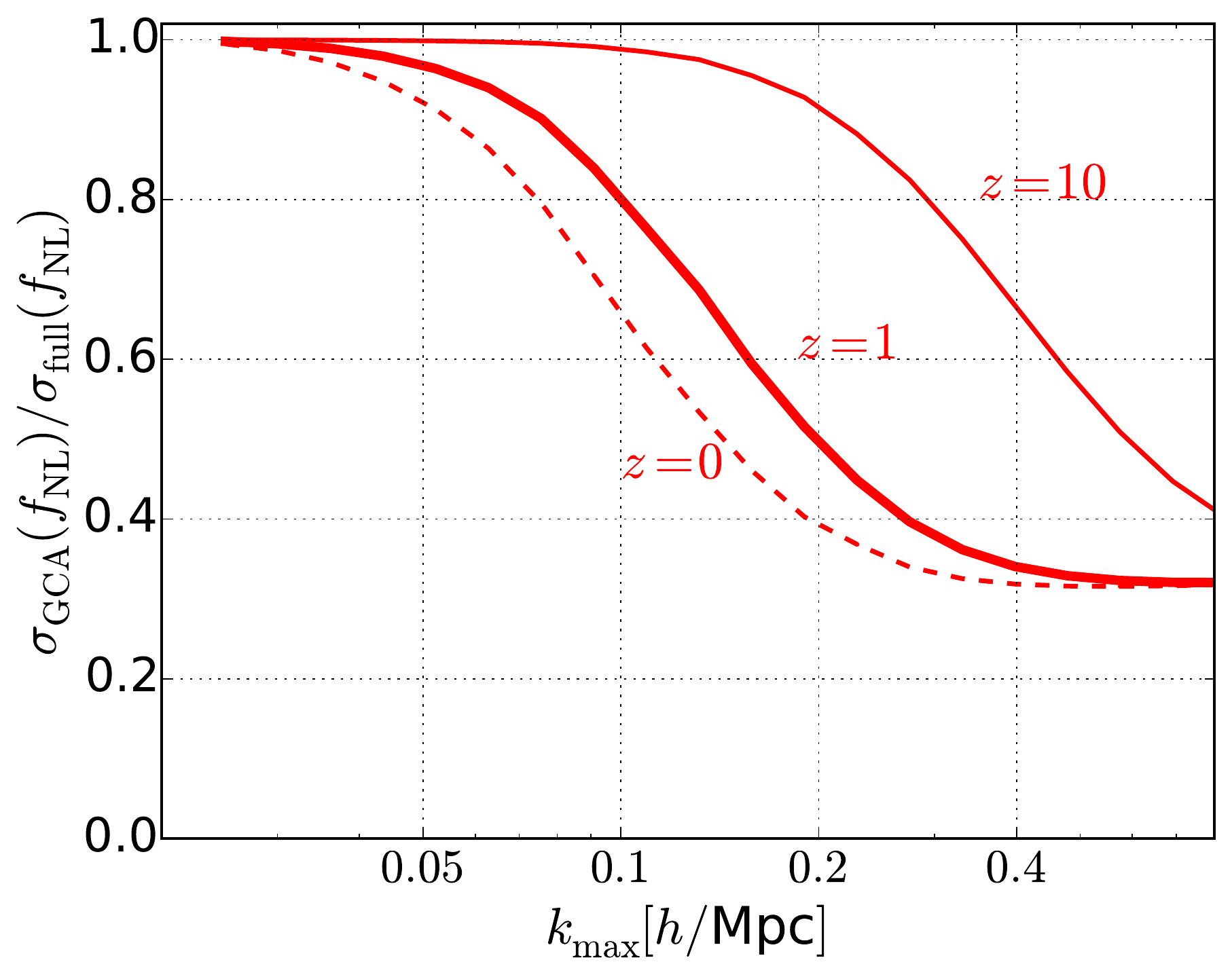} 
\caption{{\it Left:} Comparison of $\sigma(\fnl)$ from matter bispectrum using full covariance (red) and Gaussian covariance approximation (green), as function of maximum short mode wave number, $k_{\rm max}$, for fixed $q_{\rm min} = 0.001 h/$Mpc.
The yellow curve shows the information content in the bispectrum averaged over the short modes (see text).
We assume a survey with volume $100 \, (h^{-1}$Gpc$)^3$ and effective redshift $z = 1$ (thick curves), and we fixed the maximum long mode and minimum short mode wave numbers to $q_{L,{\rm max}} = 0.02 h/$Mpc, $k_{S,{\rm min}} = 0.02 h/$Mpc.
The thin curves show the same quantities but for $z=0$ (dashed) and $z=10$ (solid), with lower redshift corresponding to larger $\sigma(\fnl)$.
{\it Right:} As left panel, but ratio of the bispectrum based uncertainties with the GCA to the full uncertainty. This is the factor by which the common calculation based on GCA underestimates the uncertainty. Asymptotically, the GCA underestimates the true uncertainty by a factor $\sim 3$. Note that, at large $k_{\rm max}$, there are additional relevant non-linear/non-Gaussian contributions to the signal and noise that we have not taken into account.}
\label{fig:sigma bispec}
\end{figure*}

In the left panel of Figure \ref{fig:sigma bispec},
we show the unmarginalized uncertainty on $\fnl$ from the matter bispectrum as a function of $k_{\rm max}$ (thick curves for the default $z = 1$).
Comparing the GCA (green) to the full result (red), we indeed find that the GCA strongly underestimates the uncertainty on $\fnl$ (i.e.~overestimates the information content) once the non-linear regime is entered. The right panel of Figure \ref{fig:sigma bispec} shows that, compared to the GCA, the true error bar is $20 \%, \, 90 \%, \, 180 \%$ larger for $k_{\rm max} = 0.1, \, 0.2, \, 0.4 \, h/$Mpc, respectively.
In the left panel, we also include the constraining power of the bispectrum {\it averaged} over short modes (i.e.~the signal in the cross-power spectrum $P_{1\bar{2}}(q_L)$), which we come back to in Section \ref{subsec:cancel}.

Considering next the redshift dependence, the thin curves in Figure \ref{fig:sigma bispec} show the ``exact'' bispectrum information content for $z = 0$ (dashed thin line), and $z = 10$ (solid thin). Note that the GCA information content in the GCA, Eq.~(\ref{eq:matbispecGCA}), is manifestly redshift independent. As $k_{\rm max}$ is increased, the full bispectrum uncertainty on $\fnl$ starts to diverge from the GCA result around the non-linear scale, which can clearly be seen to lie at larger $k_{\rm max}$ (smaller scale) for higher redshifts.
Interestingly, asymptotically, the GCA underestimates the error bar by a factor $\sim 3$, independent of redshift.

\begin{figure*}[]
\centering
\includegraphics[width=0.6\textwidth]{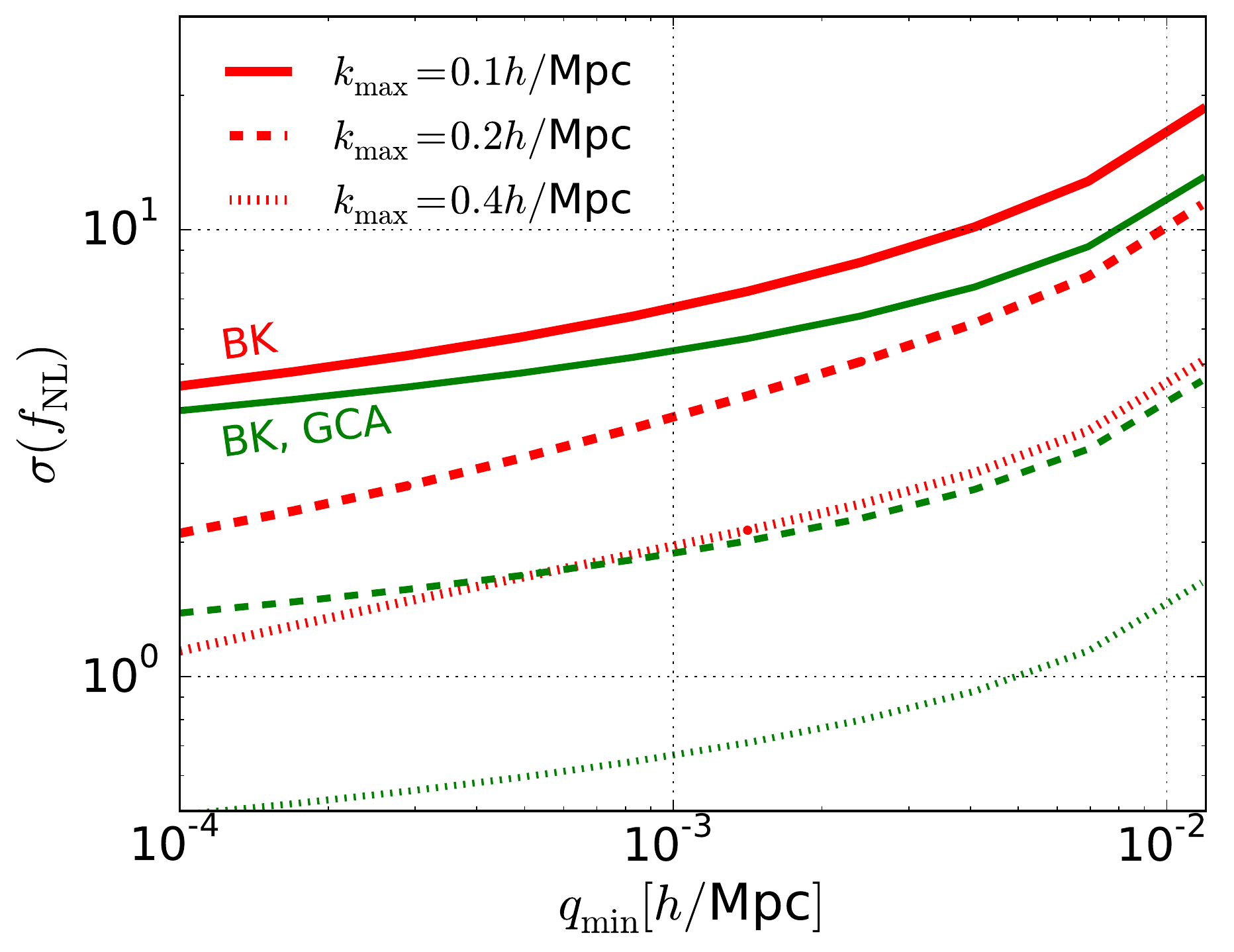} 
\caption{The uncertainty $\sigma(\fnl)$ from the matter bispectrum with the full covariance (red) and in the Gaussian covariance approximation (green), as a function of the minimum long mode wave number $q_{\rm min}$, for $k_{\rm max} = 0.1$ (solid), $0.2$ (dashed) and $0.4 h/$Mpc (dotted). Survey properties as in Figure \ref{fig:sigma bispec}.
}
\label{fig:bispec}
\end{figure*}

Figure \ref{fig:bispec} shows the Fisher information for fixed values of $k_{\rm max}$ as a function of $q_{\rm min}$ (for the default redshift, $z = 1$).
This confirms the importance of the making $q_{\rm min}$ as small as possible (including the largest modes) in the analysis.
While the $k_{\rm max}$ dependence of $\sigma(\fnl)$ might suggest that the bispectrum information content is dominated by short modes, $k_{\rm max}$ only defines the maximum short-mode wave number. Thus, in reality, a large fraction of the information comes from very squeezed triangles, with the long mode as long as possible and the short mode as short as possible.
We discuss the scale and configuration dependence of the information content in more detail in Appendix \ref{app:scaledep}.

\vskip 10pt

{\it {\bf Interpretation:}}
The information in the Gaussian covariance approximation equals the information content in the {\it primordial} field, i.e.~in the absence of non-linear evolution due to gravity. This can be seen from Eq.~(\ref{eq:fishermatbispec}) by noting that in the high-redshift limit, the quantities $\Sigma_2 \propto P(q_L)$ and $\Sigma_2' \propto \M^{-1}(q_L) \, P(q_L) = P_{\delta \phi}(q_L)$ go to zero, while $\Sigma_2'' \propto \M^{-2}(q_L) \, P(q_L) = P_{\phi \phi}(q_L)$ stays constant (remember that $\phi$ is the {\it primordial} metric perturbation). It can also be seen from the $z = 10$ (red) curves in Figure \ref{fig:sigma bispec}.
We thus treat it as the total information that is {\it in principle} available in the matter mode-coupling,
\beq
\label{eq:total info}
F(q_L) = \Sigma_2''(q_L).
\quad \text{{\bf (primordial information matter mode-coupling)}}.
\eeq
Therefore, the ratio of errors in the right panel of Figure \ref{fig:sigma bispec} on the one hand told us on the one hand how much the Gaussian covariance approximation overestimates the bispectrum information content, but on the other hand it tells us the degradation in bispectrum information content (relative to the total encoded information) due to non-linear evolution.

\subsection{Canceling cosmic variance by going beyond the bispectrum}
\label{subsec:cancel}

As stated above, in the approach of this paper, the information degradation in the non-linear regime relative to GCA is due to (long mode) cosmic variance in the tracers $\delta_{2(i)}$.
Increasing $k_{\rm max}$ corresponds to decreasing the stochastic noise $N_{2(i)}$, cf.~Eq.~(\ref{eq:bias dP}), analogous to increasing the number density of a halo sample.
Thus, for large enough $k_{\rm max}$, one ends up in the regime where cosmic variance comes to dominate over shot noise and this is where the degradation kicks in.
In studies of the clustering of multiple biased tracers, it is well known that the effect of cosmic variance can (partially) be canceled by cleverly combining the samples.
We can thus ask the same question here: can we cancel cosmic variance and recover the total information given by the dashed curve?

Before we address this question, note that, from the $k_{\rm max}$ dependence of the full result in the left panel of Figure \ref{fig:sigma bispec}, we see that, despite the degradation in information at large $k_{\rm max}$ (again, low effective shot noise), the constraint can be improved indefinitely as the shot noise $N_{2(i)}$ is lowered further.
The reason is that, since we are using the cross-correlations between $\delta_1$ and the {\it multiple} tracers, $\delta_{2(i)}$, the standard bispectrum analysis already applies a certain amount of cosmic variance cancellation.
Thus, one ends up with very similar to the cosmic variance cancellation in a multitracer analysis of multiple halo samples: as the (effective) shot noise is lowered, one first reaches a ``plateau'' and then the cosmic variance cancellation kicks in leading to significantly improved constraints.

We can see this partial cosmic variance cancellation explicitly by considering the bispectrum information in  Eq.~(\ref{eq:fishermatbispec}) in the cosmic variance dominated limit, $\Sigma_2, \Sigma_2', \Sigma_2'' \gg 1$ (i.e.~$k_{\rm max}^3 \, P(q_L) \gg 1$). In general, the equation can be slightly rewritten as,
\beq
\label{eq:fishermatbispec2}
F(q_L) = \frac{\Sigma_2''(q_L) + 2 \left(\Sigma_2(q_L) \, \Sigma_2''(q_L) - \left( \Sigma_2'(q_L) \right)^2 \right)}{1 + 2 \Sigma_2(q_L)}.
\eeq
If the bias parameters $b_{2(i)}$ were the same for all $i$, we would have $\Sigma_2 \, \Sigma_2'' = \left( \Sigma_2' \right)^2$, and in the cosmic variance dominated limit, the information per long mode would converge to,
\beq
F(q_L) \to \frac{\Sigma_2''(q_L)}{2 \Sigma_2(q_L)},
\eeq
i.e.~it would reach a maximum as $k_{\rm max}$ is increased.
It is the fact that the term in brackets in the numerator of Eq.~(\ref{eq:fishermatbispec2}) is non-zero that causes the partial cosmic variance cancellation. Concretely, rewriting (note that the $b_{2(i)}'$ parameters below are actually independent of $i$),
\bea
\Sigma_2(q_L) &=& N_{\bar{2}}^{-1} \, P(q_L) \, \bar{b_{2}^2},\, \Sigma_2'(q_L) =
N_{\bar{2}}^{-1} \, P(q_L) \, \bar{b_{2}} \, b_{2(i)}'(q_L), \nonumber \\
\Sigma_2''(q_L) &=& N_{\bar{2}}^{-1} \, P(q_L) \,  \left(b_{2(i)}'(q_L)\right)^2,
\eea
where the bars are shorthand for averages over all short modes (equivalently, over all $i$) of $b_{2(i)}^2$
and $b_{2(i)}$, and $N_{\bar{2}} \sim k_{\rm max}^{-3}$ is the {\it total} effective shot noise, cf.~Eq.~(\ref{eq:d2bar}), we then have in the cosmic variance dominated limit,
\beq
\label{eq:BK asymp}
F(q_L) \to \frac{\bar{b^2_2} - (\bar{b_2})^2}{\bar{b^2_2}} \, \Sigma_2''(q_L).
\eeq
Thus, the cosmic variance (or, in conventional parlance, the non-Gaussian contributions to the bispectrum covariance), suppresses the Fisher information by the relative difference between the average squared bias and the average bias squared. For the position-dependent matter power spectrum biases, $b_{2(i)}$, this quantity amounts to $\left(\bar{b^2_2} - (\bar{b_2})^2\right)/\bar{b^2_2} \approx 1/3$, thus explaining the behavior seen in Figure \ref{fig:sigma bispec}.

By contrast, we can replace the set of tracers $\delta_{2(i)}$ by a single averaged tracer, $\delta_{\bar{2}}$, and consider the cross-spectrum $P_{1\bar{2}}(q_L)$, which is now equal to the bispectrum {\it averaged} over short modes  (see Appendix \ref{app:d2bar}). In this case, cosmic variance cancellation should not be possible.
We show the full (i.e.~including non-Gaussian covariance) information content of this averaged bispectrum with the light green curve in Figure \ref{fig:sigma bispec} (left panel only). The thick curve is again for $z = 1$ and the two thin curves represent $z = 0$ (top), $z = 10$ (bottom).
The key point is that the information content in this averaged bispectrum is optimal in the effective shot noise dominated regime (low $k_{\rm max}$), but reaches a plateau in the zero shot noise limit (just like the single-tracer halo power spectrum information content, see Section \ref{sec:mat2 vs halo}). This is exactly what one expects due to cosmic variance in the absence of cosmic variance cancellation.

\vskip 10pt

{\bf Joint power spectrum, bispectrum and trispectrum:}
Now let us return to the full (not averaged) bispectrum and see if we can recover the total information content by further canceling cosmic variance.
A complete multitracer study of the tracers $\delta_1$ and $\delta_{2(i)}$ entering the bispectrum, would, in addition to the cross-spectrum $P_{12(i)}$ (the bispectrum), include the other correlations, $P_{11}$ (the matter power spectrum) and\footnote{Note that the standard stochastic noise subtraction in the
(cross-)power spectrum estimator,
\beq
\hat{P}_{2(i) 2(j)} \to \hat{P}_{2(i) 2(j)} - N_{2h(i)} \, \delta_{ij}^{(K)}, \nonumber
\eeq
exactly corresponds to subtracting out the disconnected part of the four-point function, so that the resulting $\hat{P}_{2(i)2(j)}$ indeed estimates the connected four-point function or trispectrum.} $P_{2(i)2(j)}$ (the collapsed limit trispectrum).
Using Eq.~(\ref{eq:Fisheralt}) for the information content of an analysis including all these combinations, and again taking advantage of the Sherman-Morrison formula, it is straightforward to derive that the full information content from a joint analysis of these quantities is,
\beq
\label{eq:fullinfo}
F(q_L) = \Sigma_2''(q_L) \quad \quad \text{{\bf (matter PK$+$BK$+$TK)}}.
\eeq
Therefore, the ``multitracer'' analysis of combining power, bi- and trispectrum recovers the full constraining power on $\fnl$, Eq.~(\ref{eq:matbispecGCA}).

\vskip 10pt

{\bf Explicit cosmic variance cancellation:}
As an alternative to a full joint analysis, the primordial information content can be recovered by creating an optimal estimator that is a linear combination of the power and bispectrum.
Specifically, since the cosmic variance degradation in the bispectrum is due to the non-zero values of the biases $b_{2(i)}(\q)$, one can define a new tracer,
\beq
\delta_{\tilde{2}(i)}(\q) \equiv \delta_{2(i)}(\q) - b_{2(i)}(\q) \, \delta_1(\q).
\eeq
By construction, this new tracer has the fiducial bias subtracted out,
\beq
b_{\tilde{2}(i)}(\q) = 0, \quad b_{\tilde{2}(i)}'(q) = b_{2(i)}'(q), \quad N_{\tilde{2}(i)} = N_{2(i)}.
\eeq
If we now consider the information in the cross-spectra, $P_{1 \tilde{2}(i)}$, we again find that it is optimal ($F = \Sigma_2''$).
In more standard language this means that the information lost due to non-Gaussian contributions to the covariance matrix can be retrieved by using the ``cosmic variance canceling'' combination,
\bea
\label{eq:bispecminusCV}
\tilde{B}(\q_L, \k_S, -\q_L - \k_S)
&\equiv& \hat{B}(\q_L, \k_S, -\q_L - \k_S) - b_{2(i)}(\q_L) \, P(k_S) \, \hat{P}(\q_L)
\nonumber \\
&=& \hat{B}(\q_L, \k_S, -\q_L - \k_S) - \left( \frac{13}{28} + \left( \frac{2}{7} - \frac{1}{4} \, \frac{d\ln P(k_S)}{d \ln k} \right) \, \left(\hat{\q}_L \cdot \hat{\k}_S \right)^2 \right) \, P(k_S) \, \hat{P}(\q_L).
\eea

\vskip 10pt

{\bf Matter trispectrum only:}
Since we have considered the constraining power of the bispectrum and of a joint power, bi- and trispectrum analysis, it is also interesting to consider the information in the trispectrum alone. The (collapsed-limit) trispectrum is equivalent to the set of cross-spectra $P_{2(i)2(j)}(\q_L)$ so that we can use Eq.~(\ref{eq:Fisheralt}) combined with the Sherman-Morrison formula to obtain,
\beq
\label{eq:fishertri}
F(q_L) = \frac{\Sigma_2(q_L) }{1 + \Sigma_2(q_L)} \, \Sigma_2''(q_L) + \frac{1 - \Sigma_2(q_L)}{\left( 1 + \Sigma_2(q_L)\right)^2} \, \left( \Sigma_2'(q_L)\right)^2
\quad \text{{\bf (matter trispectrum, TK)}}
\eeq
(cf.~Eq.~(21) of \cite{hamausetal12}, applied to the case of multiple halo samples).

\begin{figure*}[]
\centering
\includegraphics[width=0.74\textwidth]{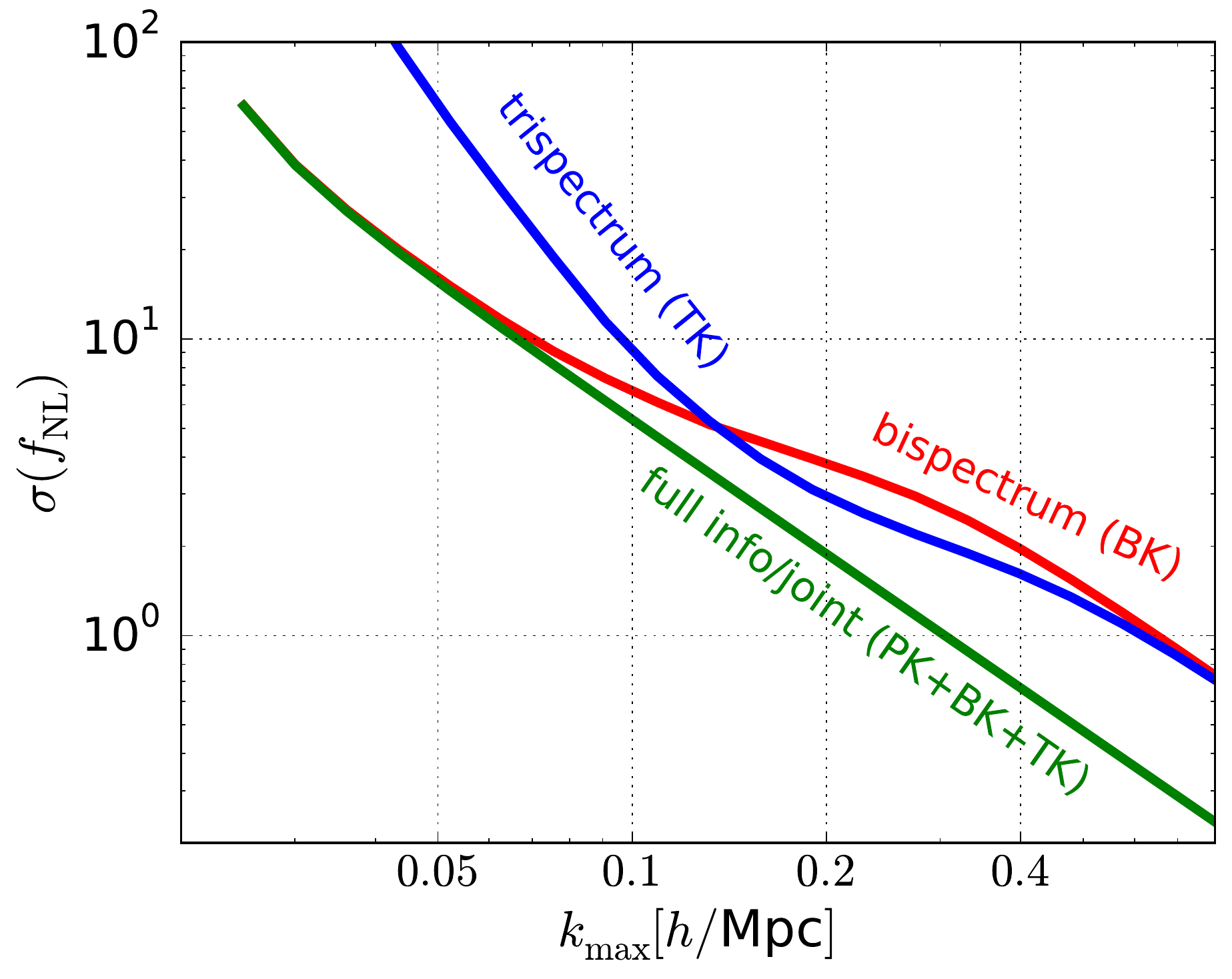} 
\caption{The uncertainty on $\fnl$ from the squeezed-limit matter bispectrum, the collapsed matter trispectrum, and
a joint analysis combining the bi- and trispectrum with the matter power spectrum (which on its own does not constrain $\fnl$). We assume a survey with volume $V = 100 \, (h^{-1} $Gpc$)^3$ at effective redshift $z = 1$, but note that future surveys may probe even larger volumes.
Long modes in the bi- and trispectra analyses range from $q_{\rm min} = 0.001 h/$Mpc to $q_{L,{\rm max}} = 0.02 h/$Mpc and short modes from $k_{S,{\rm min}} = 0.02 h/$Mpc to $k_{\rm max}$ on the horizontal axis.
For $k_{\rm max} \gtrsim 0.1 h/$Mpc, the matter bispectrum (red) does not contain the full/primordial mode-coupling information content of the matter density (green) because of the cosmic variance due to non-linear evolution (see text). The trispectrum on its own (blue) also does not capture the full information content. The full information can be recovered by canceling cosmic variance with a joint analysis of bispectrum/trispectrum and the matter power spectrum.}
\label{fig:sigma allmat}
\end{figure*}

\subsection{Information content matter statistics - Results}

We compare the information in the bispectrum, trispectrum, and the joint analysis in Figure \ref{fig:sigma allmat} (the matter power spectrum alone does not contain $\fnl$ information at leading order).

The full information on $\fnl$ contained in primordial long-short mode coupling, as given by Eq.~(\ref{eq:total info}), is shown in green.
As discussed above, this quantity is equivalent to the information content in the {\it primordial} (squeezed-limit) bispectrum, i.e.~before non-linear evolution. In our approach, it also corresponds to the information in the low-redshift bispectrum if we ignore the long-mode cosmic variance in $\delta_{2(i)}(\q)$, i.e.~the low-redshift bispectrum information in the GCA.
The bispectrum information including the cosmic variance contribution is given by Eq.~(\ref{eq:fishermatbispec}) and shown in red. For $k_{\rm max}$ in the linear regime (large effective shot noise in the tracer $\delta_{2(i)}(\q)$ as compared to the cosmic variance), the bispectrum retrieves the full information content, but as $k_{\rm max}$ enters the mildly non-linear regime (low shot noise), information is lost from the bispectrum due to cosmic variance. In the zero shot noise limit ($k_{\rm max} \to \infty$), the bispectrum information content is unbounded due to partial cosmic variance cancellation intrinsic in the bispectrum approach, but always less than the total information content by about a factor three).
The collapsed trispectrum information, given by Eq.~(\ref{eq:fishertri}), is shown in blue. For small $k_{\rm max}$, the trispectrum contains very little information on $\fnl$. The reason is that, for zero $b_{2(i)}$, the derivative of the trispectrum w.r.t.~$\fnl$ is proportional to the fiducial $\fnl$ value. This means that in the limit of zero mode-coupling from non-linear evolution, and in the absence of primordial non-Gaussianity in the fiducial model ($f_{\rm NL, fid} = 0$), the Fisher information tends to zero.
In the opposite limit, $k_{\rm max} \to \infty$, the trispectrum gives the same $\fnl$ constraint as the bispectrum.

This behavior in the cosmic variance dominated regime, $\Sigma_2, \Sigma_2', \Sigma_2'' \gg 1$ can be understood along the same lines as the discussion in the beginning of Section \ref{subsec:cancel} for the bispectrum.
Rewriting the Fisher information for the trispectrum, Eq.~(\ref{eq:fishertri}) as,
\beq
F(q_L) = \frac{\Sigma_2(q_L) \, \Sigma_2''(q_L) + \left( \Sigma_2'(q_L) \right)^2 + \Sigma_2(q_L) \, \left(\Sigma_2(q_L) \, \Sigma_2''(q_L) - \left( \Sigma_2'(q_L) \right)^2 \right)}{\left( 1 + \Sigma_2(q_L) \right)^2},
\eeq
we see that if $\Sigma_2 \, \Sigma_2'' = \left(\Sigma_2' \right)^2$ (equivalently, $\bar{b_2^2} = (\bar{b_2})^2$), then the two first terms in the denominator would dominate and the Fisher information would reach a maximum in the $k_{\rm max} \to \infty$ limit. However, since $\Sigma_2 \, \Sigma_2'' \ne \left(\Sigma_2' \right)^2$, the term in the brackets dominates, and leads to the exact same expression in the cosmic variance dominated limit as for the bispectrum, Eq.~(\ref{eq:BK asymp}).

The total information can be recovered by applying cosmic variance cancellation in the same way as is done in a multitracer analysis of multiple halo samples. In this case, either a joint analysis of the matter power spectrum, bispectrum and trispectrum, or even just a specific cosmic variance-free combination of the bispectrum and power spectrum (Eq.~(\ref{eq:bispecminusCV})), would return the optimal constraint shown in the green curve.

\section{$\fnl$ from scale-dependent halo bias}
\label{sec:mat2 vs halo}

Let us now compare the information content in measurements of the squeezed-limit higher order statistics of the matter field discussed in the previous section to that in scale-dependent halo bias.
It is clear at this point that these signals are formally extremely similar.
Both approaches exploit biased tracers of long mode matter perturbations, where the bias depends on $\fnl$ due to sensitivity of the tracer to the local primordial amplitude of perturbations.
In the former case, the tracers is effectively the position dependent small-scale matter power spectrum, with relative fluctuations $\delta \ln \hat{P}(\k_i;\x)$, while in the case of scale-dependent bias, it is the position-dependent number density of halos, with fluctuations $\delta \ln n_h(\x)$.
In the separate-Universe picture, both quantities have expectation values determined by the local primordial small-scale power spectrum, which explains their modulation by long wavelength primordial potential fluctuations $\phi_L(\x)$ in the presence of primordial non-Gaussianity, and both quantities are sensitive to the local spatial curvature, which explains their modulation by the long-mode matter overdensity $\delta_L(\x)$. For a given realization of the long modes, both quantities also have a random scatter (the effective shot noise) due to variance in the realization of the small-scale modes.

To first order, the halo overdensity can be written as (we refer to \cite{DJSreview16} for a review on halo biasing),
\beq
\label{eq:deltahlong}
\delta_h \equiv \frac{\delta n_h}{\bar{n}_h} = b^{(h)}_{10} \, \delta + \fnl \, b^{(h)}_{01} \, \phi + \epsilon_h.
\eeq
Here\footnote{We use the superscript $(h)$ to distinguish the halo bias parameters from the general bias parameters describing the tracers that enter our forecasts.} $b^{(h)}_{10}$ is the linear, Eulerian halo bias
and $b^{(h)}_{01}$ describes the response to the primordial potential fluctuation due to primordial non-Gaussianity\footnote{Technically, the non-Gaussian linear bias is proportional to $\fnl - \fnl^{1-{\rm field}}$, where $\fnl^{1-{\rm field}} = -5/12 (n_s - 1)$ is the single-field prediction. In particular, there is exactly zero physical scale-dependent bias in single-field inflation \cite{RdPDoreGreen15,daietal15}. For $\fnl$ values within near-future observational reach, the above correction is small and we will ignore it in this work.} (with $\fnl$ factored out). This bias can be written,
\beq
\label{eq:dh long}
b^{(h)}_{01} = 4  \, \frac{d\ln \bar{n}_h}{d\ln \sigma^2_R} = 2  \, (b^{(h)}_{10} - 1) \, \delta_c,
\eeq
where
$d\ln \bar{n}_h/d\ln \sigma^2_R$ is the response of the background halo number density $\bar{n}_h$ to a variation in the initial variance of fluctuations $\sigma^2_R$ on some scale $R$ characteristic of those halos.
To obtain the second equality, in which $\delta_c \approx 1.686$ is the critical overdensity for spherical collapse, we have implicitly assumed a universal halo mass function (see Appendix \ref{app:biasparams}).
We will assume the above expression as our fiducial value in the following.

Eq.~(\ref{eq:deltahlong}) also includes a stochastic noise, which we will treat as a simple Poissonian shot noise due to the finite number of halos.
The cross-spectrum between the shot noise and the matter overdensity $\delta$ is equal to zero.
Finally, we do not include redshift space distortions.

\subsection{Scale-dependent bias information content - Formalism}

We consider the information in the modulation of halo density by the {\it long-mode} primordial potential fluctuation.
Based on the above, in the general notation for biased tracers used in this paper, the halo overdensity (subscript $h$), is thus characterized by,
\beq
\label{eq:haloprops}
b_h(\q_L) = b^{(h)}_{10},\quad b_h'(\q_L) = 2 \fnl \, (b^{(h)}_{10} - 1) \, \delta_c \, \M^{-1}(q_L),
\quad N_h = \frac{1}{\bar{n}_h}.
\eeq

\renewcommand{\arraystretch}{1.4}
\begin{table}[t]
\small
\begin{center}
\begin{adjustbox}{max width=\textwidth}
\begin{tabular}{|l||l|}
\hline
Spectrum &  Description    \\
\hline \hline
$\hat{P}_{11}(\q_L)$ & Matter power spectrum ($mm$ or PK)   \\
$\hat{P}_{h1}(\q_L)$ & Halo-matter cross-spectrum ($hm$)   \\
$\hat{P}_{hh}(\q_L)$ & Halo power spectrum ($hh$ or PK$h$) \\
\hline
\end{tabular}
\end{adjustbox}
\end{center}
\caption{The power and cross-spectrum estimators of long-mode perturbations considered in Section \ref{sec:mat2 vs halo}, along with shorthand notation. }
\label{table:t3}
\end{table}

For the position dependent power spectrum, the effective shot noise was determined by the shortest included short mode, $k_{\rm max}$, while for scale-dependent bias it is (approximately) given by the Poisson noise due to finite number of sources, determined by $\bar{n}_h$, the {\it comoving} halo number density.
An important difference is that, at least in this paper, we will only consider the halo overdensity of a single sample, whereas the position-dependent power spectrum constitutes a set of multiple tracers with different biases. It is straightforward to generalize our analysis to the case of multiple halo samples, in which case the analogy is even more complete.

Analogously to the treatment of the position-dependent power spectrum in the previous section, we will consider the halo power spectrum ($P_{hh}(q_L)$, or $hh$ in short), the halo-matter cross-spectrum ($P_{h1}(q_L)$, $hm$ in short), and a joint analysis of $P_{hh}(q_L)$, $P_{h1}(q_L)$ and the matter power spectrum $P_{11}(q_L)$ (see Table \ref{table:t3}). We summarize the analytic expressions for the information content in these probes below. The derivations use the same tools as discussed earlier for the matter statistics so we will not spell them out.

The Fisher information on $\fnl$ in the halo power spectrum is analogous to that in the matter trispectrum, and is given by,
\beq
\label{eq:info halo power}
F(q_L) = \frac{2 \Sigma_h(q_L) \, \Sigma_h''(q_L)}{\left( 1 + \Sigma_h(q_L) \right)^2} \quad  \text{{\bf (halo power spectrum, $hh$)}},
\eeq
with,
\bea
\label{eq:Sh defs}
\Sigma_h(q_L) &\equiv& N_h^{-1} \, b_h^2 \, P(q_L) \nonumber \\
\Sigma_h'(q_L) &\equiv& N_h^{-1} \, b_h \, b_h'(q_L) \, P(q_L) \nonumber \\
\Sigma_h''(q_L) &\equiv& N_h^{-1} \, \left(b_h'(q_L) \right)^2 \, P(q_L).
\eea
The information in the halo-matter cross-spectrum is analogous to the matter bispectrum, and is given by,
\beq
F(q_L) = \frac{\Sigma_h''(q_L)}{1 + 2 \Sigma_h(q_L)}
 \quad \text{{\bf (halo-matter cross-spectrum, $hm$)}}.
\eeq
Finally, the joint information from a ``multitracer analysis'' of $hh$, $hm$ and $mm$ is,
\beq
\label{eq:halos all}
F(q_L) = \Sigma_h''(q_L)
\quad \text{{\bf (halo-matter multitracer combi, $hh + hm + mm$)}}.
\eeq
We again consider this latter quantity the {\it total} information per mode $\q_L$ available in scale-dependent bias for a given number density $\bar{n}_h$.
The $hh$ or $hm$ signals separately do not achieve this constraining power due to the long-mode cosmic variance caused by $b_h$, which is cancelled out in the multitracer approach.
Note that, in analogy with the matter bispectrum, the above information would also be obtained from the halo-matter cross-spectrum, $P_{hm}(\q_L)$, if $b_h$ is set to zero (i.e.~the halo equivalent of the Gaussian covariance approximation).

\begin{figure*}[]
\centering
\includegraphics[width=0.47\textwidth]{sfnl_vs_kmax_allmat.pdf} 
\includegraphics[width=0.47\textwidth]{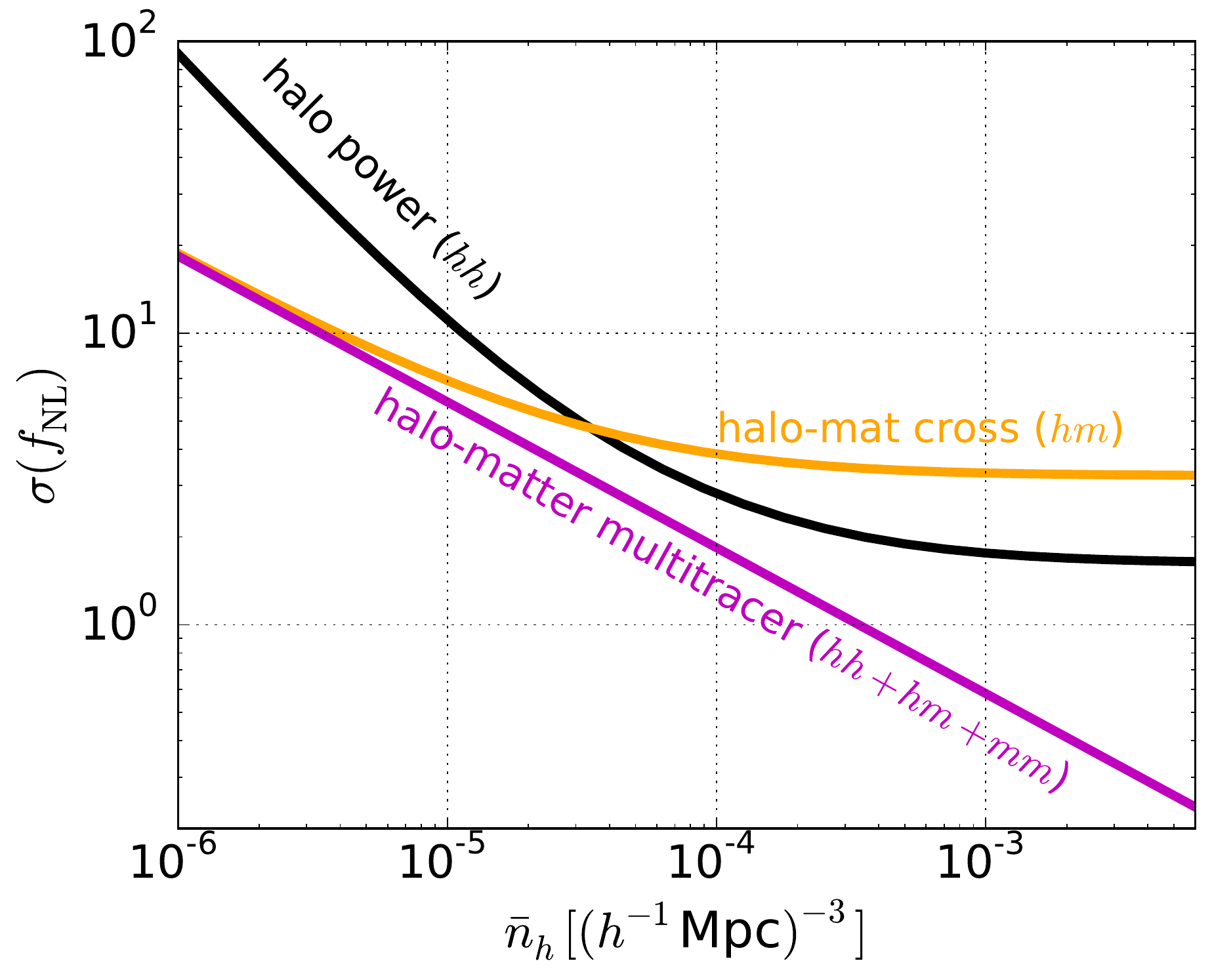} 
\caption{{\it Left:} As Figure \ref{fig:sigma allmat} (repeated for ease of comparison). {\it Right:} Uncertainty on $\fnl$ from halo power spectrum (black) as a function of the halo number density $\bar{n}$. The magenta curve shows the total information available in the large-scale halo overdensity, obtained by performing a cosmic variance canceling multitracer analysis of the long-mode matter overdensity {\it and} the halo overdensity. The observables in this case are the halo power spectrum, halo-matter cross-spectrum, and matter power spectrum. The halo power spectrum only case is akin to the trispectrum in the left panel in the sense that it is an autocorrelation of biased tracers of the long-mode matter overdensity. However, in the matter trispectrum case, one has multiple tracers (one for each short $\k$ mode), which allows for a degree of cosmic variance cancellation in the low shot noise limit (right hand side of plots), while for the single-tracer halo power spectrum a plateau of minimum uncertainty is reached in this limit.}
\label{fig:cf halo pk}
\end{figure*}

\subsection{Scale-dependent bias information content - Results}

We quantitatively compare the scale-dependent bias approach to the higher order matter statistics in Figure \ref{fig:cf halo pk}. We again use a fiducial survey volume $V = 100 \, (h^{-1} $Gpc$)^3$, effective redshift $z = 1$, and for the halo sample assume a fiducial bias $b^{(h)}_{10} = 2$. For convenience of comparison, the left panel repeats Figure \ref{fig:sigma allmat}, showing $\sigma(\fnl)$ from direct measurement of the matter density statistics. The right panel shows the constraining power of the halo power spectrum (black), halo-matter cross-spectrum (orange) and the combination of the two plus the matter power spectrum (magenta).
In both panels, we include the same range of long modes, $q_L = 0.001 h/$Mpc$ \, - \, 0.02 h/$Mpc.

Comparing the two panels of Figure \ref{fig:cf halo pk}, we clearly see the same behavior, according to the analogies spelled out above ($hh\leftrightarrow$TK, $hm\leftrightarrow$BK, etc). The only qualitative difference is that $\sigma(\fnl)$ from the halo power spectrum and from the halo-matter cross-spectrum reaches a plateau in the cosmic variance dominate regime (large $\bar{n}_h$). This is because we are only assuming a single halo sample so that cosmic variance cancellation is not possible, while the effective multitracer approach of the bi- and/or trispectrum does partially cancel cosmic variance.

Comparing more quantitatively, for our fiducial volume $V = 100 (h^{-1}$Gpc$)^3$, redshift $z = 1$ and minimum long-mode wave number $q_{\rm min} = 10^{-3} h/$Mpc, the matter bispectrum can reach $\sigma(\fnl) = 5$ (the current precision from Planck) and $\sigma(\fnl) = 1$ (a target for the next generation of experiments) if we probe short modes down to scales $k_{\rm max} = 0.16 h/$Mpc and $0.72 h/$Mpc respectively. Note that the latter scale is deep into the non-linear regime at this redshift, at which point non-linear corrections beyond the ones included in this work are very important, and our forecasts are no longer a good approximation.
For the trispectrum, one requires similar values, $k_{\rm max} = 0.15, \, 0.68 h/$Mpc, and the joint analysis (matter PK$+$BK$+$TK) would require $k_{\rm max} = 0.12, \, 0.34 h/$Mpc to reach $\sigma(\fnl) = 5, 1$.
Using halos on the other hand, for our fiducial bias $b^{(h)}_{10} = 2$, the halo power spectrum (to be compared to the matter trispectrum) requires $\bar{n}_h = 4.2 \cdot 10^{-5} (h^{-1}$Mpc$)^{-3}$
to obtain $\sigma(\fnl) = 5$, while the minimum uncertainty is $\sigma(\fnl) = 1.7$ due to the cosmic variance limit. The halo-matter cross-spectrum (analogous to the matter bispectrum) requires $\bar{n}_h = 3.9 \cdot 10^{-5} (h^{-1}$Mpc$)^{-3}$ for $\sigma(\fnl) = 5$ and reaches its plateau at $\sigma(\fnl) = 3.4$.
Note that, for fixed $q_{\rm min}$, these uncertainties scale as $\sim V^{-1/2}$, so that, for example,  more than (because of the improvement in $q_{\rm min}$) a factor two improvement in $\sigma(\fnl)$ would be obtained for a $400 \, (h^{-1}$Gpc$)^3$ survey. Equivalently, for such a larger survey, one can reach the same uncertainties $\sigma(\fnl) = 5, 1$  with smaller $k_{\rm max}$ or $\bar{n}_h$.

\begin{figure*}[]
\centering
\includegraphics[width=0.55\textwidth]{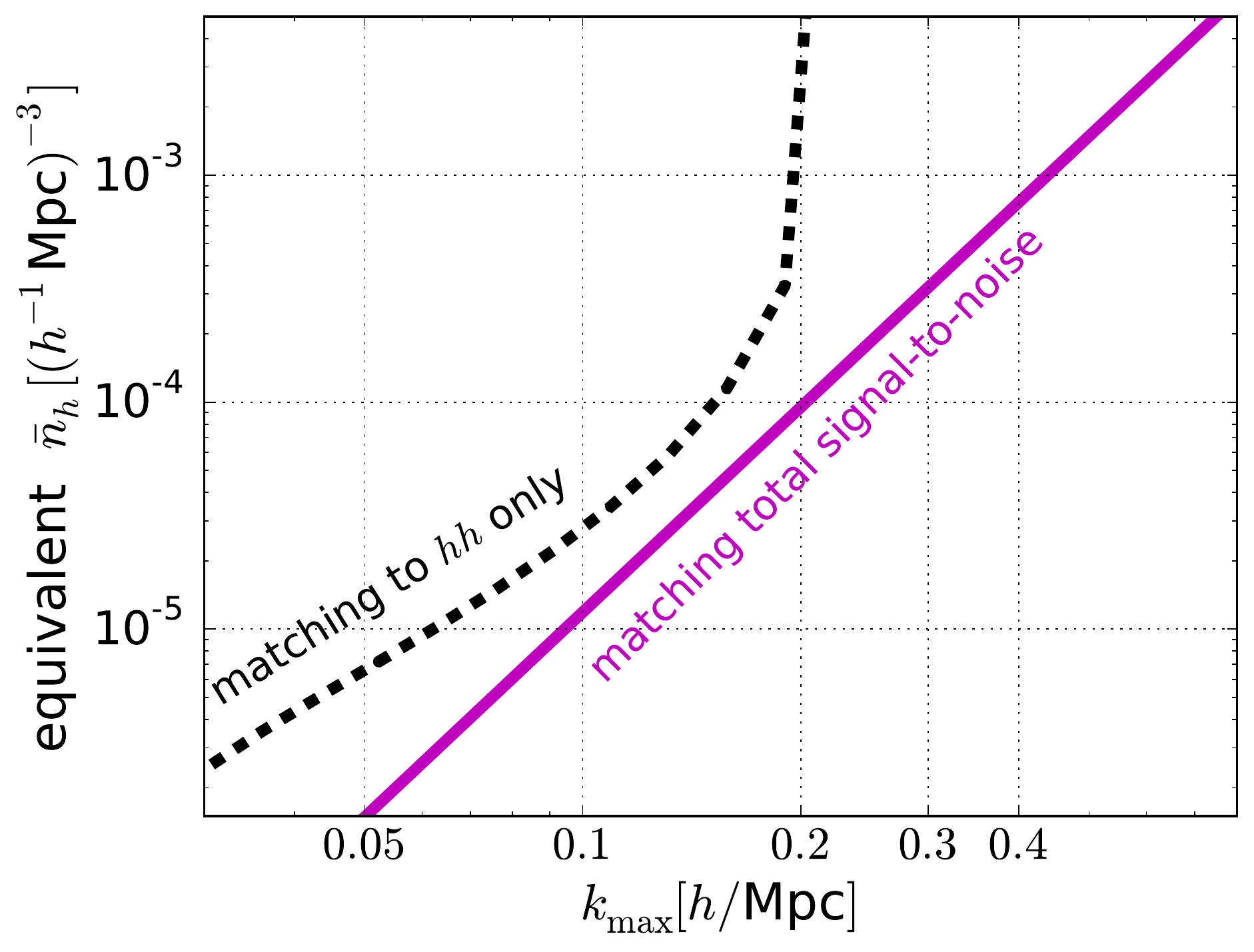} 
\caption{Halo number density $\bar{n}_h$ for which scale-dependent halo bias of a sample with number density $\bar{n}_h$ (assuming a fixed bias $b_{10}^{(h)} = 2$) contains the same information as the
matter mode-coupling itself down to $k_{\rm max}$.
For the magenta curve, we equate the {\it total} information content according to Eq.~(\ref{eq:matching}),
i.e.~PK$+$BK$+$TK (green curve in left panel Figure \ref{fig:cf halo pk})
vs.~$hh+hm+mm$ (magenta curve in right panel Figure {fig:cf halo pk}).
The dashed black curve equates the total information in the matter mode-coupling to the information from scale-dependent bias in the halo power spectrum only ($hh$). The $\sigma(\fnl)$ from $hh$ reaches an error floor at high $\bar{n}_h$ due to cosmic variance, explaining the behavior at high $k_{\rm max}$ of the black curve.
The plot does not incorporate the fact that in reality the halo bias $b_{10}^{(h)}$ is not independent of $\bar{n}_h$.}
\label{fig:cf nbar pk}
\end{figure*}

To better understand the comparison between the two general approaches, consider the relation between the $k_{\rm max}$ used in the higher order matter statistics and the number density of halos used in the scale-dependent bias analysis.
We may ask,\\ \\
{\it for a given $k_{\rm max}$, what is the value of $\bar{n}_h$ for which scale-dependent bias contains the same information as the primordial matter mode-coupling?}\\ \\
Both quantities define the effective shot noise of the respective tracers. Thus, one way of matching them is to equate,
\beq
N_h = \frac{1}{\bar{n}_h} \quad \quad \text{and} \quad \quad N_{\bar{2}} \approx \frac{12 \pi^2}{k_{\rm max}^3},
\eeq
respectively the shot noise in the halo density and the (average weighted by number of modes) shot noise  in the position dependent small-scale matter power spectrum (cf.~Eq.~(\ref{eq:bias dP})).
This would lead to the matching,
\beq
\label{eq:matching sn}
\bar{n}_h \to \frac{k_{\rm max}^3}{12 \pi^2}.
\eeq
This gives the straightforward interpretation that clustering of halos with number density $\bar{n}_h$ is similar to a direct measurement of mode-coupling in the matter down to a minimum scale equal to the mean spacing between halos, $\ell_{\rm min} \sim k_{\rm max}^{-1} \sim \bar{n}_h^{-1/3}$.

We can slightly refine this picture.
The effective shot noise represents the ``small-scale'' {\it noise} in a measurement in some local volume (cf.~the discussion in Section \ref{subsec:posdeppk}) of the power spectrum or halo number density $n_h$. However, $n_h$ also encapsulates a different {\it signal}, i.e.~response to $\fnl$, as quantified in the difference between $b_h'(q_L) = 2 (b^{(h)}_{10} - 1) \, \delta_c \, \M^{-1}(q_L)$ and $b_{2(i)}'(q_L) = 4 \M^{-1}(q_L)$ .
The signal-to-noise per unit $\fnl$ squared in the local measurement is determined by the quantities,
\beq
\left( \frac{S}{N} \right)^2_{\rm loc} \propto N_h^{-1} \, \left(b_h'(q_L)\right)^2
=  \bar{n}_h \, \left( \tfrac{1}{2} \, (b^{(h)}_{10} - 1) \, \delta_c\right)^2 \, \left( 4 \M^{-1}(q_L) \right)^2,
\eeq
and
\beq
\left( \frac{S}{N} \right)^2_{\rm loc} \propto \sum_i N^{-1}_{2(i)} \, \left( b_{2(i)}'(q_L) \right)^2
\approx \frac{k_{\rm max}^3}{12 \pi^2} \, \left( 4 \M^{-1}(q_L) \right)^2,
\eeq
leading to the improved matching,
\beq
\label{eq:matching}
\bar{n}_h \, \left( \tfrac{1}{2} (b^{(h)}_{10} - 1) \, \delta_c \right)^2 \to \frac{k_{\rm max}^3}{12 \pi^2}.
\eeq
Indeed, when the signal-to-noise in the {\it local} measurement is the same, the information content on $\fnl$ is exactly identical between the two approaches when the optimal, cosmic-variance canceling, combination of data is used (i.e.~PK$+$BK$+$TK for matter, and $hh$+$hm$+$mm$ for halos). In other words, Eq.~(\ref{eq:matching}) corresponds to matching the {\it total} Fisher information per mode scale-dependent bias and in higher order matter statistics,
\beq
\Sigma_h''(q_L) =  \Sigma_{2}''(q_L).
\eeq

We show the improved relation between $k_{\rm max}$ and its equivalent $\bar{n}_h$, Eq.~(\ref{eq:matching}), with the magenta curve in Figure \ref{fig:cf nbar pk}.
We see that the matter mode coupling information up to $k_{\rm max} = 0.1 h/$Mpc is equivalent to the halo bias information of a sample with number density $\bar{n}_h \approx 10^{-5} (h^{-1}$Mpc$)^{-3}$(assuming our fiducial bias $b_{10}^{(h)} = 2$). The scale-dependent bias in a sample with number density $\bar{n}_h = 3 \cdot 10^{-4} (h^{-1} $Mpc$)^{-3}$ (approximately the number density of the BOSS CMASS galaxy sample \cite{reidetal16}), has the same total information content as the primordial matter mode-coupling down to $k_{\rm max} \approx 0.3 h/$Mpc. Note however that in the absence of a direct, noiseless measurement of the matter density, the full scale-dependent bias information is not accessible and we only have access to the halo power spectrum ($hh$). The resulting matching of the Fisher information (again to the {\it total} information in the higher order matter statistics) is shown with the black dashed curve.
It shows for instance that the information available in the power spectrum of a halo sample with $\bar{n}_h = 3 \cdot 10^{-4} \, (h^{-1} $Mpc$)^{-3}$ is comparable to the {\it total} information in the higher order matter statistics up to $k_{\rm max} \approx 0.2 h/$Mpc.
The same information can be obtained from the matter bispectrum if $k_{\rm max} \approx 0.4 h/$Mpc.

\vskip 10pt

\subsection{Scale-dependent bias vs.~direct measurements of matter mode-coupling}

Finally, we comment on whether one of the two signals is fundamentally more optimal (leaving aside practical issues, such as the fact that it is more difficult to directly measure the matter field statistics than the halo density, but see Section \ref{sec:halos}).
The $\fnl$ signal we are after is the response of the primordial, small-scale power spectrum in some local volume to a background potential fluctuation. Under our assumption that the short modes are Gaussian for fixed realization of the background mode (both primordially and at late times, since we ignore non-Gaussianity due to non-linear evolution that is not captured by the long-short mode-coupling), the small-scale power spectrum estimator contains {\it all} the information on the true spectrum and thus on the signal, at least for the range of short modes considered, $k < k_{\rm max}$.
Like the power spectrum, the local halo density is simply a function of the initial realization of the short modes and is thus another probe of the short-scale power spectrum. In particular, the expectation value of the number density of halos with some Lagrangian radius $R$ will approximately be determined by the power spectrum on all scales $k < R^{-1}$. Since the direct power spectrum estimator is optimal, the halo density can never contain more information than the power spectrum estimator over the same range of scales relevant to the halo density. Thus, the ``local signal-to-noise'' quantities defined above should strictly be larger for the matter statistics, provided all scales that the halos are sensitive to are included.

In practice, however, in a bispectrum/trispectrum analysis, one typically avoids probing deep into the non-linear regime, partially because of the difficulty of modeling the non-linearities and baryonic effects, and partially because in practice the short-mode matter perturbations may be estimated using a tracer (halos or weak lensing) which introduces its own shot noise limiting the observable range of scales. On the other hand, for the scale-dependent halo bias analysis, one could in principle use a sample of very low mass halos (small Lagrangian radius) with high number density and thus non-optimally probe a larger range of scales than with the matter statistics. This is why, as we have seen above, the $hh$ information from a {\it moderate} number density $\bar{n}_h \approx 3 \cdot 10^{-4} (h^{-1}$Mpc$)^{-3}$ (assuming $b_{10}^{(h)} = 2$) already captures as much information as all the primordial matter mode coupling down to scale $k_{\rm max} \approx 0.2 h/$Mpc, which is pushing significantly into the non-linear regime. Since we have argued that for the {\it same} range of scales, the matter statistics should always be optimal, this means that our supposed halo sample with $b_{10}^{(h)} = 2$ and $\bar{n}_h = 3 \cdot 10^{-4} (h^{-1}$Mpc$)^{-3}$ (sub-optimally) probes a significantly larger range of scales, $k_{\rm max, eff} > 0.2 h/$Mpc.

\subsection{Pushing the bispectrum into the non-linear regime}
\label{subsec:BK general}

The discussion above suggests that in the matter bispectrum and trispectrum analysis, one does not necessarily have to avoid the non-linear regime (for the short modes). For halos, we are perfectly happy to consider the long-wavelength modulation of the number density of (very small) halos, an extremely non-linear quantity impossible to fully model with perturbation theory. In this case, we simply parametrize our ignorance with free linear bias parameters $b_{10}^{(h)}$ and\footnote{In forecasts, $b_{01}^{(h)}$ is commonly fixed in terms of $b_{10}^{(h)}$ by the assumption of a universal mass function, but future precision measurements need to take into account deviations from this relation.} $b_{01}^{(h)}$, which is sufficient when the long mode is much larger than the non-linear scale. We see no reason why the same approach is not in principle possible also for the position-dependent (small-scale) power spectrum, i.e.~one could treat the quantities $b_{2(i)}$ and $b_{2(i)}'$ as free parameters. In principle this gives a large number of free parameters to marginalize over, but the number is likely constrained by symmetry considerations.
In any case, parametrizing our ignorance\footnote{One may also try to {\it measure} the response/bias parameters from simulations using the separate-Universe approach, although this would still rely on the ability to accurately simulate very small scales.} should then allow us to model the bispectrum and trispectrum for configurations with short modes deep into the non-linear regime.

A complication of taking the bispectrum (and/or trispectrum) short modes far into the non-linear regime
is that the statistics of the short modes for fixed long mode become increasingly non-Gaussian.
This means that the mode-coupling between short modes becomes important. Indeed, we know that this leads to correlations between the short-scale power spectrum at different wave vectors, leading in turn to a suppression of information in the small-scale power spectrum relative to the Gaussian case. In the position-dependent power spectrum approach this would manifest itself as the stochastic noise matrix ${\bf N}_{2(i),2(j)}$ containing off-diagonal elements, and the total stochastic noise averaged over bins no longer decreasing like $\propto k_{\rm max}^{-3}$. Thus, even if the approach suggested in the previous paragraph allows us to model the squeezed-limit bispectrum for short modes deep into the non-linear regime,
the information content would not be as great as suggested by the $k_{\rm max}$ dependence plotted in this paper, as our plots ignore this mode-coupling between the short modes.

One way of thinking of the loss of information due to mode-coupling between short modes is that information on these small scales leaks into higher order statistics (of short modes), which suggest
that perhaps one can recover information by considering the position-dependent small-scale bispectrum, etc.
We plan to further develop the ideas above in future work.

\section{$\fnl$ from the halo power spectrum, bispectrum and trispectrum}
\label{sec:halos}

In Section \ref{sec:matter}, we considered the information contained in a direct measurement through higher order matter statistics of mode-coupling in the matter density.
However, since realistically one cannot directly measure the matter field, we in this section study
the information content in the
power, bi- and trispectrum of {\it halos}.
These halo statistics combine the two probes of primordial mode-coupling discussed in the previous two sections:
they contain both the long-short mode-coupling of the halo overdensity field (arising from the previously discussed matter mode-coupling {\it and} from the new effect of non-Gaussian, non-linear halo biasing), and scale-dependent halo bias.
We will again ignore redshift space distortions in our treatment of halo statistics, leaving their inclusion and forecasts of constraints for realistic surveys for future work.

We again describe the {\it long-wavelength} halo density perturbation $\delta_{h,L}$ to linear order (cf.~Eq.~(\ref{eq:haloprops})) in underlying perturbations. To capture the modulation of the {\it short} modes by the long mode, however, we need to expand our (Eulerian) halo biasing model to second order (see e.g.~\cite{mcdonaldroy09,assassietal14, sena15,DJSreview16}),
\beq
\label{eq:halos 2nd}
\delta_h = b^{(h)}_{10} \, \delta + b^{(h)}_{01} \, \phi + \epsilon_h
+ b^{(h)}_{20} \, \delta^2 + \fnl \, b^{(h)}_{11} \, \delta \, \phi + \fnl^2 \, b^{(h)}_{02} \, \phi^2 + b^{(h)}_{s^2} \, s^2.
\eeq
Here, we have introduced the quadratic and tidal-tensor biases $b^{(h)}_{20}$ and $b^{(h)}_{s^2}$ \cite{Baldauf:2012hs},
as well as the non-Gaussian biases $b^{(h)}_{11}$ and $b^{(h)}_{02}$ (with factors $\fnl$ factored out). The effect of the latter on the mode-coupling is strongly suppressed relative to the other primordial non-Gaussianity contributions (and of order $\fnl^2$) so we will neglect it in the following.
For simplicity, we have included stochastic noise only up to first order and we ignore mode-coupling in $\epsilon_h$, i.e.~we treat the long- and short-mode components of $\epsilon_h$ as independent.
We refer to Appendix \ref{app:stochnoise} for a discussion of the effect of keeping the mode-coupling terms involving the halo shot noise $\epsilon_h$.
We have also neglected a ``convection term'', $\delta_h \supset {\bf \Psi} \cdot {\bf \nabla} \phi$ (with ${\bf \Psi}$ the displacement vector between Lagrangian and Eulerian coordinates), that is in principle present in a second order bias expansion \cite{tellarinietal15,assassietal15}.
The primordial mode-coupling due to this term is suppressed in the squeezed limit relative to the leading order contributions (see Appendix \ref{app:biasparams}).

\renewcommand{\arraystretch}{1.4}
\begin{table*}[t]
\small
\begin{center}
\begin{adjustbox}{max width=\textwidth}
\begin{tabular}{|l||l|l|}
\hline
Spectrum &  Description & Equivalent expression   \\
\hline \hline
$\hat{P}_{hh}(\q_L)$ & Halo power spectrum (PK$h$ or $hh$) & N/A   \\
$\hat{P}_{h2h(i)}(\q_L)$ & Sq.-lim.~halo bispectrum (BK$h$)  & $\hat{B}_{h}(\q_L, \k_i, -\k_i - \q_L)/\left( b^{(h)\,2}_{10} \, P(k_i) \right)$ \\
$\hat{P}_{2h(i)2h(j)}(\q_L)$ & Collapsed halo trispectrum (TK$h$) & $\hat{T}_{h}(\k_i, -\k_i + \q_L, \k_j, -\k_j - \q_L)/\left( b^{(h)\,2}_{10} \, P(k_i) \right)/\left( b^{(h)\,2}_{10} \, P(k_j) \right)$ \\
\hline
\end{tabular}
\end{adjustbox}
\end{center}
\caption{The power and cross-spectra of long-mode perturbations considered in Section \ref{sec:halos}, and their equivalent expressions in terms of halo bi- and trispectra.
The bispectrum and trispectrum in the third column are averages over bins of short modes $\k_{S}$ labeled by the index $i$ (with central wave vector $\k_i$). See text for more details.}
\label{table:t4}
\end{table*}

To describe the short halo mode $\delta_{h,S}$, of the ``signal terms''(i.e.~those proportional to $\fnl$), we will keep only those proportional to the long-wavelength potential mode, $\phi_L$,
neglecting contributions proportional to $\phi_S$.
This is in keeping with previous sections: we specifically are after the modulation by the long-wavelength primordial potential.  Moreover, terms proportional to $\phi_S$ are strongly suppressed.
Schematically, we thus use the following expression for the short-mode halo overdensity (using the tidal-tensor bias mode-coupling kernel
$\mathcal{S}(\k, \k') = b_{s^2}^{(h)} \, \left( \hat{\k} \cdot \hat{\k}' - \frac{1}{3} \right)$),
\beq
\delta_{h,S} = b^{(h)}_{10} \, \delta_S + \epsilon_{h,S} + 2 b^{(h)}_{20} \, \delta_L \, \delta_S + 2 b^{(h)}_{s^2} \, \left( \mu^2 - \tfrac{1}{3} \right) \, \delta_L \, \delta_S + \fnl \, b^{(h)}_{11} \, \phi_L \, \delta_S, \quad \mu \equiv \hat{\q}_L \cdot \hat{\k}_S,
\eeq
where $\delta_S$ is the short-mode matter overdensity at low redshift (i.e.~$\delta_S$ contains mode-coupling itself).

As for the position-dependent power spectrum of {\it matter} perturbations,
the modulation of the position-dependent small-scale power spectrum of halos is obtained from the averaged mode-coupling kernel. Specifically,
if the mode-coupling in the halo overdensity is written in terms of a {\it symmetrized} kernel $\mathcal{F}({\bf k}_1, {\bf k}_2)$ (relative to the Gaussian matter density $\tilde{\delta}$),
\beq
\delta_h(\k) = \int \frac{d^3 \k'}{(2 \pi)^3} \, \mathcal{F}(\k', \k - \k') \, \tilde{\delta}(\k') \, \tilde{\delta}(\k - \k'),
\eeq
then the modulation of the small-scale halo power spectrum is (cf.~Eq.~(\ref{eq:meankernelmat})),
\beq
\label{eq:modulationfromkernel}
\delta \hat{P}_{hh}(\k; \q_L) = 2 \, b_{10}^{(h)} \, \delta(\q_L) \, \left[  \mathcal{F}_2(\q,-\k) \, P(k)
+ \mathcal{F}_2(\q,\k-\q) \, P(|\k - \q|) \right].
\eeq

\subsection{Halo information content - Formalism}
\label{subsec:halos theory}

We now define relative fluctuations in the position-dependent small-scale {\it halo} power spectrum (subscript $2h(i)$ for a bin of short modes centered at $\k_i$) by,
\beq
\label{eq:def d2hi}
\delta_{2h(i)}(\q_L) \equiv \delta \ln \hat{P}_{hh}(\k_i; \q_L) = \frac{\delta \hat{P}_{hh}(\k_i; \q_L)}{b_{10}^{(h) \, 2} \, P(k_i)},
\eeq
so that,
\bea
\label{eq:d2h}
b_{2h(i)}(\q_L) &=& 4 \left( \bar{F}_2(\mu, k_i) + \frac{b_{20}^{(h)}}{b_{10}^{(h)}} + \frac{b^{(h)}_{s^2}}{b_{10}^{(h)}} \, \left(  \mu^2 - \tfrac{1}{3} \right) \right),
\quad b_{2h(i)}'(\q_L) = 4 \M^{-1}(q_L) \left( 1 + \frac{1}{2} \, \frac{b^{(h)}_{11}}{b^{(h)}_{10}} \right) \nonumber \\
N_{2h(i)} &=& \frac{2 (2 \pi)^3}{V_{\k,i}} \, \left( \frac{1 + \bar{n}_h \, b_{10}^{(h) \, 2} \, P(k_i)}{\bar{n}_h \, b_{10}^{(h) \, 2} \, P(k_i)} \right)^2,  \quad \text{with} \quad \mu \equiv \hat{k}_i \cdot \hat{q}_L.
\eea
The power spectrum, $N_{2h(i)}$, of the stochastic noise, $\epsilon_{2h(i)}$, is derived in the same way as that of the position-dependent matter spectrum, $N_{2(i)}$ (cf.~discussion above Eq.~(\ref{eq:mat sn})),
and is again associated with variance in the realization of the short modes (for fixed realization of the long modes).
The difference with the matter case is
that for halos the stochastic is increased due to the $\sim 1/\bar{n}_h$ Poisson noise in the measurement of the small-scale halo power spectrum. Equivalently, the effective volume is decreased.
We take the effective shot noise $\epsilon_{2h(i)}(\q_L)$ to be uncorrelated with the long-mode halo shot noise itself, i.e.~$\langle \epsilon_{2h(i)} \, \epsilon_h \rangle = 0$. While
this should be a reasonable approximation, in reality one does
expect some level of correlation between the small-scale halo power spectrum and the local halo number density.
This could in principle be quantified with N-body simulations, but here we neglect this effect.

\renewcommand{\arraystretch}{1.4}
\begin{table*}[t]
\small
\begin{center}
\begin{adjustbox}{max width=\textwidth}
\begin{tabular}{|l||l|l|l|l|l|l|}
\hline
Parameter &  $b_{10}^{(h)}$ &  $b_{01}^{(h)}$ & $b_{20}^{(h)}$ & $b_{s^2}^{(h)}$ &  $b_{11}^{(h)}$ & $\bar{n}_h$ \\
\hline \hline
Fiducial value & 2 & 3.4 & -0.05 & -0.29 & -0.23 & $3 \cdot 10^{-4} \, (h^{-1}$Mpc$)^{-3}$ \\
\hline
\end{tabular}
\end{adjustbox}
\end{center}
\caption{Fiducial parameters describing default halo sample. Given $b_{10}^{(h)} = 2$, the other bias parameters were derived based on the fitting function Eq.~(5.2) of \cite{lazetal16}, and on Eqs (\ref{eq:bs2}) and (\ref{eq:b01 b11}) (see also Figure \ref{fig:biases2}). All biases here are Eulerian.}
\label{table:fiducial}
\end{table*}

Finally, in Eq.~(\ref{eq:d2h}) we have neglected the effect of mode-coupling terms in $\delta_h$ involving the halo shot noise $\epsilon_h$. These terms would lead to additional contributions to the position-dependent halo spectrum $\delta_{2h(i)}$ proportional to $\delta_L$ and to $\epsilon_{h,L}$. As discussed in Appendix \ref{app:stochnoise}, such terms are responsible for the shot noise contributions to the halo bispectrum (and trispectrum) that arise because the halo shot noise is Poissonian, not Gaussian.
Note, however, that the leading order shot noise contributions to the covariance are already included without adding the non-Gaussian shot noise terms.

The higher order halo statistics probe $\fnl$ both through halo mode-coupling, i.e.~$\delta_{2h(i)}$,
{\it and} through scale-dependent bias, $\delta_h$.
Regarding the former,
we see from Eq.~(\ref{eq:d2h}), that in addition to the direct contribution due to primordial {\it matter} mode-coupling, the halo-mode coupling also contains a contribution from non-linear, non-Gaussian biasing, $b_{11}^{(h)}$. Depending on the value of $b_{11}^{h}$, this effect can either enhance or weaken the halo mode-coupling relative to the matter mode-coupling.
The {\it non-primordial} modulation of the position-dependent halo power spectrum (characterized by $b_{2h(i)}$), also receives additional contributions, relative to the position-dependent matter spectrum, from the non-linear biasing parameters $b_{20}^{(h)}$ and $b_{s^2}^{(h)}$.

\vskip 10pt

We now consider the information content of the {\it halo} power spectrum (PK$h$), squeezed-limit bispectrum (BK$h$), collapsed trispectrum (TK$h$), and that of a joint analysis (PK$h$ + BK$h$ + TK$h$), see Table \ref{table:t4}.
We will below first
introduce the analytic treatment of these and a few more halo statistics of interest,
and we will discuss results in Section \ref{subsec:halo results}.

{\bf Halo power spectrum:} We have already covered the constraining power of the halo power spectrum, $P_{hh}(\q_L)$ in the previous section, see Eq.~(\ref{eq:info halo power}).

\vskip 10pt

{\bf Halo bispectrum:}
The halo bispectrum is equivalent to the set of cross-spectra $P_{h, 2h(i)}(\q_L)$. The signal is given by,
\beq
\label{eq:BKh signal}
\frac{\pa P_{h, 2h(i)}(\q_L)}{\pa \fnl} = \left( b_h \, b'_{2h(i)}(q_L) + b_h'(q_L) \,  b_{2h(i)}(\q_L) \right) \, P(q_L).
\eeq
The first term in the parentheses contains the halo mode-coupling signal and the second term contains the signal from scale-dependent halo bias.
An interesting question we will address is which of these contributions contains more information.

Using the methods from the previous sections, it is straightforward to derive the Fisher information per long mode in the halo bispectrum,
\bea
\label{eq:BKh}
F(q_L) &=& \frac{\Sigma_h(q_L) \, \Sigma_{2h}''(q_L) + 2 \, \Sigma_h'(q_L) \, \Sigma_{2h}'(q_L) + \Sigma_h''(q_L) \, \Sigma_{2h}(q_L)}{1 + \Sigma_h(q_L)}
 \\
&-& \frac{\left( 1 + 2 \, \Sigma_h(q_L) \right) \, \left(\Sigma_h(q_L) \, (\Sigma_{2h}'(q_L))^2 + \Sigma_h''(q_L) \, \Sigma_{2h}^2(q_L) + 2 \Sigma_h'(q_L) \, \Sigma_{2h}(q_L) \, \Sigma_{2h}'(q_L) \right)}{\left( 1 + \Sigma_h(q_L) + (1 + 2 \Sigma_h(q_L)) \, \Sigma_{2h}(q_L) \right) \, \left( 1 + \Sigma_h(q_L) \right)}  \quad \text{{\bf (halo bispectrum, BK$h$)}}, \nonumber
\eea
where we have defined the ``signal-to-shot noise'' quantities,
\bea
\label{eq:S2h def}
\Sigma_{2h}(q_L) &\equiv& P(q_L) \, \sum_i N^{-1}_{2h(i)} \, b_{2h(i)}^2(\q_L)  \nonumber \\
\Sigma_{2h}'(q) &\equiv& P(q_L) \, \sum_i N^{-1}_{2h(i)} \, b_{2h(i)}(\q_L) \, b_{2h(i)}'(\q_L)\nonumber \\
\Sigma_{2h}''(q) &\equiv& P(q_L) \, \sum_i N^{-1}_{2h(i)} \, \left( b_{2h(i)}'(\q_L) \right)^2.
\eea
By substituting the expressions (\ref{eq:d2h}), the above quantities can all be written as integrals over all short modes.

In the Gaussian covariance approximation (setting $b_{2h(i)} = 0$ in the bispectrum covariance), the Fisher information is given by only the first line in Eq.~(\ref{eq:BKh}). We have checked that the covariance in the GCA reproduces the standard squeezed-limit result including halo shot noise. We discuss this in detail in Appendix \ref{app:F GCA}.

\begin{figure*}[]
\centering
\includegraphics[width=0.47\textwidth]{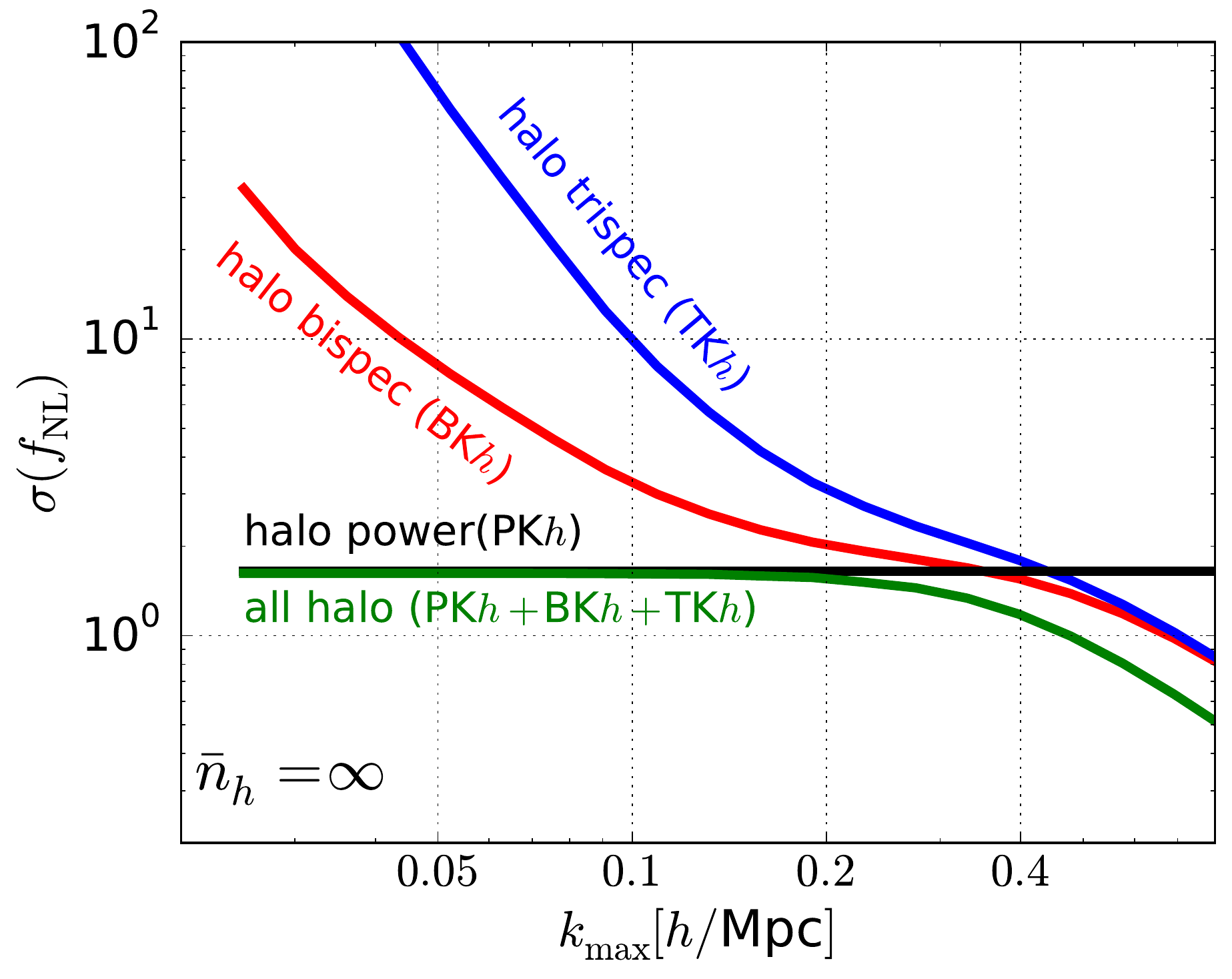} 
\includegraphics[width=0.47\textwidth]{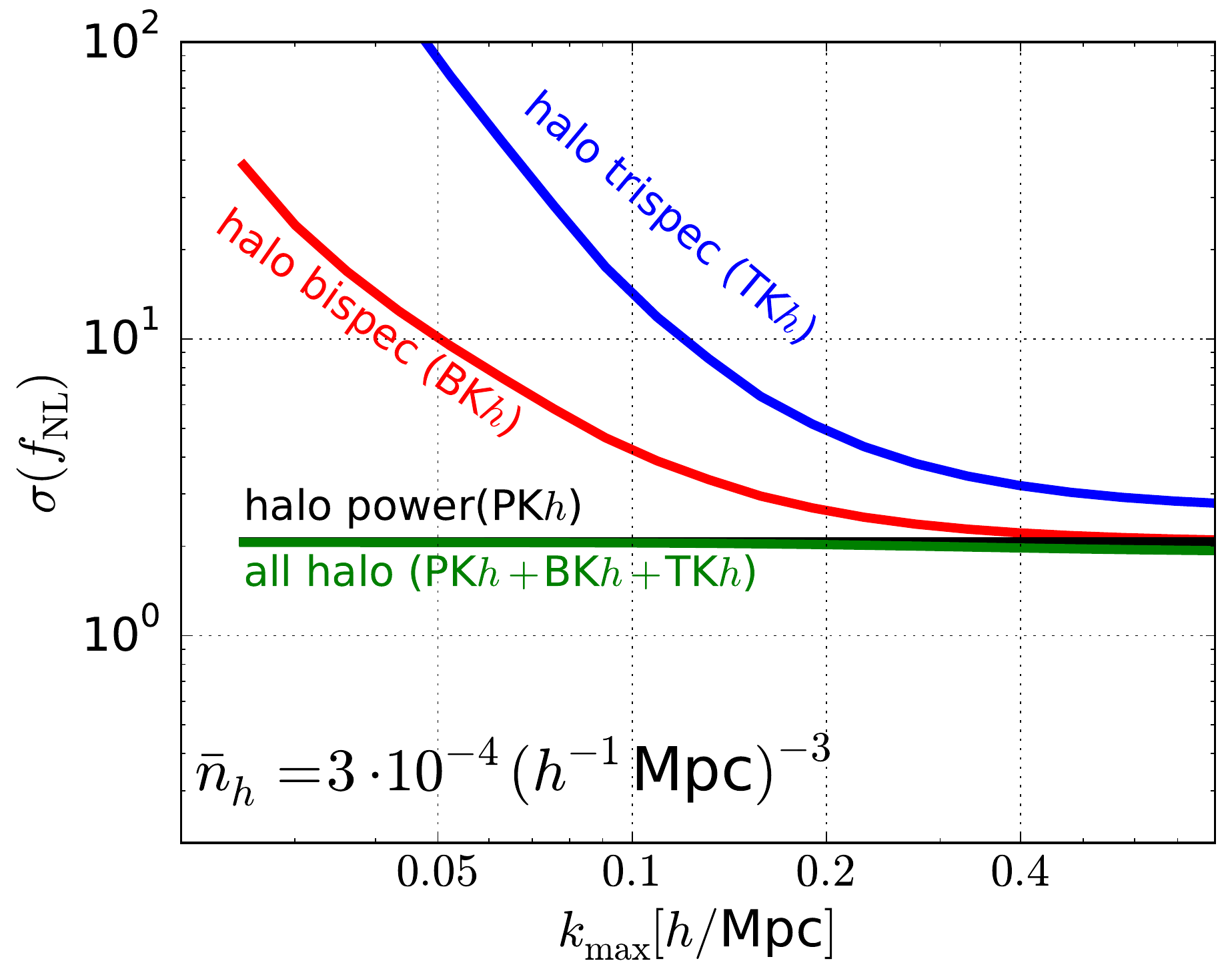} 
\caption{Uncertainty on $\fnl$ from (squeezed-limit) halo statistics as a function of maximum short-mode wave number, $k_{\rm max}$. For the long modes, we use the default range $q_L = 0.001 - 0.02 \, h/$Mpc and we assume the fiducial bias parameters given in Table \ref{table:fiducial}.
We assume a survey with volume $V = 100 \, (h^{-1} $Gpc$)^3$ at effective redshift $z = 1$, but note that future surveys may probe even larger volumes.
{\it Left:} The case without halo shot noise, i.e.~the limit of infinite halo number density.
{\it Right:} Default, ``moderate'' halo number density, $\bar{n}_h = 3 \cdot 10^{-4} \, (h^{-1}$Mpc$)^{-3}$.
}
\label{fig:halos vary nbarh}
\end{figure*}

\vskip 10pt

{\bf Joint halo power spectrum, bispectrum and trispectrum:}
Combining the halo power spectrum, bispectrum and trispectrum (green curves) again corresponds to a complete multitracer analysis of the tracers $\{\delta_h(\q_L), \delta_{2h(i)}(\q_L) \}$. We find for the Fisher information (cf.~Eq.~(\ref{eq:fishertri})),
\beq
\label{eq:joint halo}
F(q_L)
= \frac{\Sigma_{h,{\rm tot}}(q_L)}{1 + \Sigma_{h,{\rm tot}}(q_L)} \, \Sigma_{h,{\rm tot}}''(q_L)
+ \frac{1 - \Sigma_{h,{\rm tot}}(q_L)}{\left( 1 + \Sigma_{h,{\rm tot}}(q_L)\right)^2} \, \left( \Sigma_{h,{\rm tot}}'(q_L) \right)^2  \quad \text{{\bf (halo PK$h+$BK$h+$TK$h$)}},
\eeq
with
\beq
\Sigma_{h,{\rm tot}}(q_L) \equiv \Sigma_h(q_L) + \Sigma_{2h}(q_L), \quad \text{etc}.
\eeq

\vskip 10pt

{\bf Halo trispectrum only:}
Finally, the halo (collapsed) trispectrum (blue curves) employs all possible correlations between the set of position-dependent halo power spectra, $\{ \delta_{2h(i)}(\q_L) \}$, i.e.~the set of spectra $P_{2h(i)2h(j)}(\q_L)$. This leads to the same expressions as for the combination of all data and for the matter trispectrum, except now using $\Sigma_{2h}$, etc. Specifically, we find,
\beq
F(q_L) = \frac{\Sigma_{2h}(q_L)}{1 + \Sigma_{2h}(q_L)} \, \Sigma_{2h}''(q_L)
+ \frac{1 - \Sigma_{2h}(q_L)}{\left( 1 + \Sigma_{2h}(q_L) \right)^2} \, \left( \Sigma_{2h}'(q_L) \right)^2  \quad \text{{\bf (halo trispectrum, TK$h$)}}.
\eeq

\subsection{Halo information content - Results}
\label{subsec:halo results}

\begin{figure*}[]
\centering
\includegraphics[width=0.47\textwidth]{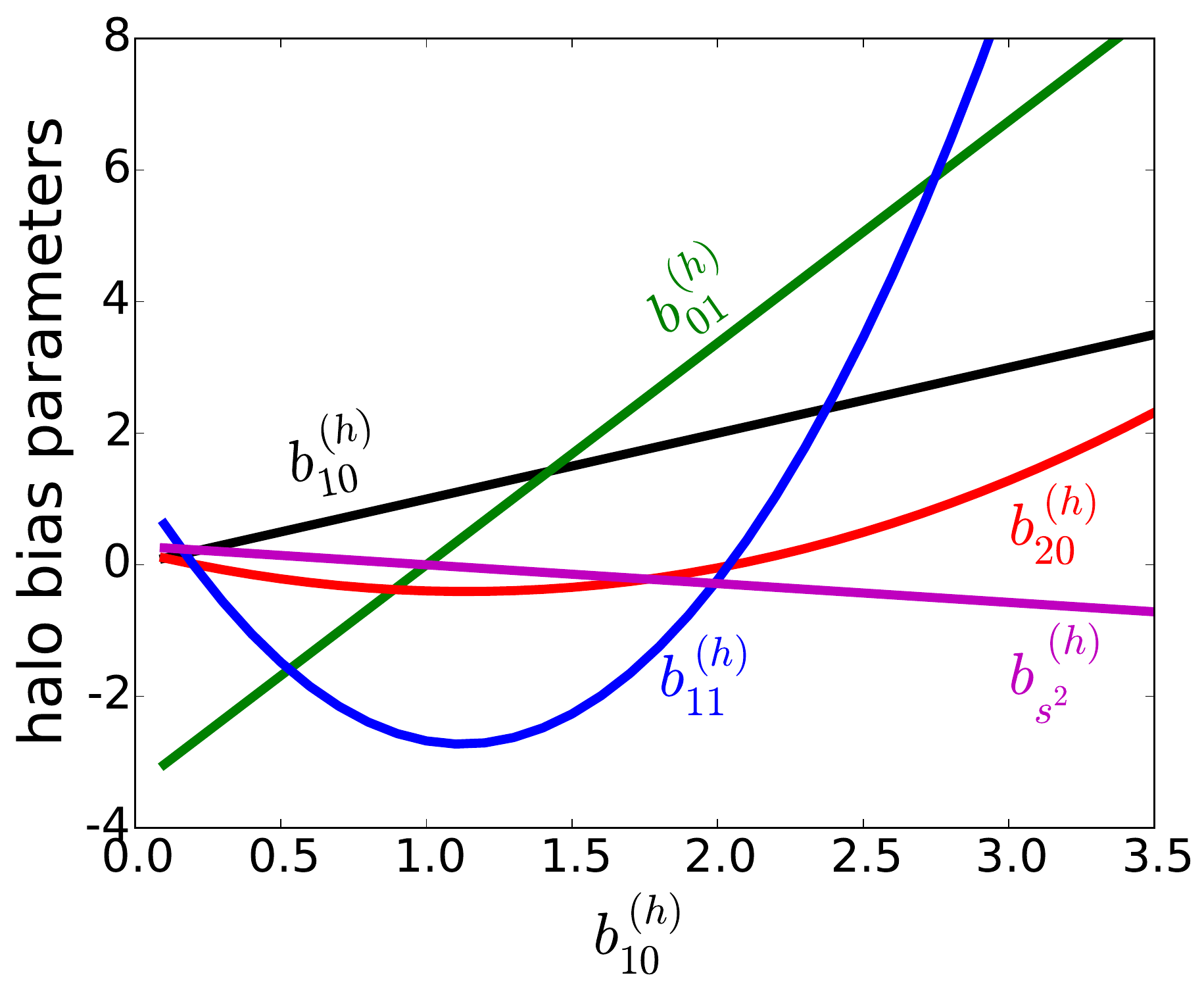} 
\includegraphics[width=0.47\textwidth]{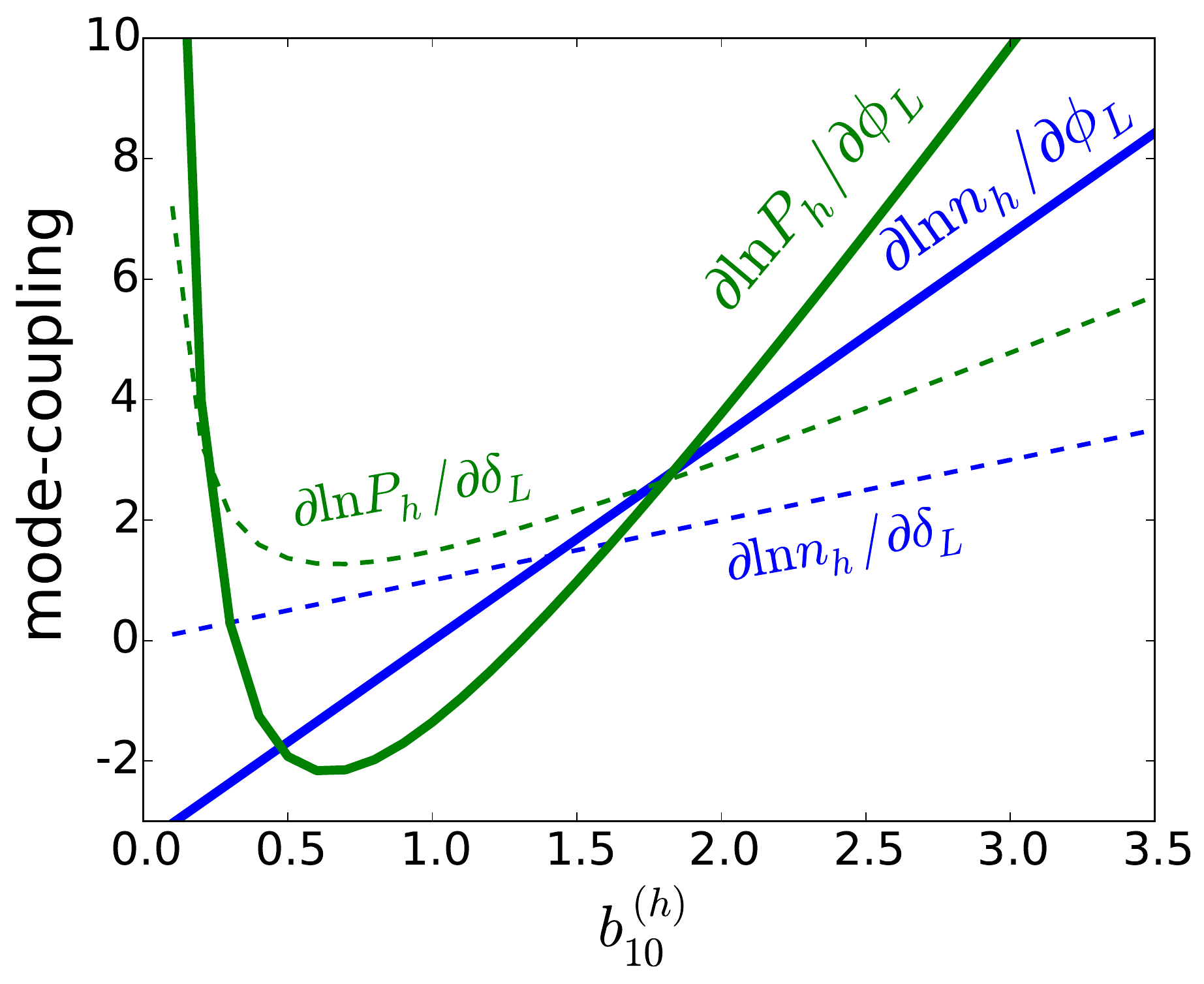} 
\caption{{\it Left:} Halo bias parameters as a function of linear bias, $b_{10}^{(h)}$.
All biases are Eulerian. $b_{20}^{(h)}$ (red) is based on simulations (Eq.~(5.2) of \cite{lazetal16}),
and $b_{01}^{(h)}$ (green) and $b_{11}^{(h)}$ (blue) are derived assuming a universal halo mass function (Eq.~(\ref{eq:b01 b11})). The tidal-tensor bias $b_{s^2}^{(h)}$ assumes zero Lagrangian tidal-tensor bias (Eq.~(\ref{eq:bs2})).
{\it Right:} The response of the halo overdensity ($\delta_h = \delta \ln n_h$) and the position-dependent halo power spectrum ($\delta_{2h(i)} = \delta \ln P_h$) to long-mode perturbations.
Solid curves show the response to $\fnl \, \phi_L$ due to primordial non-Gaussianity, i.e.~the quantities $b_h'(q_L) \, \M(q_L)$ and $b_{2h(i)}'(q_L) \, \M(q_L)$. Dashed curves show the non-primordial responses to $\delta_L$, i.e.~$b_h$ and $b_{\bar{2h}}$. The latter is the bias/response of the position-dependent halo spectrum {\it averaged} over short modes, $\delta_{\bar{2h}}$ (see Appendix \ref{app:d2bar}).
}
\label{fig:biases2}
\includegraphics[width=0.47\textwidth]{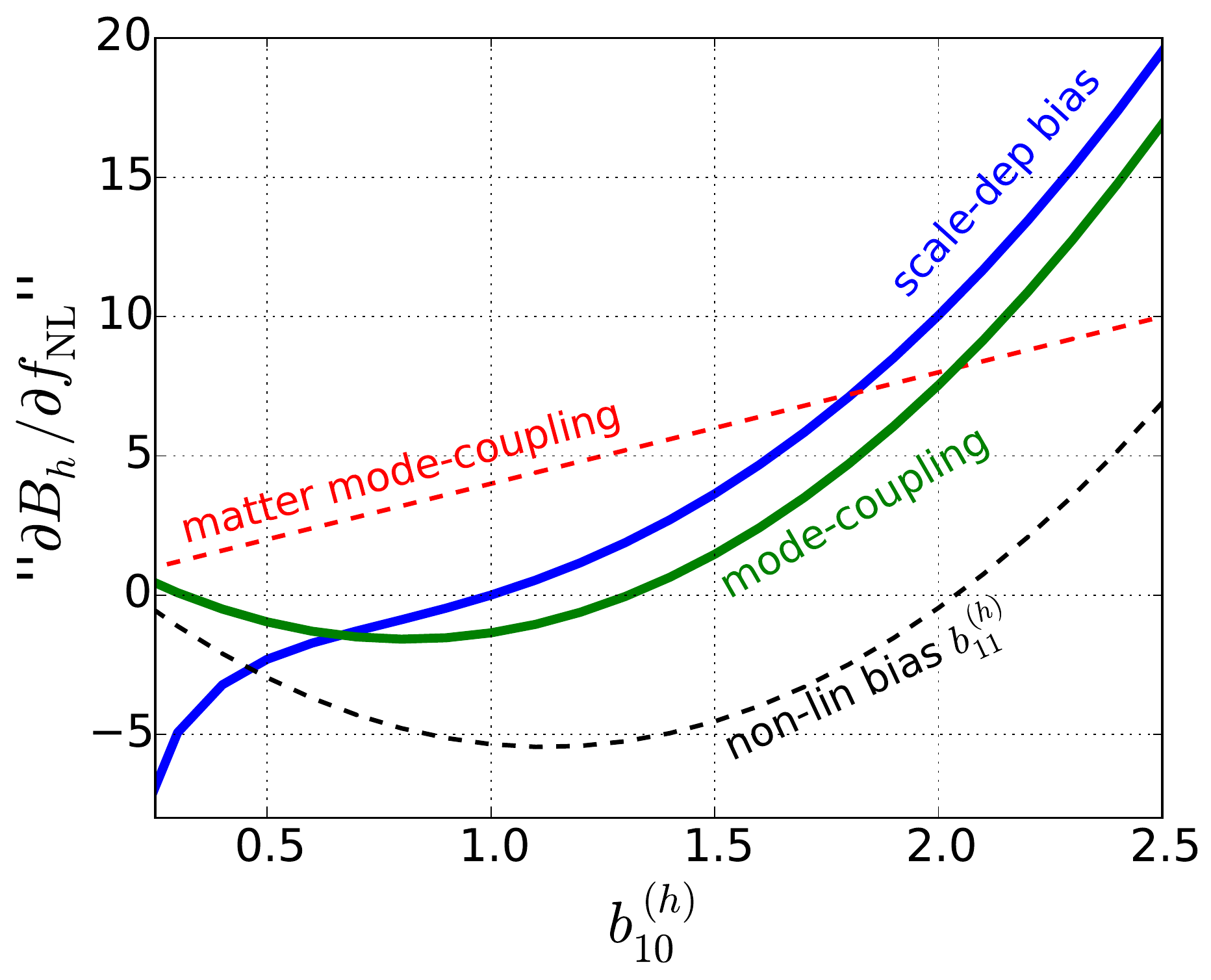} 
\caption{The signal from primordial mode-coupling in the squeezed-limit halo bispectrum, broken down into its seperate contributions. We assume the relations between halo bias parameters shown in Figure \ref{fig:biases2}.
We quantify the signal as the derivative of the halo bispectrum with respect to $\fnl$, averaged over short modes,
and normalized by $P_{\phi \delta}(q_L) \, b_{10}^{(h) \, 2} \, P(k_i)$ (see text for details).
It contains two distinct contributions: explicit halo mode-coupling (green), i.e.~the modulation of the position-dependent halo power spectrum, explicitly $b_h \, b_{\bar{2h}}'(q_L) \, \M(q_L)$,
and scale-dependent halo bias (blue), explicitly $b_{\bar{2h}} \, b_{h}'(q_L) \, \M(q_L)$. The former signal is the sum of a contribution due to intrinsic matter mode-coupling (red dashed), and a contribution from the non-linear halo bias $b_{11}^{(h)}$ (black dashed).
}
\label{fig:BKhcontributions}
\end{figure*}

To forecast constraints, we as before assume a survey of volume $V = 100 \, (h^{-1}$Gpc$)^3$ (remember that for a fixed range of scales/configurations, $\sigma(\fnl) \propto V^{-1/2}$) and an effective redshift $z = 1$.
We remind the reader that $V = 100 \, (h^{-1} $Gpc$)^3$ is less than the volume covered by galaxy samples expected from some planned/proposed surveys and, accordingly, it will in principle be possible to reach lower $\sigma(\fnl)$ values than discussed here.
For the halo sample, in addition to the default number density,  $\bar{n}_h = 3 \cdot 10^{-4} \, (h^{-1}$Mpc$)^{-3}$, and fiducial linear bias, $b_{10}^{(h)} = 2$, we now also need to specify the higher order bias parameters.
To do this, for $b_{20}^{(h)}$, we use the fitting formula, obtained from N-body simulations,
in Eq.~(5.2) of\footnote{Note the difference in convention with \cite{lazetal16}, $b_{2}^{\rm here} = \tfrac{1}{2} \, b_{2}^{\rm there}$} \cite{lazetal16}. For the non-Gaussian bias parameters $b_{01}^{(h)}$ and $b_{11}^{(h)}$, we
make the assumption of a universal halo mass function, leading to Eq.~(\ref{eq:b01 b11}).
Finally, we assume zero Lagrangian tidal-tensor bias, leading to the expression for $b_{s^2}^{(h)}$
given in Eq.~(\ref{eq:bs2}).
We refer to Appendix \ref{app:biasparams} for more details.
Figure \ref{fig:biases2} (left) shows the resulting halo bias parameters as a function of $b_{10}^{(h)}$.
The fiducial values ($b_{10}^{(h)} = 2$) are summarized in Table \ref{table:fiducial}.

\begin{figure*}[]
\centering
\includegraphics[width=0.47\textwidth]{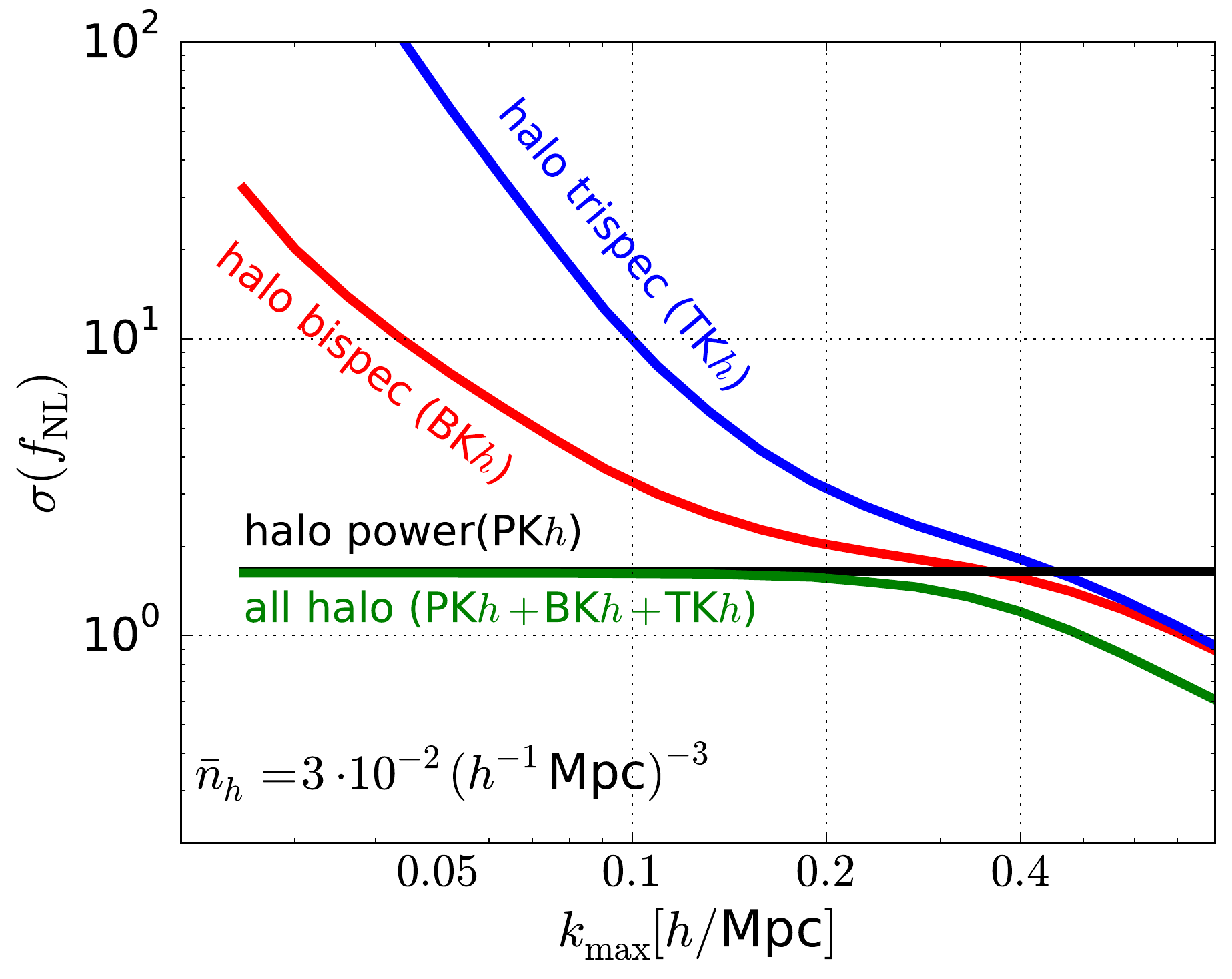} 
\includegraphics[width=0.47\textwidth]{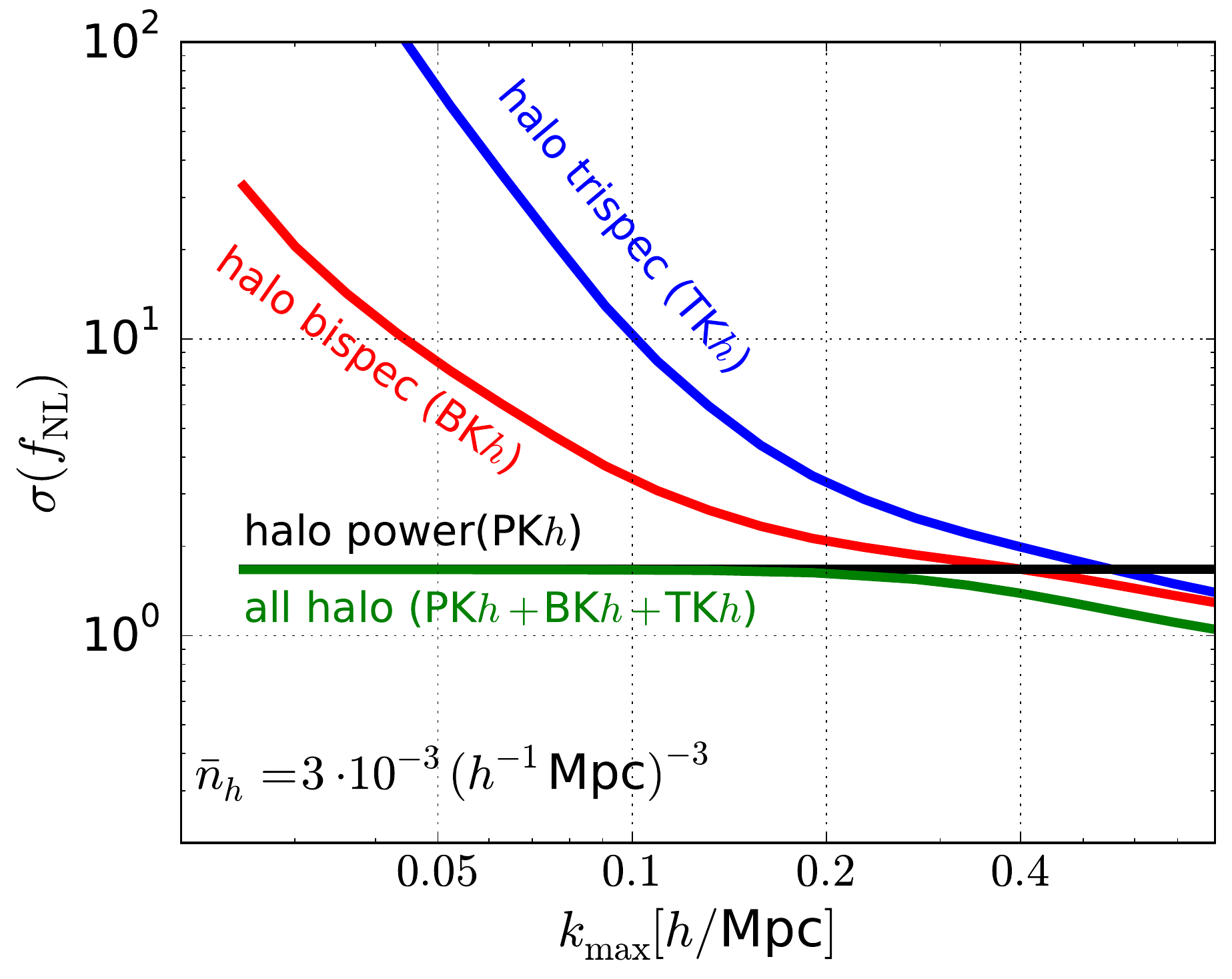} 
\caption{As Figure \ref{fig:halos vary nbarh}, but showing intermediate halo number densities.
Higher order statistics may be competitive with the halo power spectrum (and a joint analysis may improve over the halo power spectrum only) at high number densities,
$\bar{n}_h \gtrsim 3 \cdot 10^{-3} \, (h^{-1}$Mpc$)^{-3}$ (for our fiducial halo sample with $b_{10}^{(h)} = 2$, $z = 1$, etc.).}
\label{fig:halos intermediate nbarh}
\end{figure*}

For the fiducial bias model discussed above, Figure \ref{fig:halos vary nbarh} shows $\sigma(\fnl)$ vs.~$k_{\rm max}$ for the main halo statistics (see Figure \ref{fig:halos intermediate nbarh} for intermediate halo number densities).
It is instructive to first consider the limit of zero shot noise in the halo overdensity, $\bar{n}_h \to \infty$, shown in the left panel.
In this case, the halo power spectrum constraint (black horizontal line) reaches its cosmic variance limited
value. Except for $k_{\rm max} > 0.2 h/$Mpc, the higher order halo statistics are not competitive with the
power spectrum.
In the large $k_{\rm max}$ limit, however, the stochastic noise in the position-dependent halo power spectrum goes to zero, and the higher order statistics gain in constraining power. In particular, the ``multitracer'' cosmic variance cancellation becomes activated at large $k_{\rm max}$ so that, at the right end of the plot, the constraints with higher order statistics keep improving with $k_{\rm max}$. The halo bispectrum on its own always does better than the halo trispectrum only, and it surpasses the halo power spectrum at $k_{\rm max} \approx 0.4 h/$Mpc. A joint analysis of all statistics starts to improve on the PK$h$ case already at $k_{\rm max} > 0.2 h/$Mpc.
It is worth noting, however, that one can in principle do much better than PK$h$ with $2$-point functions alone
if one can perform a multitracer analysis of multiple halo samples with different biases (or if one can combine with a direct measurement of the matter overdensity).

\begin{figure*}[]
\centering
\includegraphics[width=0.55\textwidth]{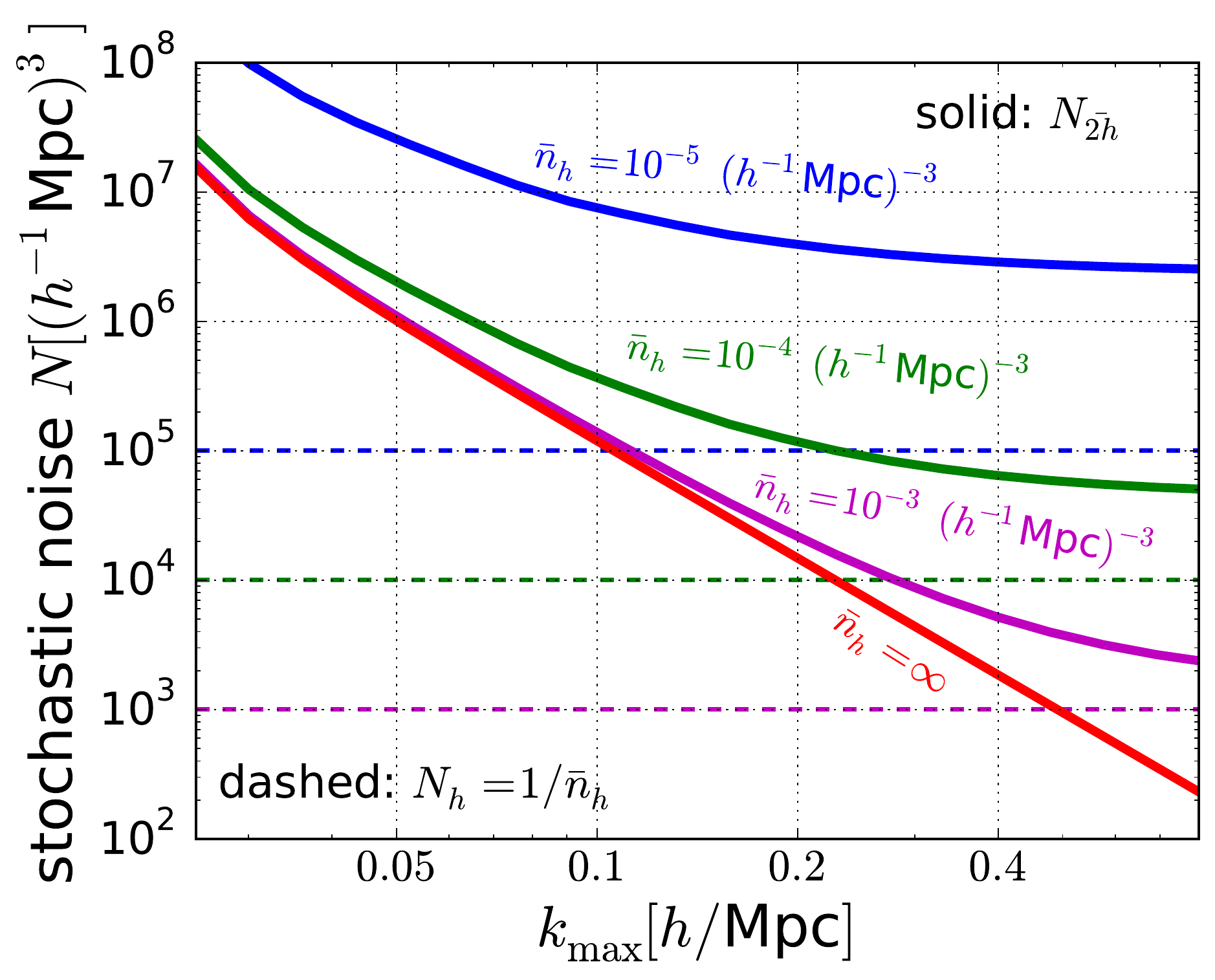} 
\caption{Solid curves show $N_{\bar{2h}}$, the stochastic noise in the position-dependent halo power spectrum averaged over short modes (see Appendix \ref{app:d2bar}), for various halo number densities $\bar{n}_h$ (assuming the fiducial halo bias model with $b_{10}^{(h)} = 2$, etc.). The result in the limit $\bar{n}_h = \infty$ is the same as the effective shot noise for the position-dependent {\it matter} power spectrum.
Dashed curves show the stochastic noise in the halo overdensity field itself, $N_h = 1/\bar{n}_h$. For a given halo number density, the latter is typically significantly lower than the former, even at high $k_{\rm max}$. This partially explains why, except at high number densities, it is hard for higher order halo statistics to compete with scale-dependent halo bias in the power spectrum.}
\label{fig:shotnoise}
\end{figure*}

Next consider including the effect of halo shot noise, i.e.~using a finite halo number density (right panel). The effect of this is not just to introduce shot noise in the measurement of the halo overdensity $\delta_h$, but also to increase the ($k_{\rm max}$ dependent) stochastic noise in the position-dependent halo power spectrum $\delta_{2h(i)}$. This is illustrated in Figure \ref{fig:shotnoise}, which shows the stochastic noise in both quantities as a function of $k_{\rm max}$.
To visualize the stochastic noise in the set of $\delta_{2h(i)}$ tracers, the figure shows the noise power spectrum of the position-dependent power spectrum {\it averaged} over short modes, $\delta_{\bar{2h}}$ (see Appendix \ref{app:d2bar} for details).
The plot clearly shows that a finite halo shot noise leads to a minimum possible value (a floor) for the stochastic noise $N_{\bar{2h}}$, effectively setting a maximum $k_{\rm max}$ beyond which $N_{\bar{2h}}$ cannot be lowered. This minimum stochastic noise in the position-dependent power spectrum is typically significantly above the shot noise in the halo overdensity.

The effect of these two manifestations of halo shot noise on the information content is shown in the right panel of Figure \ref{fig:halos vary nbarh}. First, we see that the constraint from PK$h$ is slightly weakened.
More dramatically, for the higher order statistics, the stochastic noise essentially cuts off the improvement with increasing $k_{\rm max}$ (as compared to the left panel) at the point where the halo shot noise becomes important.
For this particular number density, this happens well before the bispectrum has a chance to surpass the power spectrum. Even the joint constraint is not notably better than PK$h$ alone, regardless of how large $k_{\rm max}$ is made.
The above discussion shows that, in order for the higher order statistics to improve over PK$h$ alone, one wants to be in the large $\bar{n}_h$, large $k_{\rm max}$ regime.
In that case, $\sigma(\fnl)$ from PK$h$ reaches an error floor due to cosmic variance, whereas higher order statistics apply cosmic variance cancellation to improve $\sigma(\fnl)$ below the naive cosmic variance limit. Thus, even though Figure \ref{fig:shotnoise} shows that the stochastic noise $N_{\bar{2h}}$ will always be higher than the shot noise in the halo overdensity, $N_h$, in this scenario, more information can be extracted from the tracers $\delta_{2h(i)}$, than from PK$h$ only.
We emphasize that for this cosmic variance cancellation to become effective, one wants the $\delta_{2h(i)}$ to be deep into the cosmic variance dominated regime. In other words, one needs very low stochastic noise, which can only be reached for high number densities and large $k_{\rm max}$.

To make this more quantitative, in our fiducial halo bias model at $z = 1$, in the absence of halo shot noise, Figure \ref{fig:halos vary nbarh} (left) shows that higher order statistics start improving appreciably over the power spectrum at $k_{\rm max} \gtrsim 0.4 h/$Mpc. Comparing Figure \ref{fig:halos vary nbarh} with Figure \ref{fig:halos intermediate nbarh} shows that, once the number density drops below $\bar{n}_h \approx 3 \cdot 10^{-3} \, (h^{-1}$Mpc$)^{-3}$, this is no longer the case, and at that point the higher order statistics do not significantly improve over the $2$-point function at any $k_{\rm max}$ (as is clearly the case for $\bar{n}_h \approx 3 \cdot 10^{-4} \, (h^{-1}$Mpc$)^{-3}$ in the right panel of Figure \ref{fig:halos vary nbarh}).

\begin{figure*}[]
\centering
\includegraphics[width=0.55\textwidth]{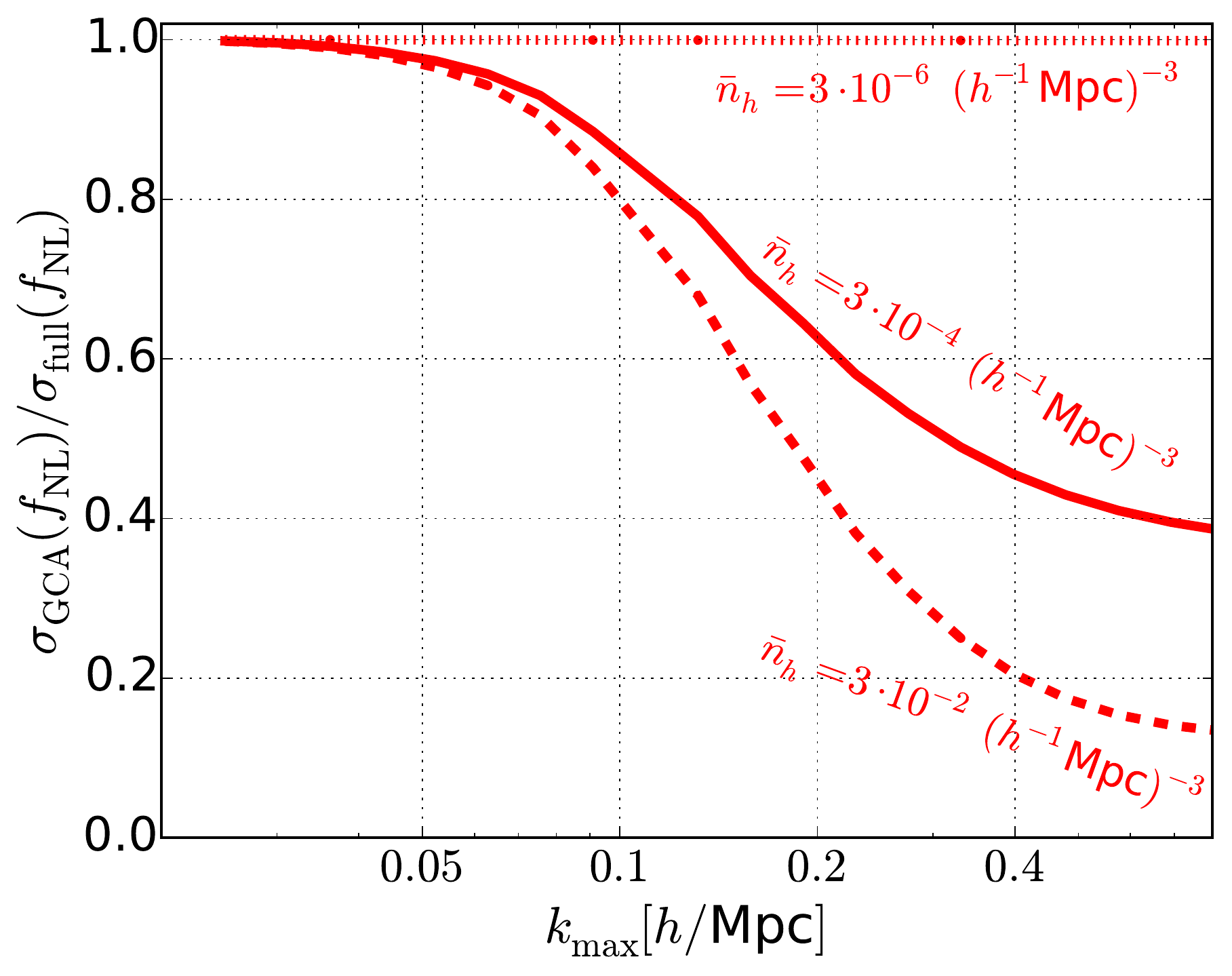} 
\caption{Ratio of $\sigma(\fnl)$ from halo bispectrum in the Gaussian covariance approximation to the uncertainty including non-Gaussian covariance due to non-linear evolution (cf.~right panel of Figure \ref{fig:sigma bispec} for the case of the matter bispectrum). The solid curve shows the result for the default halo number density, $\bar{n}_h = 3 \cdot 10^{-4} \, (h^{-1}$Mpc$)^{-3}$ and the other two curves show the cases of very high and very low number density.}
\label{fig:BKh GCA}
\end{figure*}

We next compare the halo bispectrum constraint to that obtained making the Gaussian covariance approximation.
Figure \ref{fig:BKh GCA} shows the ratio of $\sigma(\fnl)$ under the GCA to the full result (assuming the fiducial bias parameters, etc), cf.~Figure \ref{fig:sigma bispec}.
The GCA strongly underestimates the uncertainty on $\fnl$ in the regime of low stochastic noise in $\delta_{2h(i)}$, which is reached for high $k_{\rm max}$ and $\bar{n}_h$. For $\bar{n}_h = 3 \cdot 10^{-4} \, (h^{-1} $Mpc$)^{-3}$, the GCA underestimates $\sigma(\fnl)$ by up to a factor $\sim 2.5$.
Whereas for the {\it matter} bispectrum, the GCA underestimates $\sigma(\fnl)$ by about a factor of three in the high $k_{\rm max}$ limit, for the halo bispectrum, in the zero stochastic noise limit (large $k_{\rm max}$ and $\bar{n}_h$), it can be up to a factor $\sim 10$.
In conclusion, one has to be very careful not to overestimate the constraining power of the halo bispectrum, especially when taking advantage of the information on very small scales (high $\bar{n}_h, k_{\rm max}$).

\begin{figure*}[]
\centering
\includegraphics[width=0.47\textwidth]{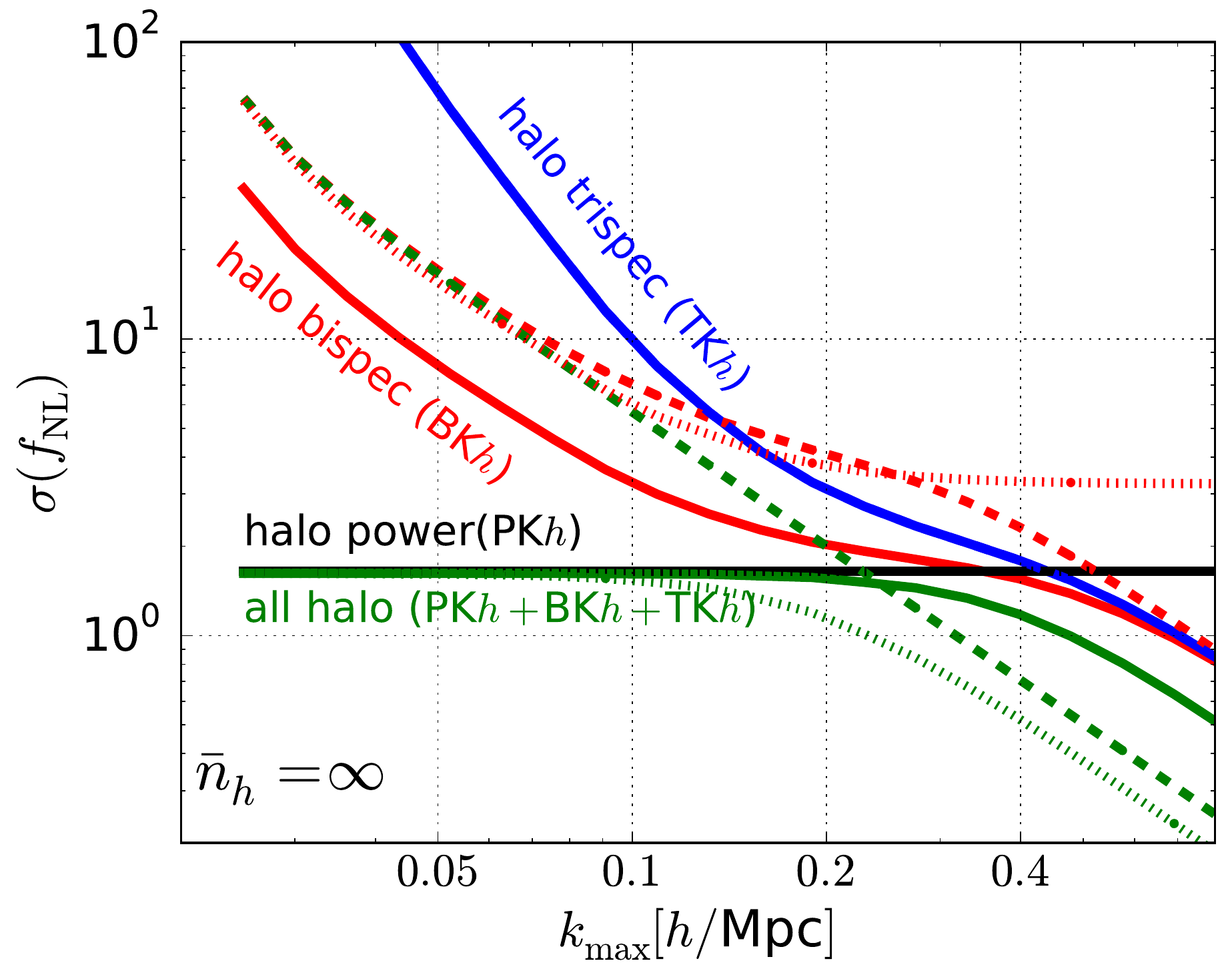} 
\includegraphics[width=0.47\textwidth]{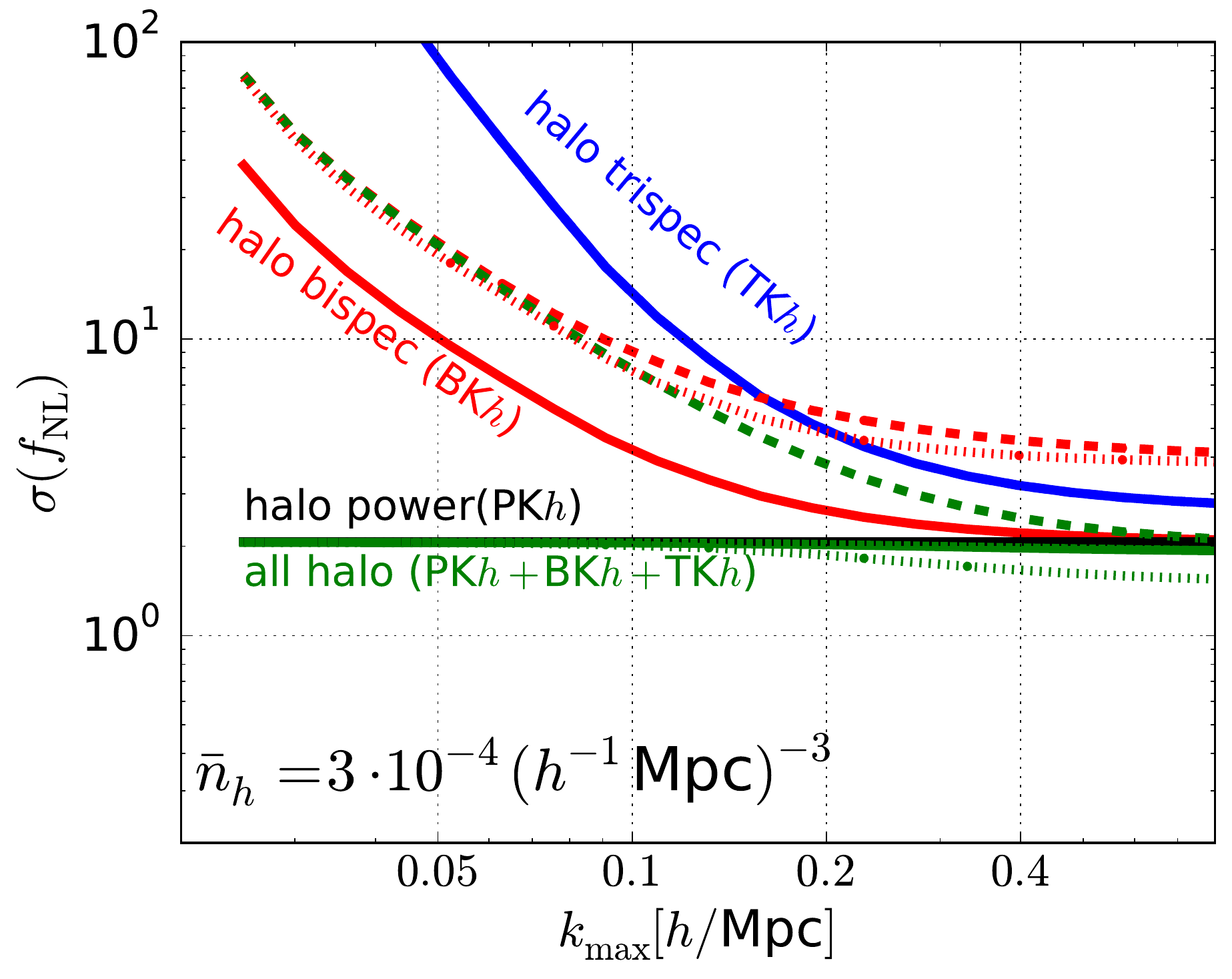} 
\caption{As Figure \ref{fig:halos vary nbarh} (solid curves), but now adding curves showing the information content when only the signal (i.e.~the $\fnl$ dependence) from scale-dependent halo bias/$\delta_h$ (dotted) or only that from halo-mode coupling/$\delta_{2h(i)}$ (dashed) is included (by setting $b_{2h(i)}'=0$ and $b_h' = 0$, respectively). For the halo trispectrum, the dashed and solid curves coincide and the dotted curve is absent (there is no information from long-mode scale-dependent halo bias in the collapsed halo trispectrum). For the halo power spectrum, the dotted and solid curves coincide and the dashed curve is absent.}
\label{fig:db vs mc}
\end{figure*}

\vskip 7pt

Let us now examine in more detail the information content of the halo bispectrum, and in particular its dependence on the halo bias parameters.
The halo bispectrum contains two types of primordial non-Gaussianity signals: the modulation of the halo overdensity, $\delta_h$, and the modulation of the position-dependent halo power spectra, $\delta_{2h(i)}$.
The importance of these two signals depends on the halo bias parameters. Assuming the relations discussed above,  the left panel of Figure \ref{fig:biases2} shows all halo bias parameters
as a function of $b_{10}^{(h)}$, and the right panel shows the resulting signals in $\delta_h$ and
$\delta_{2h(i)}$.
In particular, the solid curves in the right panel give the coefficients of $\fnl \, \phi_L$, or in other words $b_{h}'(q_L) \, \M(q_L)$ (blue) and $b_{2h(i)}'(q_L) \, \M(q_L)$ (green).
The scale-dependent bias signal in $\delta_h$ is proportional to $(b_{10}^{(h)} - 1)$.
The signal in $\delta_{2h(i)}$ is the sum of two terms: the primordial modulation of the position-dependent matter power spectrum, with amplitude $\pa \ln P_h/\pa (\fnl \phi_L) = 4$, and a contribution from $b_{11}^{(h)}$,
giving $\pa \ln P_h/\pa (\fnl \phi_L) = 2 b_{11}^{(h)}/b_{10}^{(h)}$.
For our fiducial model, $b_{10}^{(h)} = 2$, the left panel shows that the $b_{11}^{(h)}$ contribution is minimal.
For larger $b_{10}^{(h)}$, the effect of the non-linear, non-Gaussian halo bias
is to add to the modulation signal. However, it is interesting to note that for $b_{10}^{(h)} < 2$, it actually
negatively interferes with the primordial matter mode-coupling, lowering the signal in $\delta_{2h(i)}$.
Comparing the signals in $\delta_h$ and $\delta_{2h(i)}$ in the right panel for our fiducial bias $b_{10}^{(h)} = 2$, we see that the two primordial modulations are of very similar amplitude, with that of the position-dependent halo power spectrum the larger of the two.

The halo bispectrum contains both of the above signals, but weighted differently,
i.e.~the derivative of the bispectrum w.r.t.~$\fnl$ is proportional to
$b_h \, b_{2h(i)}' + b_{2h(i)} \, b_h'$.
Figure \ref{fig:BKhcontributions} shows these contributions
separately for the halo bispectrum averaged over short modes (see Appendix \ref{app:d2bar}),
concretely, $b_h \, b_{\bar{2h}}'(q_L) \, \M(q_L)$ (green) and $b_{\bar{2h}} \, b_{h}'(q_L) \, \M(q_L)$ (blue).
For $b_{10}^{(h)} = 2$, the previous conclusion that both types of signal yield comparable contributions stands, but with the weighting included, we now find the scale-dependent bias contribution to be slightly more important.
The dashed curves indicate the individual contributions to the halo mode-coupling/position-dependent halo power spectrum from the position-dependent matter power spectrum (red) and from $b_{11}^{(h)}$ (black).
We again see that for $b_{10}^{(h)} < 2$, the effect on the halo mode-coupling from non-linear biasing is actually negative (compared to the effect from the matter mode-coupling).

Now that we have discussed the contributions to the halo bispectrum signal from the primordial mode-coupling signals in $\delta_{2h(i)}$ and $\delta_h$, we next turn to how $\sigma(\fnl)$ depends on these two signals.
Figure \ref{fig:db vs mc} shows $\sigma(\fnl)$ for the same data combinations as in Figure \ref{fig:halos vary nbarh} (solid curves), but now also showing the constraints from the signal in $\delta_{2h(i)}$ only (dashed)
and in $\delta_h$ only (dotted). These cases are obtained from the previously discussed expressions for the Fisher information by setting $b_h' = 0$ and $b_{2h(i)}' = 0$ respectively.

Let us first consider the left panel ($\bar{n}_h = \infty$).
For the halo bispectrum (red), we see that at low $k_{\rm max}$, the constraints from halo mode-coupling only and scale-dependent halo bias only are similar to each other, as expected, but weaker than the full constraint that takes into account both signals.
Thus, in particular, due to the scale-dependent bias contribution, the halo bispectrum performs significantly better than if only halo mode-coupling had been present
(and better than the matter bispectrum).
At high $k_{\rm max}$, the case of only halo mode-coupling incorporates multitracer cosmic variance cancellation, leading to decreasing $\sigma(\fnl)$ with increasing $k_{\rm max}$, whereas the case of only scale-dependent halo bias reaches a cosmic variance limited plateau.
In the limit of very large $k_{\rm max}$ ($N_{\bar{2h}} \to 0$), the $\sigma(\fnl)$ curves from the full halo bispectrum and that from only the halo mode-coupling signal in the bispectrum overlap.
This can be shown explicitly from Eq.~(\ref{eq:BKh}) (taking the limit $\bar{n}_h \to \infty$ and using $\Sigma_{2h} \gg 1$), which in this limit gives $F \to \Sigma_{2h}'' - \left( \Sigma_{2h}' \right)^2/\Sigma_{2h}$. This asymptotic result is equivalent to that from the {\it matter} bispectrum, Eq.~(\ref{eq:fishermatbispec}), in the high $k_{\rm max}$ limit, with the replacements $b_{2(i)} \to b_{2h(i)}$ and $b_{2(i)}' \to b_{2h(i)}'$.

The halo trispectrum contains only information from halo mode-coupling in the first place so that the solid and dashed curves overlap and there is no constraint from scale-dependent bias information only.
The converse holds for the halo power spectrum.
For the joint information from all three probes, at low $k_{\rm max}$, the case of $\delta_h$ signal only is equivalent to the information from PK$h$, while the case of $\delta_{2h(i)}$ signal only is equivalent to BK$h$. At high $k_{\rm max}$, both cases take advantage of cosmic variance cancellation.

An interesting and perhaps counterintuitive feature is that, in this limit, the joint information content including only the scale-dependent bias signal or only the halo mode-coupling is {\it larger} than the actual information content, which contains both signals.
Indeed, taking the low stochastic noise limit (at $\bar{n}_h = \infty$) of Eq.~(\ref{eq:joint halo}) gives
$F \to \Sigma_{2h}'' - 2 \Sigma_h' \, \Sigma_{2h}'/\Sigma_h$, which is lower than $\Sigma_{2h}''$, the joint analysis information (again at $\bar{n}_h = \infty$) in the absence of the scale-dependent bias signal (i.e.~the dashed green curve).
Since the latter quantity is approximately equal to the information from
a joint analysis of {\it matter} statistics (which has Fisher information $F = \Sigma_2''$, the same as $\Sigma_{2h}''$ up to the change $b_{2(i)}' \to b_{2h(i)}'$),
the joint halo analysis contains less information (at least for the chosen bias parameters) than a joint analysis of the higher order {\it matter} statistics.
The physical reason is that, at high $k_{\rm max}$, the cosmic variance cancellation becomes less efficient when both signals are included.
This statement depends strongly on the bias parameters. In particular, for $b_{10}^{(h)} < 1$, $b_{01}^{(h)}$ becomes negative so that the sign of the term $2 \Sigma_h' \, \Sigma_{2h}'/\Sigma_h$ above changes. In this case (not shown), the joint analysis of halo statistics performs better than that of matter statistics.

\begin{figure*}[]
\centering
\includegraphics[width=0.47\textwidth]{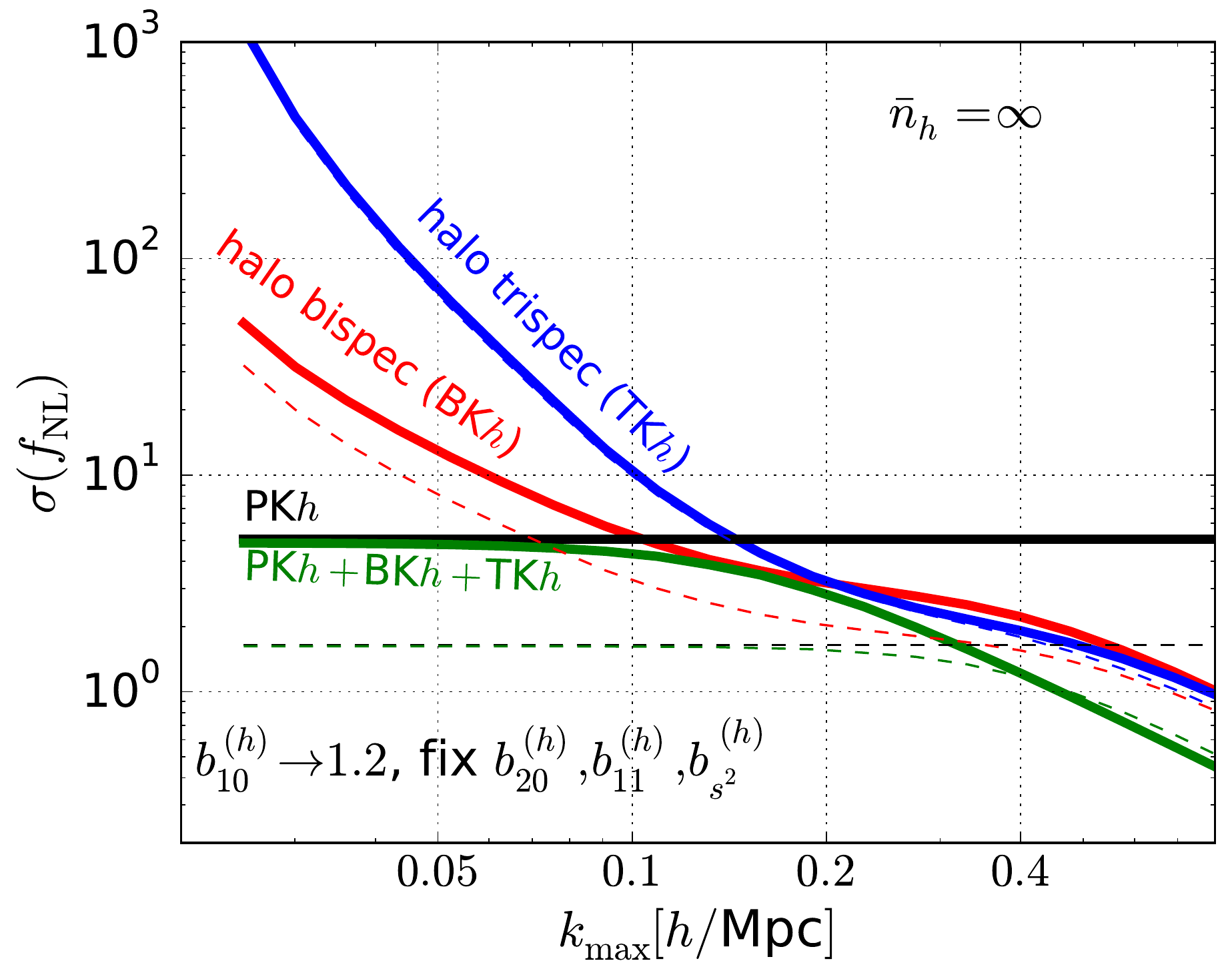} 
\includegraphics[width=0.47\textwidth]{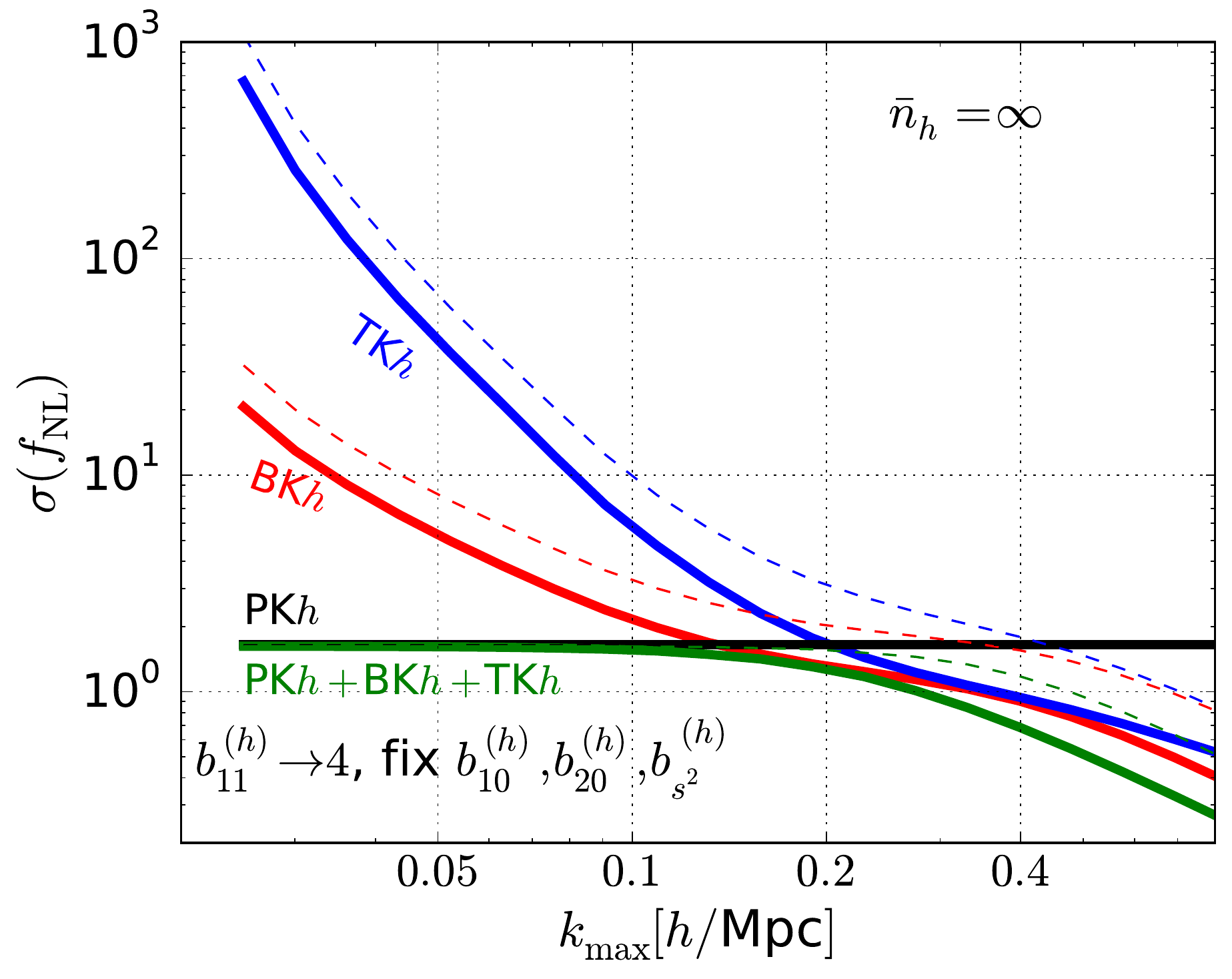} 
\caption{As Figure \ref{fig:halos vary nbarh}, but now varying fiducial bias parameters.
We here show only the limit of zero shot noise in the halo overdensity ($\bar{n}_h = \infty$).
For ease of comparison, the thin dashed curves repeat the results with the fiducial bias parameters (left panel of Figure \ref{fig:halos vary nbarh}).
{\it Left:} The linear bias is decreased to $b_{10}^{(h)} = 1.2$. We adjust $b_{01}^{(h)} \propto (b_{10}^{(h)} - 1)$ accordingly, but keep all other bias parameters fixed to their fiducial values. The lower scale-dependent bias signal lowers the information from the halo power spectrum and, relative to PK$h$, the higher order statistics become more powerful. However, in an absolute sense, all statistics perform worse than in the fiducial bias model ($b_{10}^{(h)} = 2$).
{\it Right:} The non-Gaussian, non-linear bias is increased to the (extremely large) value $b_{11}^{(h)} = 4$ (cf.~the fiducial value $b_{11}^{(h)} = -0.23$), while keeping all other parameters fixed. The result is to strongly boost the halo mode-coupling signal, thus improving the performance of the higher order statistics.}
\label{fig:halos vary bias1}
\end{figure*}

The case with non-zero halo shot noise ($\bar{n}_h = 3 \cdot 10^{-4} \, (h^{-1} $Mpc$)^{-3}$, right panel)
can again be understood starting from the $\bar{n}_h = \infty$ case, but cutting off improvement in $\sigma(\fnl)$ when shot noise becomes important ($k_{\rm max} \gtrsim 0.2 h/$Mpc).
We reiterate that, for moderate halo number density, this stochastic noise strongly weakens the performance of the higher order statistics.

Finally, we find that the tidal-tensor bias, $b_{s^2}^{(h)}$, which does not directly contribute to the modulation of $\delta_{h,L}$ or $\delta_{2h(i),L}$ by $\fnl \, \phi_L$ (i.e.~to the signal) has minimal effect on the constraints (note that the fiducial value is rather small). We will not discuss $b_{s^2}^{(h)}$ further.

\subsection{Varying the halo bias parameters}

Let us now consider how the information content is affected by changing the fiducial halo bias parameters.
As an example, Figure \ref{fig:halos vary bias1} shows the results of lowering the scale-dependent bias signal (left) and increasing the halo mode-coupling signal (right).
In both cases, we show only the zero halo shot noise limit, $\bar{n}_h = \infty$.
In particular, in the left panel we have lowered the linear bias $b_{10}^{(h)} = 2 \to 1.2$, changing $b_{01}^{(h)} \propto (b_{10}^{(h)} - 1)$ accordingly, while keeping all other halo bias parameters fixed.
The dominant effect is to lower the scale-dependent halo bias signal. This leads to weakened constraints from PK$h$ compared to the default halo bias model. Since the information content of the higher order statistics is weakened less, their relative performance compared to the halo power spectrum only improves.

In the right panel of Figure \ref{fig:halos vary bias1}, we have changed the value of the non-Gaussian, non-linear bias $b_{11}^{(h)} = -0.23 \to 4$ (note that this may be an unrealistically large value, cf.~left panel of Figure \ref{fig:biases2}), keeping all other bias parameters fixed.
This significantly increases the halo mode-coupling signal, $\pa P_h/\pa (\fnl \phi_L) \approx 4 \to 8$ and we indeed see a significant improvement in the performance of the higher order halo statistics compared to the fiducial halo bias model.

\begin{figure*}[]
\centering
\includegraphics[width=0.47\textwidth]{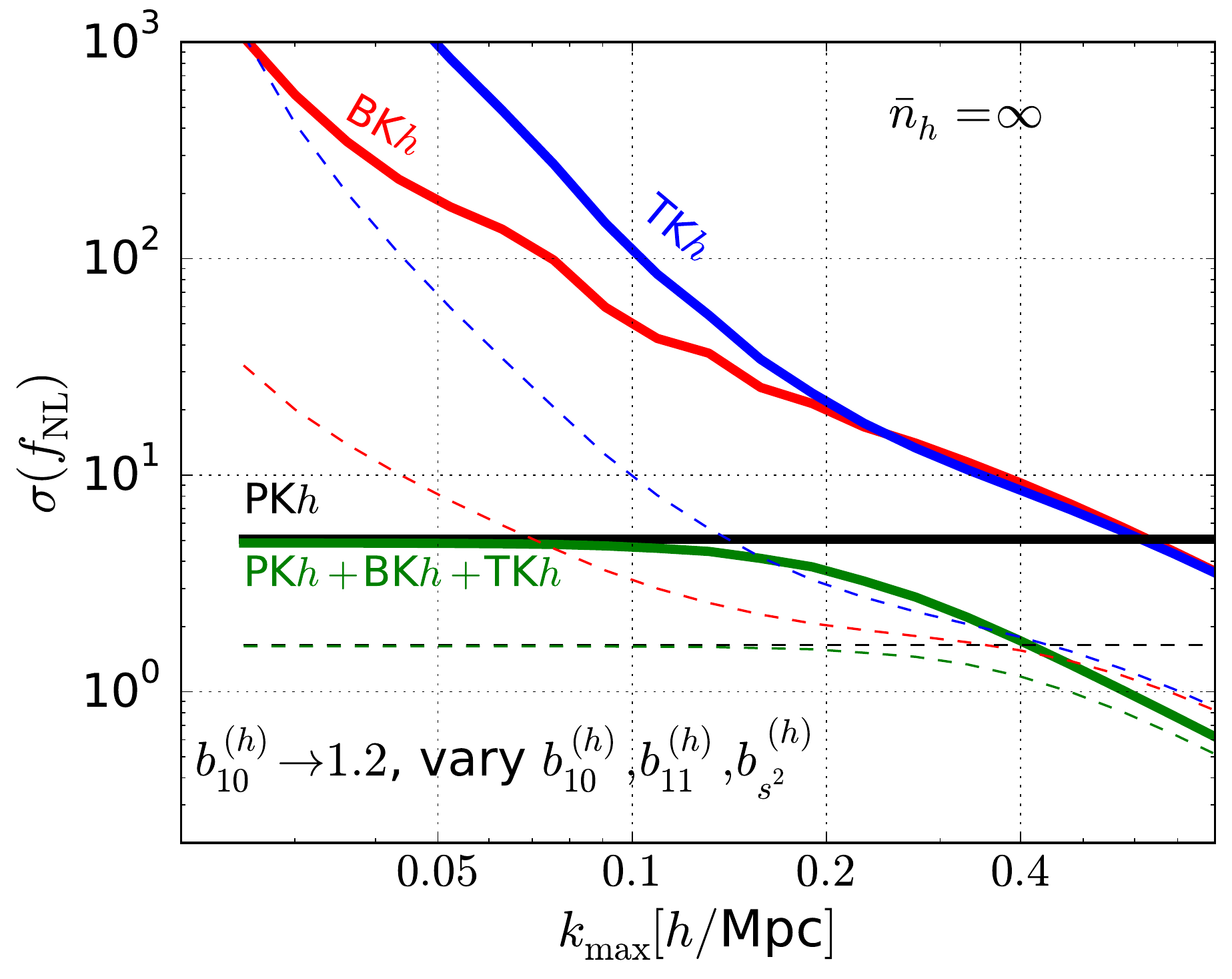} 
\includegraphics[width=0.47\textwidth]{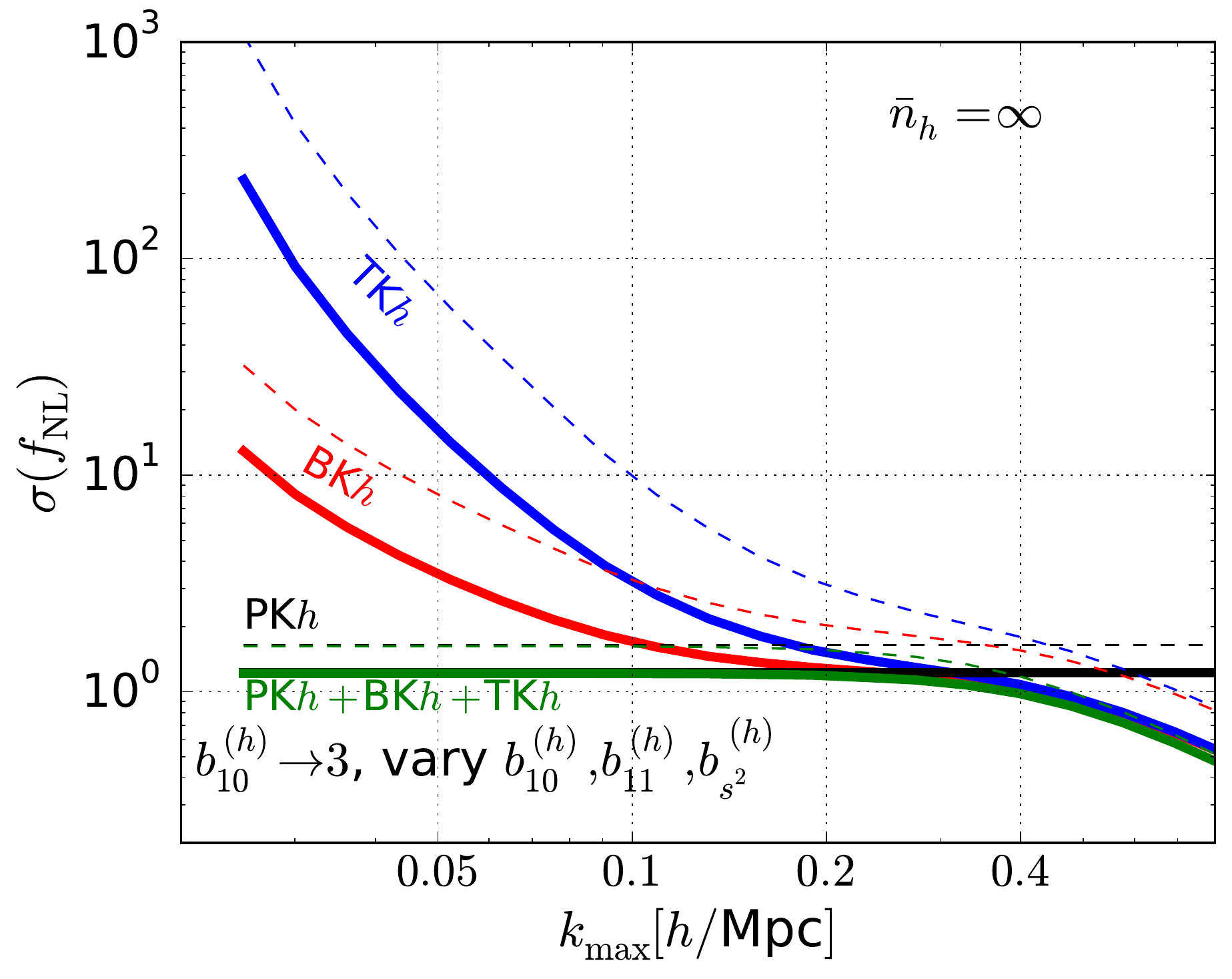} 
\caption{As Figure \ref{fig:halos vary nbarh}, but varying all fiducial halo bias parameters simultaneously according to the relations discussed in the text (and shown in left panel of Figure \ref{fig:biases2}).
We show only the zero halo shot noise limit ($\bar{n}_h = \infty$). For comparison, thin dashed curves depict the results assuming our default halo bias model ($b_{10}^{(h)} = 2$, as in left panel of \ref{fig:halos vary nbarh}).
{\it Left:} Low linear bias, $b_{10}^{(h)} = 1.2$. Both the scale-dependent bias and halo mode-coupling signals are strongly suppressed (the latter due to $b_{11}^{(h)}$) relative to the fiducial bias model, leading to weaker constraints from all probes (see text).
{\it Right:} High linear bias, $b_{10}^{(h)} = 3$. Now both the scale-dependent bias and halo mode-coupling signals are larger than in the fiducial bias model, leading to stronger constraints.}
\label{fig:halos vary bias2}
\includegraphics[width=0.47\textwidth]{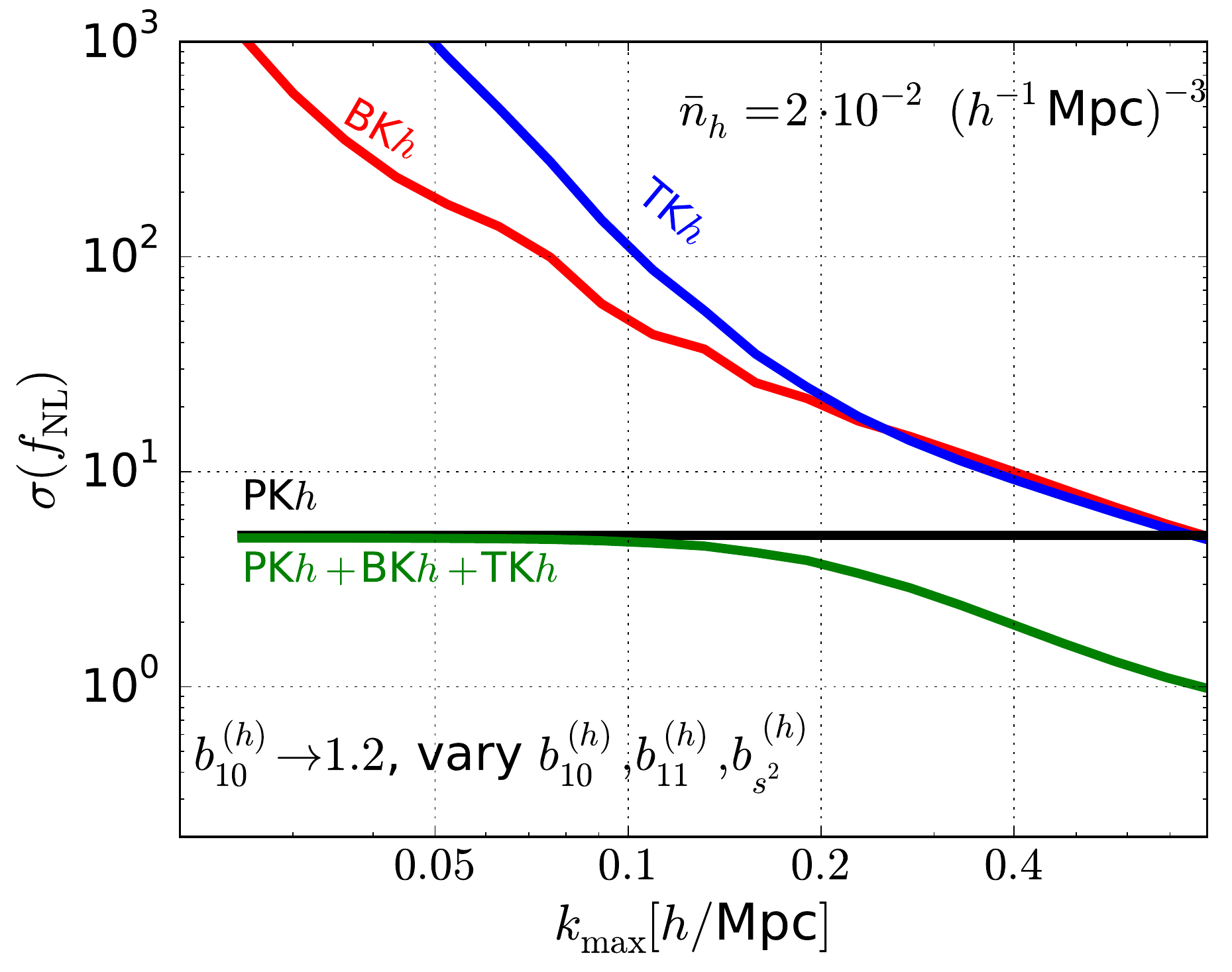} 
\includegraphics[width=0.47\textwidth]{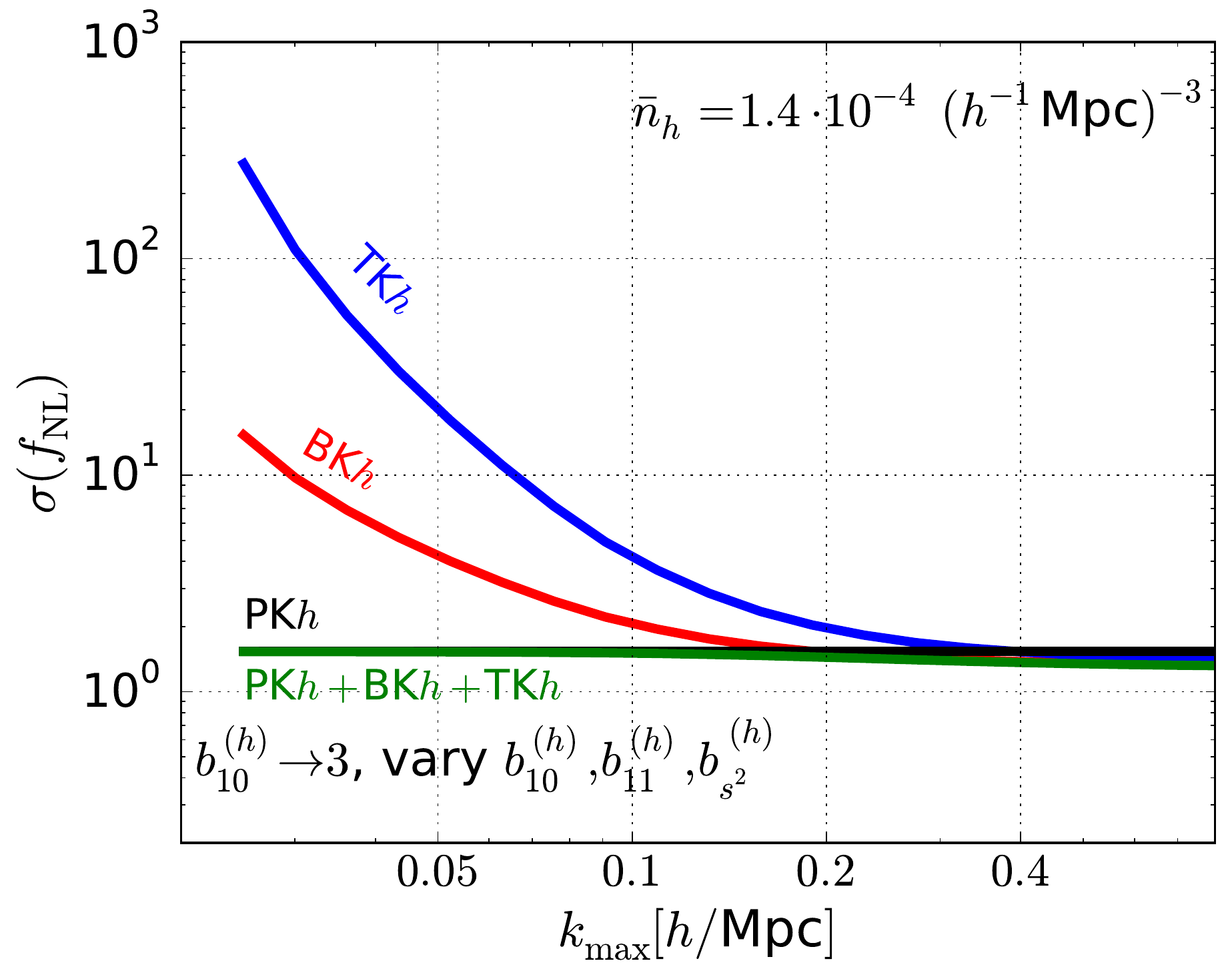} 
\caption{As Figure \ref{fig:halos vary bias2} (without the dashed curves for comparison), but using halo bias-dependent halo number densities (see text).
{\it Left:} Low bias sample: $b_{10}^{(h)} = 1.2$ and $\bar{n}_h = 1.4 \cdot 10^{-4} \, (h^{-1} $Mpc$)^{-3}$.
{\it Right:} High bias sample: $b_{10}^{(h)} = 3$ and $\bar{n}_h = 2 \cdot 10^{-2} \, (h^{-1} $Mpc$)^{-3}$.}
\label{fig:halos vary bias3}
\end{figure*}

In reality, the bias parameters are not independent, but related to each other.
Assuming the relations plotted in the left panel of Figure \ref{fig:biases2}, once a linear bias $b_{10}^{(h)}$ is chosen,
all other bias parameters are fully specified.
Enforcing these relations, in Figure \ref{fig:halos vary bias2}, we now consider a low linear bias (again $b_{10}^{(h)} = 1.2$)
and high linear bias ($b_{10}^{(h)} = 3$) case.
We again show results in the case of zero halo shot noise, $\bar{n}_h = \infty$.

In the low linear bias case, the scale-dependent halo bias signal is suppressed. However, compared to the case discussed above where we left other bias parameters fixed, an important additional effect is the change in $b_{11}^{(h)}$. As can be seen in Figure \ref{fig:biases2}, for $b_{10}^{(h)} = 1.2$, the effect on halo mode-coupling of $b_{11}^{(h)}$ is to almost cancel out the contribution due to primordial matter mode-coupling. Thus, both the signals in $\delta_h$ {\it and} in $\delta_{2h}$ are strongly suppressed, as can be seen for the halo bispectrum in Figure \ref{fig:BKhcontributions}. As a consequence, the left panel of Figure \ref{fig:halos vary bias2} shows a large increase in $\sigma(\fnl)$ for all probes relative to the default halo bias model (thin dashed curves).
Note, however, that for large $k_{\rm max}$, due to cosmic variance cancellation, the joint analysis (PK$h+$BK$h+$TK$h$) is not weakened as much as the individual statistics. Interestingly, combining all probes can in principle lead to significant improvements over individual probes. On the other hand, in the presence of realistic halo shot noise due to finite $\bar{n}_h$, the constraints will again be weakened at large $k_{\rm max}$, even for PK$h+$BK$h+$TK$h$.

In the high linear bias case, both the modulations of the halo number density and of the position-dependent halo power spectrum are increased relative to the default halo bias model, leading to an increased signal in all statistics, cf.~Figure \ref{fig:biases2}, \ref{fig:BKhcontributions}.
Indeed, the right panel of Figure \ref{fig:halos vary bias2} shows significantly improved values of $\sigma(\fnl)$, especially for the higher order halo statistics.

\vskip 7pt

In the zero halo number density limit, the above discussion, as well as an inspection of Figures
\ref{fig:biases2} and \ref{fig:BKhcontributions}, show that samples with large bias $b_{10}^{(h)}$
are optimal for constraining $\fnl$ (at least with the assumed bias relations here).
However, in reality the galaxy or halo number density is not independent of the bias parameters.
In particular, based on the clustering of halos, the maximum number density of a sample with large $b_{10}^{(h)}$ is smaller than that of a sample with low bias. This counteracts the trend that larger bias is better.
As an illustration of this phenomenon, Figure \ref{fig:halos vary bias3} shows the same low and high bias models as in Figure \ref{fig:halos vary bias2}, but now in each case choosing an approximately appropriate halo number density. Specifically, we consider mass limited halo samples, containing all halos above a minimum mass $M_{h,{\rm min}}$, chosen such that {\it mean} linear bias equals $b_{10}^{(h)}$. Using the fitting functions in \cite{tinkeretal08,tinkeretal10}, we then find that the number densities of these samples are $\bar{n}_h \approx 1.4 \cdot 10^{-4}, 2 \cdot 10^{-2} \, (h^{-1} $Mpc$)^{-3}$ for $b_{10}^{(h)} = 1.2$ and $4$ respectively.
While the low bias sample is ``intrinsically'' less constraining, its high number density allows one to further improve constraints by going to high $k_{\rm max}$, whereas stochastic noise precludes further improvements at $k_{\rm max} \gtrsim 0.2 h/$Mpc for the high bias sample.
We see that, if one can push a joint analysis of Pk$h+$BK$h+$TK$h$ of the low bias sample to $k_{\rm max} \gtrsim 0.5 h/$Mpc, there is in principle as much information to be obtained as from the high bias sample.

It would be very interesting to further explore the effect of the relation between halo bias and number density and to perform a full trade study of how to optimize the halo sample for non-Gaussianity constraints. We leave such a study, along with analysis of the expected performance of samples targeted by specific future missions, for future work.

\section{Discussion \& Conclusions}
\label{sec:disc}

Two key large-scale structure probes of primordial mode-coupling are scale-dependent bias in the $2$-point function(s) of  halos on the one hand, and higher order matter or halo statistics, specifically the squeezed-limit bispectrum (and trispectrum), on the other hand.
In this paper, we have performed a comprehensive analysis of the information content of these classes of probes
by taking advantage of the position-dependent power spectrum picture.
The position-dependent power spectra in bins of short modes constitute a set of multiple biased tracers of underlying long-wavelength matter density perturbations,
where the biases of these tracers arise both from long-short mode-coupling generated by non-linear evolution (and biasing) and from primordial mode-coupling. Stochastic noise in the position-dependent power spectra is due to variance in the realizations of the short modes so that the aggregate stochastic noise decreases with the number of short modes included and therefore with $k_{\rm max}$.
The power spectrum, squeezed-limit bispectrum and collapsed trispectrum
can then all be described in terms of simple ``multitracer'' $2$-point functions of long-wavelength fluctuations in the matter/halo density and in the matter/halo position-dependent power spectrum.
In particular, the multitracer property allows for the well known phenomenon of cosmic variance cancellation.

The position-dependent power spectrum approach allowed us to derive simple analytic expressions for the Fisher information of various higher order statistics and combinations thereof, making the calculations fast and easy.
Moreover, this method makes it possible to describe the information in scale-dependent bias in the halo $2$-point functions and that in higher order statistics within the same framework,
which enabled us to perform a fair comparison of the constraining power of the two classes of probes and to study joint constraints of different statistics while properly taking into account their mutual covariance.

\vskip 7pt

For the matter bispectrum, we found that cosmic variance in the position-dependent power spectrum (which arises from non-linear evolution), leads to known non-diagonal, non-Gaussian contributions to the bispectrum covariance matrix that are commonly (though not always) neglected in conventional bispectrum forecasts.
In the position-dependent power spectrum approach, these cosmic variance terms take on a simple form, which allowed us to analytically invert the covariance matrix, leading to simple analytic expressions for the information content including cosmic variance (analogous to the standard expressions for the information content in, e.g., the halo power spectrum, which as a rule include the cosmic variance contribution).
We have quantified the degradation, due to the above cosmic variance, in the matter bispectrum information content relative to its primordial information content, and found that it can be up to a factor $\sim 3$ in the uncertainty on $\fnl$ at high $k_{\rm max}$.
An alternative formulation is that calculations neglecting the non-Gaussian contributions to the bispectrum covariance underestimate $\sigma(\fnl)$ by up to that same factor $\sim 3$.
The importance of non-Gaussian convariance in the large-scale structure bispectrum in the non-linear regime has previously been pointed out in e.g.~\cite{Sefusatti:2007ih,kayotaka13,chanblot17,byunetal17}.

Interestingly, despite cosmic variance, the matter bispectrum performs better than one might naively expect
and $\sigma(\fnl)$ sharply decreases with increasing $k_{\rm max}$ (i.e.~decreasing stochastic noise) even in the low stochastic noise limit.
The reason is
that the combination of multiple tracers in the bispectrum (i.e.~the position-dependent power spectra in different bins of short modes)
leads to partial cosmic variance cancellation.
Moreover, we have shown that the primordial information can in principle be regained by using the full power of multitracer analyses: by combining the matter bispectrum with the power spectrum and trispectrum, full cosmic variance cancellation is achieved and $\sigma(\fnl)$ from the primordial bispectrum (equivalently $\sigma(\fnl)$ in the Gaussian covariance approximation) is recovered.
The matter trispectrum alone leads to the same $\sigma(\fnl)$ as the bispectrum in the large $k_{\rm max}$ limit ($k_{\rm max} \gg 0.4 h/$Mpc at $z = 1$). It performs less well at low $k_{\rm max}$, but slightly outperforms the bispectrum at intermediate $k_{\rm max} \approx 0.2 - 0.4 h/$Mpc.

\vskip 7pt

We next compared the constraining power of the higher order matter statistics to the information contained in scale-dependent halo bias as measured from $2$-point functions involving the halo overdensity.
We found that scale-dependent bias is a very competitive probe. For instance, to match the constraint on $\fnl$ from the halo power spectrum for a sample (at $z = 1$) with moderate number density $\bar{n}_h = 3 \cdot 10^{-4} \, (h^{-1} $Mpc$)^{-3}$, and linear bias $b_{10}^{(h)} = 2$, one would need to measure the matter bispectrum to about $k_{\rm max} = 0.4 h/$Mpc. To improve on the bound from the halo power spectrum
by combining it with the matter bispectrum and trispectrum, one needs $k_{\rm max} \gtrsim 0.2 h/$Mpc.
One would thus have to probe the higher order statistics down to rather small scales, where modeling of non-linear effects becomes more challenging and where shot noise in whatever proxy for the matter density one actually observes may become important.
The reason for the relative strength of scale-dependent halo bias is that it (sub-optimally) probes mode-coupling involving short modes well into the very small-scale, non-linear regime, approximately down to scales of order the size of the halo. It thus measures mode-coupling with a large effective $k_{\rm max}$.

\vskip 7pt

Finally, we studied constraints on primordial mode-coupling from the {\it halo} power spectrum, bispectrum and trispectrum, the latter two providing
a realistic way (along with weak gravitational lensing) to indirectly measure the higher order matter statistics in the low redshift Universe.
We again found that the commonly made Gaussian covariance approximation may strongly underestimate $\sigma(\fnl)$ from the halo bispectrum, e.g.~by a factor $\sim 5$ for (extremely large) $\bar{n}_h = 0.03 \, (h^{-1} $Mpc$)^{-3}$, and $k_{\rm max} = 0.4 h/$Mpc.
However, for more ``moderate'' survey specifications, say $k_{\rm max} \sim 0.2 h/$Mpc and $\bar{n}_h \sim 10^{-3} \, (h^{-1}$Mpc$)^{-3}$, it is typically a factor $\sim 2$.
While it depends on the halo bias parameters, for our fiducial model ($b_{10}^{(h)} = 2$),
we found that scale-dependent halo bias and long-short halo mode-coupling contribute approximately equally to the $\fnl$ constraint from the halo bispectrum.
In addition to the effect of the mode-coupling of the underlying matter density, the halo mode-coupling contains a contribution due to non-linear, non-Gaussian halo bias, described by $b_{11}^{(h)}$.
For our fiducial model, this latter term is subdominant. However, it can in principle be important and depending on the sign of $b_{11}^{(h)}$, we have shown that it can either strengthen or weaken the halo mode-coupling signal, in the latter scenario partially canceling the contribution from primordial matter mode-coupling.

For halos, we found that the bispectrum typically outperforms the trispectrum.
We also showed that it is not easy for the higher order halo statistics to improve over the halo power spectrum.
The dominant reason for this is the floor at large $k_{\rm max}$ in the aggregate stochastic noise in the position-dependent halo power spectrum caused by shot noise in the halo overdensity (due to finite number density of halos).
While in the absence of halo shot noise, the stochastic noise in the position-dependent halo spectrum scales like $\propto k_{\rm max}^{-3}$ (inversely proportional to the number of short modes),
halo shot noise cuts off the information obtained from the small-scale power spectrum at large $k$,
making it impossible to lower the stochastic noise below a minimum value/floor. This minimum value is typically larger than the halo shot noise itself. For moderate number density, e.g.~$\bar{n}_h = 3 \cdot 10^{-4} \, (h^{-1} $Mpc$)^{-3}$ (and at $z = 1$), the halo shot noise cuts off the improvement with $k_{\rm max}$ of the constraints from higher order statistics before the $k_{\rm max}$ is reached where the higher order statistics otherwise would have outperformed the halo power spectrum.

Despite the large stochastic noise in the position-dependent power spectrum (even in the large $k_{\rm max}$ limit) as compared to that in the halo overdensity itself, the higher order statistics can still lead to improvements over the halo power spectrum thanks to cosmic variance cancellation. In the low shot noise limit, $\sigma(\fnl)$ from the halo power spectrum reaches a cosmic variance limited plateau\footnote{If we employ multiple halo samples with different biases, even the analysis of the halo $2$-point function(s) takes advantage of cosmic variance cancellation. In this sense, in the low shot noise limit one can do significantly better than the halo power spectrum of a single sample, even without resorting to higher order statistics.}, while the higher order statistics inherently apply cosmic variance cancellation, thus improving indefinitely, as stochastic noise is lowered.
This means that the regime where the higher order halo statistics are most powerful relative to the power spectrum (and where constraints in general are the strongest) is that of large halo number density and large $k_{\rm max}$.
In particular, this means that the regime in which the halo bispectrum (and trispectrum) are strong probes of $\fnl$ is exactly the cosmic variance dominated regime where the non-Gaussian contributions to the covariance, which were a main focus of this paper, are crucial.

We have also performed a preliminary study of what would be an optimal halo sample in terms of its halo bias parameters, and found that, if we are free to choose a large halo number density independent of halo bias, a sample with large linear bias $b_{10}^{(h)}$ maximizes both the constraining power from the power spectrum and from the higher order statistics.
However, in reality, the bias of the sample places restrictions on the number density and we found that this effect may (depending on $k_{\rm max}$) cancel out the advantage of having large $b_{10}^{(h)}$.

\vskip 7pt

Our results suggest that, when it comes to constraining local-type primordial mode-coupling, it will be difficult for the halo bispectrum (and/or trispectrum) to improve constraints by an order of magnitude or more relative to the halo power spectrum.
However, even a bispectrum $\fnl$ constraint at similar precision to the power spectrum would be extremely useful, as they are two very different types of measurements with distinct systematics.
Before drawing definite conclusions about expected constraints from future surveys, further study is required, which we will leave for a future publication.
In particular, the following are potential extensions/improvements of the treatment in the present paper.
\begin{itemize}
\item
A multitracer analysis involving multiple halo samples and/or the inclusion of (CMB) lensing, may strongly improve constraints from the bispectrum (see e.g.~\cite{yamaetal17}) and trispectrum, as it is known to do for the power spectrum.
\item
We have neglected redshift space distortions in this work and they should be included in a more realistic forecast.
\item
While we expect the effect to be modest, an improved analysis would also include marginalization over other parameters than $\fnl$. In this work, we have solely quantified the unmarginalized information that is in principle there.
\item
We have here restricted analysis to squeezed/collapsed configurations with a limited range of long modes $q_L$ and short modes $k_S$. We have also neglected terms of order $\phi_S$ in the halo statistics (as well as non-linear mode-coupling corrections to the long mode).
While these choices are well motivated (see Appendix \ref{app:scaledep}, \ref{app:stochnoise} and the main text), and we have checked that relaxing these assumptions does not make a significant qualitative difference, preliminary calculations suggest that lifting the above restrictions might improve the Fisher information by a factor $\sim 2$. It would be interesting to investigate this in more detail.
\item
We have shown results that probe quite far into the non-linear regime (up to $k_{\rm max} = 0.8 h/$Mpc, mostly at $z = 1$), which motivates extending the treatment of non-linearities beyond just the long-short mode-coupling in standard perturbation theory used here.
One example important for both matter and halos is that for large short-mode wave numbers, say $k_S \gg 0.1 h/$Mpc, the mode-coupling {\it between} short modes becomes important, leading to loss of information in the small-scale power spectrum so that the stochastic noise in the position-dependent power spectrum does not decrease as rapidly as the $\propto k_{\rm max}^{-3}$ scaling used here.
\item
More generally, it would be interesting to study non-Gaussian terms in the (e.g.~bispectrum) covariance matrix beyond the tree-level, long-short mode-coupling ones studied here, and to study the effect of
additional (higher order) stochastic noise contributions to the position-dependent power spectra.
\item
We would like to perform a more thorough trade study of the performance of the various statistics as a function of the properties of the halo sample and of other survey properties, taking into account for instance the relation between halo bias parameters and the number density of the halo sample. Another interesting question is that of the tradeoff between survey volume and number density: is it always preferable to maximize the volume? or might cosmic variance cancellation make a smaller volume with very high number density more optimal?
\item
We would like to forecast constraints for the specific samples expected to be observed by planned and proposed future surveys such as EUCLID \cite{euclid} and SPHEREx \cite{Spherexweb,spherex_wp,Dore:2016tfs}.
\end{itemize}

We wish to point out two other potentially interesting directions for further study motivated by the insights in this work.
First, as discussed in Section \ref{subsec:BK general}, it should in principle be possible to push the analysis of the squeezed-limit bispectrum (and beyond) deep into the non-linear regime (if stochastic noise allows it) by, instead of modeling the signal from first principles, treating the bias parameters introduced in this paper describing mode-coupling as free parameters to be fitted for.
This is completely analogous to the standard treatment of halo biasing. There, instead of fully modeling the complicated, highly non-linear physics of the halos themselves, we parameterize our ignorance of small-scale physics by free halo bias parameters that we then measure from the data.
Second, our analytic inversion of the bispectrum covariance matrix (including non-Gaussian covariance)
may be used to derive a more optimal estimator of $\fnl$ than one would obtain using the bispectrum covariance in the Gaussian approximation. This might in particular find a useful application in the context of the cosmic microwave background.

\vskip 7pt

To conclude, it is generally challenging to obtain intuitive, physical insight into the information content of higher order statistics beyond the power spectrum
due to their complicated nature and in particular due to the high dimensionality of the space of configurations of the former compared to the latter.
Closely related to this issue is that bispectrum forecasts, etc., tend to be computationally intensive and time consuming, especially when including covariance beyond the Gaussian approximation.
On the other hand, physical insight is crucial to avoid that studies of higher order statistics become ``black box'' analyses: one needs a deeper understanding in order to optimize survey design, the choice of analysis techniques, and for the proper interpretation of results.
Applying concepts such as the position-dependent power spectrum, multitracer analysis and cosmic variance cancellation, we believe we have provided a step forward in obtaining this necessary physical insight,
at least for the primordial mode-coupling information contained by squeezed-limit higher order statistics.

\vskip.1cm
\emph{Acknowledgements:}
I would like to thank Tobias Baldauf, Phil Bull, Tzu-Ching Chang, Olivier Dor\'e, Jerome Gleyzes, Daniel Green,  Elisabeth Krause, Emmanuel Schaan, Fabian Schmidt and Uros Seljak
for helpful discussions and gratefully acknowledge support by the Heising-Simons foundation.

\appendix

\section{Derivation of relation $\delta P(\k_S; \q_L)$ and pairs of short-wavelength perturbations}
\label{app:app1}

We here derive the approximate identity given in Eq.~(\ref{eq:dPk est}) regarding long-wavelength Fourier modes of the binned position-dependent power spectrum,
\beq
\label{eq:dPk app}
\delta \hat{P}(\k_{i}; \q_L) \approx \int_{\k_{i}} \frac{d^3 \k_S}{V_{\k,i}} \, \delta(\k_S) \, \delta(-\k_S + \q_L),
\eeq
where the integral is over the bin, $\k_i$, of short modes included in the estimator,
and $\delta(\k)$ is the Fourier transform of the matter overdensity $\delta(\x)$.

The position-dependent small-scale power spectrum estimator for a single short mode $\k$, in some volume
$V$ (to be identified with the subvolume $V_S$ from the main text, Section \ref{subsec:posdeppk}, not with the full survey volume) centered on a position $\x_0$, can be defined as \cite{FKP},
\beq
\hat{P}(\k; \x_0) \equiv |F_V(\k; \x_0)|^2,
\eeq
with
\beq
F_{V}(\k; \x_0) \equiv \int d^3 \x \, e^{\iu \k \cdot \x} \, W_{V}(\x - \x_0) \, \delta(\x).
\eeq
We assume the short mode $\k$ to be small compared to the volume over which the spectrum is estimated,
$k \gg V^{-1/3}$.
The integral above is over all of space and
we understand the window function $W_V(\x)$ to be constant (or at least non-zero) inside a volume $V$ centered on $\x = 0$, and zero outside this volume.
It is normalized by
\beq
\int d^3 \x \, W_V^2(\x) = 1.
\eeq
For definiteness, we may choose the window function to be a spherical top-hat function,
\beq
W_V(\x) =
\begin{cases}
1/\sqrt{V}  \quad \quad \text{if} \, |\x| < R\\
0 \quad \quad \quad \quad \text{if} \, |\x| \geq R,
\end{cases}
\eeq
where $R$ is the radius of the sphere, i.e.~$V = \tfrac{4}{3} \pi \, R^3$.

We can write the overdensity estimator as a convolution in Fourier space,
\beq
F_{V}(\k; \x_0) = \int \frac{d^3 \q}{(2 \pi)^3} \, \delta(\k - \q) \, W_V(\q) \, e^{\iu \q \cdot \x_0}.
\eeq
The Fourier transform of the window function, $W_V(\q)$, is a localized function, peaked at $\q = 0$, with a width $(\Delta q)_V \sim R^{-1} \sim V^{-1/3}$. For the spherical top-hat choice, it can at low $q$ be expanded as,
\beq
W_V(\q) = \sqrt{V} \, \left[1 - \tfrac{1}{10} q^2 \, R^2 + \mathcal{O}(q^4 \, R^4) \right].
\eeq

We can now write the position-dependent power spectrum as,
\beq
\hat{P}(\k ; \x_0) = \int \frac{d^3 \q}{(2 \pi)^3} \, \int \frac{d^3 \q'}{(2 \pi)^3} \,
W_V(\q) \, W_V(-\q') \, e^{\iu \x_0 \cdot (\q - \q')} \, \delta(\k - \q) \, \delta(-\k + \q').
\eeq
Fourier transforming the dependence on $\x_0$ then gives,
\beq
\delta \hat{P}(\k; \q_L) = \hat{P}(\k; \q_L) \equiv \int d^3 \x_0 \, e^{\iu \x_0 \cdot \q_L} \, \hat{P}(\k; \x_0)
= \int \frac{d^3 \q}{(2 \pi)^3} \, W_V(\q) \, W_V(-\q - \q_L) \, \delta(\k - \q) \,
\delta(-(\k - \q) + \q_L).
\eeq
We are interested in modes $\q_L$ much larger than the local volume $V = V_S$, i.e.~$q_L \ll V^{-1/3}$.
We then see already from the above expression that the long-mode position-dependent power spectrum perturbation is equal to a weighted integral over matter overdensity pairs of the form $\delta(\k') \, \delta(-\k' + \q_L)$. Using that $q_L$ is small compared to the Fourier-space width of the window functions $W_V$, the weight function picks out a range of short modes $\k'$ centered on $\k$, with a spread $(\Delta q)_V$.

We reach the final result, Eq.~(\ref{eq:dPk app}), by considering the power spectrum averaged over a bin of short modes. We take the bin to be centered on wave vector $\k_i$, to have Fourier-space volume $V_{\k,i}$, and to be wide compared to the fundamental wave number of the volume $V$, i.e.~$V_{\k, i} \gg (\Delta q)^3_V \sim V^{-1}$.
If we define an additional weight function $W_i(\k)$ that equals unity inside the $i$-th bin, and zero outside, we can conveniently write the binned power spectrum estimator as an integral over all $\k$,
\bea
\delta \hat{P}(\k_i; \q_L) &\equiv& \int \frac{d^3 \k}{V_{\k,i}} \, W_i(\k) \, \delta \hat{P}(\k_i; \q_L)
= \int \frac{d^3 \k}{V_{\k,i}} \, \int \frac{d^3 \q}{(2 \pi)^3} \, W_i(\k) \, W_V(\q) \, W_V(-\q - \q_L) \, \delta(\k - \q) \,
\delta(-(\k - \q) + \q_L) \nonumber \\
&=& \int \frac{d^3 \k'}{V_{\k,i}} \, \left[ \int \frac{d^3 \q}{(2 \pi)^3} \, W_i(\k' + \q) \, W_V(\q) \, W_V(-\q - \q_L) \right] \, \delta(\k') \,
\delta(-\k' + \q_L).
\eea

We now indeed see that in the limit that the $\k_i$ bin is much larger than the width of the kernel, $R^{-1}$, the weight function in square brackets, equals
\beq
\left[ \int \frac{d^3 \q}{(2 \pi)^3} \, W_i(\k' + \q) \, \tilde{W}_V(\q) \, \tilde{W}_V(-\q - \q_L) \right] \approx
\begin{cases}
\tilde{(W^2)}_V(-\q_L) = \frac{1}{\sqrt{V}} \, \tilde{W}_V(-\q_L) \quad \, \text{if} \, \k' \, \text{inside} \, V_{\k,i} \\
0 \quad \quad \quad \quad \quad \quad \quad \quad \quad \quad \quad  \quad \quad  \quad \text{else}
\end{cases}
\eeq
Furthermore, using our assumption that $q_L \, R \ll 1$, the value for $\k'$ inside the $i$-th bin
equals,
\beq
\frac{1}{\sqrt{V}} \, \tilde{W}_V(-\q_L) \approx 1.
\eeq
Thus, we have confirmed the simple weighted average over the bin $\k_i$ given in Eq.~(\ref{eq:dPk app}).

Finally we note that the same derivation holds for the position-dependent {\it halo} power spectrum.
For simplicity, one may imagine a halo number density field $n_h(\x)$ that extends to infinity and has position-independent expectation value, $\bar{n}_h$.
Then, we simply need to replace the matter overdensity $\delta(\x)$ in the above derivation by the halo overdensity,
$\delta_h(\x) \equiv (n_h(\x) - \bar{n}_h)/\bar{n}_h$,
leading to the same expression as Eq.~(\ref{eq:dPk app}), but now for halos.
We make use of this in Section \ref{sec:halos}.

\section{Scale and configuration dependence of primordial mode-coupling information}
\label{app:scaledep}

We here study in more detail how the ability of various statistics to constrain primordial mode-coupling depends on the range of long modes included in the analysis, and in the case of higher order statistics, how it depends on which configurations are used.
The main conclusion is that the information is indeed dominated by squeezed configurations and that, while dependent on the specific statistic being used, most information is captured by the range of very long modes studied in this paper ($q_L < 0.02 h/$Mpc).

\subsection{Primordial information in matter bispectrum (BK, GCA)}

We found in Section \ref{sec:matter} that the total Fisher information on $\fnl$ from the primordial matter bispectrum (or equivalently from the low-redshift bispectrum under the Gaussian covariance approximation), is given by,
\beq
\label{eq:Ftotapp}
F = 16 V \,  \int_{q_{\rm min}}^{q_{L,{\rm max}}} \frac{d q_L \, q_L^2}{2 \pi^2} \, P_\phi(q_L) \, \int_{k_{S,{\rm min}}}^{k_{\rm max}} \frac{d k_S \, k_S^2}{2 (2 \pi^2)}.
\eeq
We have here re-written $F$ directly in terms of the nearly scale-invariant primordial potential power spectrum,
\beq
P_{\phi}(q) \propto k^{-3 + (n_s - 1)},
\eeq
where $n_s$ is the scalar spectral index.

In the analysis throughout this paper, we have chosen the long modes to cover a range $q_{\rm min} - q_{L,{\rm max}}$ (in practice $q_{\rm min} = 0.001 h/$Mpc and $q_{L,{\rm max}} = 0.02 h/$Mpc) and the short modes to cover a range $k_{S,{\rm min}} - k_{\rm max}$ ($k_{S,{\rm min}} = 0.02 h/$Mpc by default, and varying $k_{\rm max}$), {\it independent} of $q_L$.
We here loosen these restrictions and look at the scale-dependence of the information content more generally.
Among other things, we are interested in the information content of all configurations combined, i.e.~effectively setting $q_{L,{\rm max}} = k_{\rm max}$ and $k_{S,{\rm min}} = q_L$.

First, even though
Eq.~(\ref{eq:Ftotapp}) is technically only valid in the squeezed limit, $k_S \gg q$,
it shows, as expected, that the signal is dominated by the most squeezed configurations. For fixed $q_L$, the information content scales with $k_{\rm max}^3$ and is thus dominated by the shortest modes close to the cutoff.
The only exception are modes $q_L$ close to $k_{\rm max}$, but these (as we will see) contribute a small fraction of the total information.
We therefore conclude that
our calculation of the information content based on the position-dependent power spectrum (which assumes the squeezed limit) should be a good approximation for the range of scales considered.

Second, the build-up of information as a function of the long mode $q_L$ is almost scale-invariant (see also e.g.~\cite{cremetal07}),
\beq
\label{eq:scaling F}
F \sim  \int d\ln q_L \, \Delta^2_\phi(q_L) \sim \int d\ln q_L \,q_L^{n_s - 1},
\eeq
where $\Delta^2_\phi(q_L) \equiv q_L^3 P_\phi(q_L)/2 \pi^2$ is the dimensionless power spectrum of the primordial potential fluctuations.
This result makes sense if we consider the signal-to-noise (per unit $\fnl$) squared in the following picture.
If we divide the Universe into patches of volume $V_S$, the amplitude of the matter power spectrum in each patch is modulated by $\phi_i$, where $i$ here is an index labeling the subvolumes. The measurement of the local power spectrum amplitude in each patch has some noise due to variance in the local short modes (represented by the $k_S$ integral above),
and the rms {\it signal} (squared) is proportional to the variance $\sigma^2(\phi_i) \approx \int d\ln q_L \, \Delta^2_\phi(q_L)$, thus explaining the scaling of $F$.
The scale dependence, Eq.~(\ref{eq:scaling F}), implies that, while we have conservatively focused on very large-scale long modes, $q_{L,{\rm max}} = 0.02 h/$Mpc,
there in principle is additional constraining power coming from shorter ``long'' modes.
We will quantify this additional information further below.

An advantage of the case of the {\it primordial} matter bispectrum (or bispectrum in the GCA) is that we can perform a direct comparison with the information content obtained from summing over all triangles without assuming the squeezed limit. Once the non-Gaussian covariance is included, this becomes difficult to do analytically because different triangles become correlated.
Using the bispectrum variance in the GCA (e.g.~\cite{scoccetal04,baldaufetal16}), the full information content in all triangles involving modes between $q_{\rm min}$ and $k_{\rm max}$, is given by (cf.~Appendix \ref{app:F GCA}),
\bea
\label{eq:F sum triangles}
F &=& \sum_{\rm triangles} \frac{\left( \pa B/\pa \fnl \right)^2}{\sigma^2(\hat{B})}
= \frac{V}{8 \pi^4} \, \int_{q_{\rm min}}^{k_{\rm max}} dq \, q \, \int_{q}^{k_{\rm max}} dk \, k \int_{k}^{{\rm min}(k+q,k_{\rm max})} dk' \, k' \, \frac{\left( \pa B(q,k,k')/\pa \fnl \right)^2}{P(k) \, P(k') \, P(q)} \\
&=& \frac{V}{2 \pi^4} \int_{q_{\rm min}}^{k_{\rm max}} dq \, q \, P_\phi(q) \, \int_{q}^{k_{\rm max}} dk \, k \, \int_{k}^{{\rm min}(k+q,k_{\rm max})} dk' \, k' \, \left[ \sqrt{\frac{P_\phi(k)}{P_\phi(k')}} + \sqrt{\frac{P_\phi(k')}{P_\phi(k)}} + \frac{\sqrt{P_\phi(k) \, P_\phi(k')}}{P_\phi(q)}\right]^2. \nonumber
\eea
The third term inside the square brackets in the second line comes from the non-Gaussian contribution to the {\it long} mode, schematically, $\phi_L \supset \tilde{\phi}_S \, \tilde{\phi}_{S'}$. We have ignored contributions of this type throughout the paper. Given that $P_\phi(k) \propto k^{-3 + (n_s - 1)}$, we see that (except for equilateral configurations, which we had already established contribute negligibly) this contribution is indeed strongly suppressed compared to the other terms (from $\phi_S \supset \tilde{\phi}_L \, \tilde{\phi}_S$) so that we were justified in neglecting it.

It is straightforward to confirm that in the squeezed limit ($k' \to k$, $\int_{k}^{k+q} dk'\, f(k') \to q \, f(k)$), Eq.~(\ref{eq:F sum triangles}) reduces to Eq.~(\ref{eq:Ftotapp}).
For our default range of long modes, $q = 0.001 - 0.02 \, h/$Mpc, we
have numerically compared $\sigma(\fnl)$ from the full integral over triangles, Eq.~(\ref{eq:F sum triangles}),
to the result based on the position-dependent
power spectrum, Eq.~(\ref{eq:Ftotapp}).
In the former case, we have considered both the integral over triangles
with short modes in the same range as for the position-dependent power spectrum (set by $k_{S,{\rm min}} = 0.02 h/$Mpc and varying $k_{\rm max}$),
as well as the case of including all short modes up to $k_{\rm max}$ (i.e.~setting $k_{S,{\rm min}} = q_L$).
We have found agreement at the few-percent level in both cases (for $k_{\rm max} \gtrsim 0.05 h/$Mpc).

\begin{figure*}[]
\centering
\includegraphics[width=0.55\textwidth]{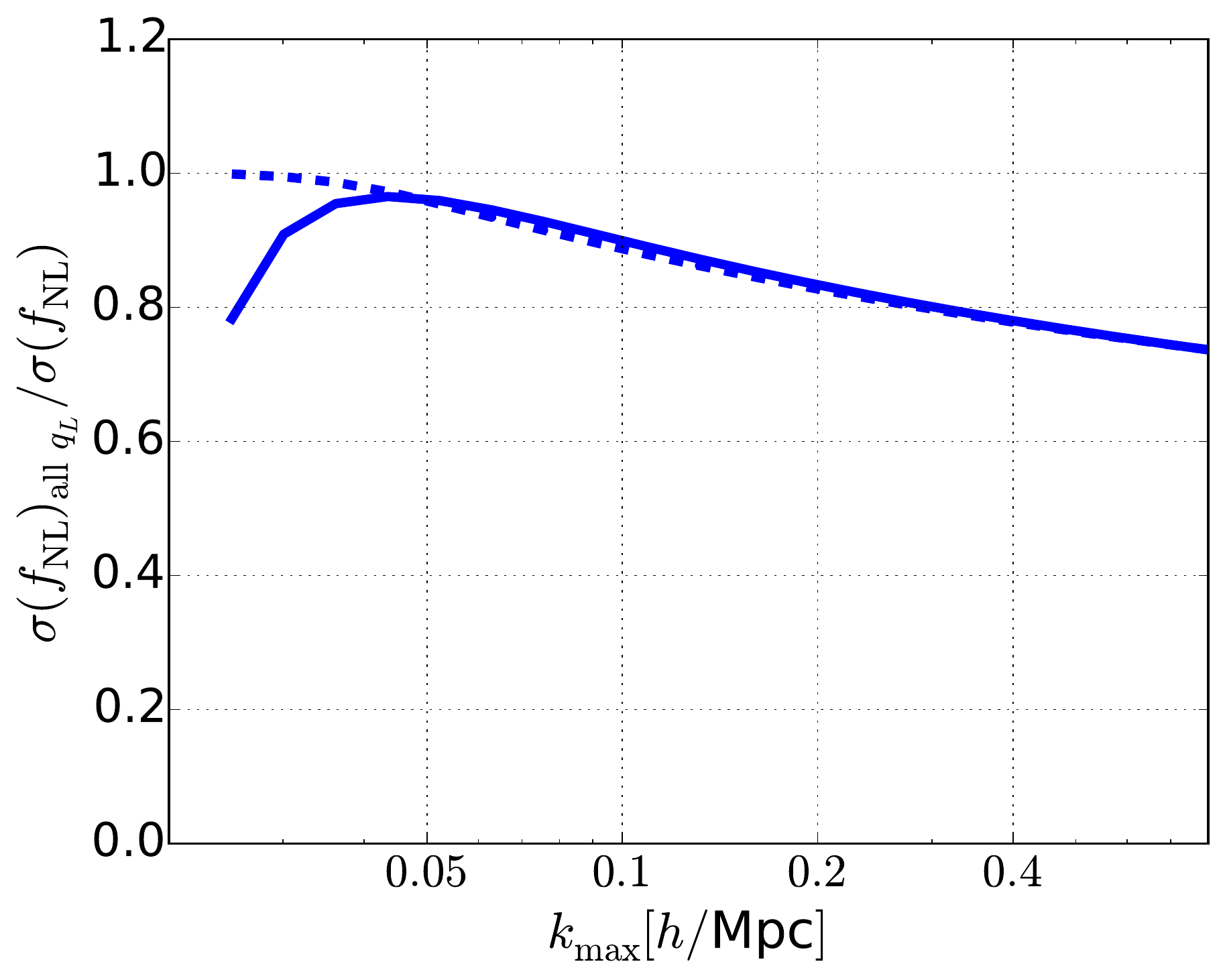} 
\caption{Ratio of $\sigma(\fnl)$ values from the matter bispectrum, both for the numerator and denominator assuming the Gaussian covariance approximation, where different triangles are uncorrelated. {\it Solid curve:} The numerator, $\sigma(\fnl)_{{\rm all}\, q_L}$, includes all configurations restricted by $q_{\rm min}$ and $k_{\rm max}$, and in particular all long modes $q_L$ down to $k_{\rm max}$.
It is computed as an integral over all triangles, without assuming the squeezed limit.
The denominator is the quantity used throughout this paper, where we restrict the long modes to $q < q_{L, {\rm max}} = 0.02 h/$Mpc (and the short modes to $k_{S} > 0.02 h/$Mpc), and we use the position dependent power spectrum method.
The main result is that one may in principle improve $\sigma(\fnl)$ by $10 - 20 \%$ by pushing $q_L$ to smaller scales.
The behavior at low $k_{\rm max}$ is due not to the difference in $q_L$ range, but because for $k_{\rm max} \approx k_{S,{\rm min}}$, setting $k_{S,{\rm min}} = 0.02 \, h/$Mpc neglects a significant fraction of short modes (i.e.~those with $q_L < k_S < k_{S,{\rm min}}$).
{\it Dashed curve:} Same ratio as in solid curve,
but now in the denominator setting $k_{S,{\rm min}} = q_L$ (and using the integral over triangles).
At low $k_{\rm max}$, the ratio of $\sigma(\fnl)$ using all $q_L$ over $\sigma(\fnl)$ using $q_{L,{\rm max}} = 0.02 h/$Mpc, now tends to one.
}
\label{fig:all q_L}
\end{figure*}

Finally, coming back to our discussion of the dependence on the range of long modes, we have also compared the information in the default range with $q_{L,{\rm max}} = 0.02 h/$Mpc to using all long modes up to $k_{\rm max}$.
In the latter case, we have used the full integral over triangles, setting $q_{L,{\rm max}} = k_{\rm max}$ (and $k_{S,{\rm min}} = q_L$).
The ratio of the uncertainty with all long modes out to $k_{\rm max}$ to the result with only our default restricted range of long modes is shown as a function of $k_{\rm max}$ in Figure \ref{fig:all q_L}.
We see that for, say, $k_{\rm max} = 0.1 - 0.4 h/$Mpc, in principle the uncertainty on $\fnl$ can be improved relative to the analysis in the body of this paper by approximately $10 - 20 \%$ by including long modes beyond our assumed $q_{L,{\rm max}} = 0.02 h/$Mpc. This is a modest, but non-negligible amount.

\subsection{Low-redshift matter bispectrum (BK)}
\label{subapp:BK}

We have above shown that for the {\it total} information content in the matter mode-coupling, most information comes from squeezed-limit configurations, and that long modes contribute approximately equally on all scales.
We now briefly ask if/how this picture is modified if instead we consider the bispectrum as measured at low redshift, i.e.~including the non-Gaussian contribution to the bispectrum covariance.

In this case the information content is given by (cf.~Eq.~(\ref{eq:fishermatbispec})),
\beq
F(q_L) = \frac{\Sigma_2''(q_L) + 2 \left( \Sigma_2(q_L) \, \Sigma_2''(q_L) - \left( \Sigma_2'(q_L) \right)^2\right)}{1 + 2 \Sigma_2(q_L)}.
\eeq
In the stochastic noise dominated regime, $\Sigma_2 \ll 1$ (i.e.~small $k_{\rm max}^3 \, P(q_L)$), this expression reverts to Eq.~(\ref{eq:Ftotapp}), $F(q_L) \approx \Sigma_2''$, so that the conclusions about the scale-dependence of the signal from the previous discussion apply.
In the cosmic variance dominated regime, $\Sigma_2 \gg 1$ (large $k_{\rm max}^3 \, P(q_L)$),
the partial cosmic variance cancellation is contained in the fact that $\Sigma_2 \, \Sigma_2'' \ne \left( \Sigma_2' \right)^2$. This can be rephrased as the inequality, $\bar{b_{2}^2} \ne \bar{b_{2}}^2$, where the bar represents averaging $b_{2(i)}^2$, $b_{2(i)}$ over short modes (or equivalently over bins $i$), see Section \ref{subsec:cancel}.
In the limit $\Sigma_2 \gg 1$, the information per mode approaches (Eq.~(\ref{eq:BK asymp})),
\beq
F(q_L) \to \frac{\bar{b^2_2} - (\bar{b_2})^2}{\bar{b^2_2}} \, \Sigma_2''(q_L).
\eeq
For fixed $q_L$, this again leads to a scaling $F(q_L) \sim k_{\rm max}^3$ so that most information comes from squeezed configurations. Considering the build-up of information as a function of the range of long modes included, the information also again scales like $\int d\ln q_L \, \Delta^2_\phi(q_L)$.
Therefore, the low-redshift matter bispectrum qualitatively has the same dependence on scale/configurations as the bispectrum in the GCA.

\subsection{Halos - total information content ($hh+hm+mm$)}

Next, consider the information in scale-dependent halo bias. Here, we are by construction already
probing the squeezed limit (halos significantly smaller than the long modes) so that the remaining question is that of the $q_L$ dependence of the information content.
First consider the {\it total} information in scale-dependent halo bias, defined as the information content of the combination of the halo power spectrum ($hh$), halo-matter cross-spectrum ($hm$) and matter power spectrum ($mm$). In this case, the combination of the halo overdensity with the matter overdensity serves to cancel cosmic variance in the measurement.
From Eq.~(\ref{eq:halos all}) in Section \ref{sec:mat2 vs halo},
\beq
F(q_L) = \Sigma_h''(q_L) = \bar{n}_h \, P(q_L) \, \left(b_h'(q_L)\right)^2 \propto P_\phi(q_L).
\eeq
This expression is analogous to the case of the total matter mode-coupling information.
We see that the information again approximately receives equal contributions from each decade in long modes, $F \sim \int d\ln q_L \, \Delta^2_\phi(q_L)$.

\subsection{Halos - halo power spectrum ($hh$)}

For the halo power spectrum (of a single sample), there is no cosmic variance cancellation, and the information content is significantly suppressed for large $\Sigma_h$ (i.e.~$\bar{n}_h \, P(q_L) \gg 1$). From Eq.(\ref{eq:info halo power}),
\beq
F(q_L) = \frac{2 \Sigma_h(q_L) \, \Sigma_h''(q_L)}{\left( 1 + \Sigma_h(q_L) \right)^2}.
\eeq
First, in the shot noise dominated regime, the above expression leads to a scaling, $F \sim \int d\ln q_L \, \Delta_\phi^2(q_L) \, P(q_L)$, which implies the information content is dominated by scales near the peak of the matter power spectrum, $q_L \sim k_{\rm eq} \approx 0.02 h/$Mpc. This makes sense, as here the shot noise suppression is lowest (the power highest).
In the more interesting cosmic variance dominated limit, the scaling is $F \sim \int d\ln q_L \, q_L^3 \, \M^{-2}(q_L)$. At low $q_L \ll k_{\rm eq}$, this leads to a scaling $F \sim \int d\ln q_L \, q_L^{-1}$, meaning the information over this range of scales (to the left of the peak of the matter power spectrum) is dominated by the longest modes in the survey. In the opposite regime, $q_L \gg k_{\rm eq}$, the scaling becomes approximately $F \sim \int d\ln q_L \, q_L^{2.5}$, meaning that to the right of the power spectrum peak, the information is in principle dominated by the shortest long modes. However, note the caveats to pushing $q_L$ to larger values discussed at the end of Section \ref{subsec:modecoupling}.

Finally, note that the halo-matter {\it cross}-spectrum is analogous to the matter bispectrum (without the GCA) discussed in Section \ref{subapp:BK}, but {\it without} the cosmic variance cancellation (since we consider only a single halo sample), leading to a different scaling of $F$ with $q_L$.

\section{Recovery of standard expression for bispectrum Fisher information (in Gaussian covariance approximation)}
\label{app:F GCA}

In this Appendix, we explicitly confirm that the bispectrum Fisher information in the Gaussian covariance approximation, as obtained from the position-dependent power spectrum formalism, reproduces the squeezed limit of the standard expression for the Fisher information \cite{scoccetal04}.
We will directly consider the halo bispectrum. The Fisher information of the matter bispectrum is then straightforwardly obtained by taking the special case $b_h = b_{10}^{(h)} \to 1, b_h' \to 0, \bar{n}_h \to \infty$, etc.

The Fisher information per long mode $\q_L$ for the halo bispectrum in the GCA was given in the first line of Eq.~(\ref{eq:BKh}),
\beq
F(q_L) = \frac{\Sigma_h(q_L) \, \Sigma_{2h}''(q_L) + 2 \, \Sigma_h'(q_L) \, \Sigma_{2h}'(q_L) + \Sigma_h''(q_L) \, \Sigma_{2h}(q_L)}{1 + \Sigma_h(q_L)}.
\eeq
Using the definitions of the $\Sigma$ quantities, this leads to the total Fisher information,
\beq
\label{eq:FGCA app}
F = \frac{1}{2} \, V \, \int_{q_{\rm min}}^{q_{L,{\rm max}}} \frac{d^3 \q}{(2\pi)^3} \, \int_{k_{S,{\rm min}}}^{k_{\rm max}} \frac{d^3 \k_i}{(2\pi)^3} \,
\frac{\left[ b_{10}^{(h) \, 2} \, P(k_i) \, \left( b_{2h(i)} \, b_h' + b_{2h(i)}' \, b_h \right) \,P(q)\right]^2}{\left( b_{10}^{(h) \, 2} \, P(q) + \tfrac{1}{\bar{n}_h} \right)\,\left( b_{10}^{(h) \, 2} \, P(k_i) + \tfrac{1}{\bar{n}_h} \right)^2}
\eeq
We now note that the term in the square brackets is the derivative of the squeezed-limit bispectrum,
\beq
\frac{\pa B_h}{\pa \fnl} = b_{10}^{(h) \, 2} \, P(k_i) \, \left( b_{2h(i)} \, b_h' + b_{2h(i)}' \, b_h \right) \,P(q).
\eeq
The denominator in Eq.~(\ref{eq:FGCA app}) is related to the (Gaussian) bispectrum covariance per triangle,
\beq
\sigma^2(\hat{B}_h) = V \, P^{\rm tot}(q) \, P^{\rm tot}(k_i) \, P^{\rm tot}(k_i')
\approx V \, \left( b_{10}^{(h) \, 2} \, P(q) + \tfrac{1}{\bar{n}_h} \right)\,\left( b_{10}^{(h) \, 2} \, P(k_i) + \tfrac{1}{\bar{n}_h} \right)^2,
\eeq
where $P^{\rm tot}(k) \equiv b_{10}^{(h) \, 2} P(k) + \tfrac{1}{\bar{n}_h}$, $\k_i' \equiv -\k_i - \q$, and the second line is the squeezed-limit approximation.
Finally, identifying $k_f^3 \equiv (2 \pi)^3/V$ as the phase-space volume per mode, the double integral performs (the continuum limit of) a sum over squeezed triangles. The integral double-counts each configuration, but this is corrected for by the factor $1/2$ in front. Therefore, the Fisher information above equals
\beq
\label{eq:F BK general}
F = \sum_T \frac{\left( \pa B_h/\pa \fnl\right)^2}{\sigma^2(\hat{B}_h)}
\eeq
in the squeezed limit,
and thus recovers the standard result.

A more common form for Eq.~(\ref{eq:FGCA app}) is obtained by writing it as an integral over the norms $q, k_i$
and $k_i'$,
\beq
\label{eq:FGCAqkkp}
F = \frac{V}{8 \pi^4} \, \int_{q_{\rm min}}^{q_{L,{\rm max}}} dq \, q \, \int_{k_{S,{\rm min}}}^{k_{\rm max}} dk_i \, k_i \, \int_{k_i}^{\text{min}(k_i + q, k_{\rm max})} dk_i' \, k_i' \,
\frac{\left( \pa B_h/\pa \fnl\right)^2}{\left( b_{10}^{(h) \, 2} \, P(q) + \tfrac{1}{\bar{n}_h} \right)\,\left( b_{10}^{(h) \, 2} \, P(k_i) + \tfrac{1}{\bar{n}_h} \right)^2},
\eeq
where we have used that $k_i \, q \, d\mu = k_i' \, d k_i'$ ($\mu \equiv \hat{q} \cdot \hat{k}_i$) and we have restricted $k_i' > k_i > q$
to avoid double-counting.
This again corresponds to the standard expression.
It indeed follows from Eq.~(\ref{eq:F BK general})
using that a bin in $q, k_i$ and $k_i'$
spans a six-dimensional phase-space volume $V_{123} \equiv 8 \pi^2 \, q \, k_i \, k_i' \, dq \, dk_i \, dk_i'$ and contains a number of independent triangles $s^{-1}_{123} \, V_{123}/k_f^6$, where the symmetry factor $s_{123} = 6,2,1$ for equilateral, isosceles and general bin configurations.
The symmetry factor accounts for the fact that equilateral and isosceles bins count the same fundamental triangles multiple times, and its effect becomes negligible in the limit that the bin widths $dq$, etc., go to zero.
Without the restriction to the specific squeezed configurations made in this paper, the Fisher information would be given by the same expression as in Eq.~(\ref{eq:FGCAqkkp}), except changing the integration bounds to integrate over all (distinct) triangles,
\beq
\int_{q_{\rm min}}^{k_{\rm max}} dq \, q \, \int_q^{k_{\rm max}} dk_i \, k_i \, \int_{k_i}^{\text{min}(k_i + q, k_{\rm max})} dk_i' \, k_i' \, \dots.
\eeq

Finally, we note that some standard references, including \cite{scoccetal04} (SSZ04), use a different power spectrum and bispectrum convention compared to the current standard, used throughout this paper.
Specifically,
\beq
P^{\rm SSZ04} = (2 \pi)^{-3} \, P \quad \quad \text{and} \quad \quad B^{\rm SSZ04} = (2 \pi)^{-6} \, B.
\eeq
We caution that, using the expression for the bispectrum variance from SSZ04 (or other articles with the same convention), without correcting for the difference in conventions, would lead one to overestimate the Fisher information from the bispectrum by a factor $(2 \pi)^3$ and thus {\it to underestimate bispectrum based parameter uncertainties by a factor $\sim 16$}.

\section{Position-dependent power spectrum averaged over short modes}
\label{app:d2bar}

At various points in this article, we consider the averaged tracer,
\beq
\delta_{\bar{2}}(\q_L) \equiv \frac{1}{V_{\k,S}} \, \sum_i V_{\k,i}\,\delta_{2(i)}(\q_L)
= \int_S \frac{d^3 \k_S}{V_{\k,S}} \, \frac{\delta \hat{P}(\k_S; \q_L)}{P(k_S)},
\eeq
where the sum is over all bins of short modes,
$V_{\k,i}$ is the Fourier volume of each $\k_i$ bin, and $V_{\k,S}$ is the integrated volume over {\it all} short modes.
The quantity $\delta_{\bar{2}}(\q_L)$ thus describes long-wavelength variations in the position-dependent, small-scale power spectrum amplitude, estimated by averaging over short modes,
and the cross spectrum with a long-mode matter perturbation is proportional to a weighted average of the matter bispectrum,
\beq
P_{1\bar{2}}(q_L) = \int_S \frac{d^3 \k_S}{V_{\k,S}} \, \frac{B(\q_L, \k_S, -\k_S - \q_L)}{P(k_S)}.
\eeq

The averaged position-dependent matter power spectrum can be seen as the inverse noise weighted average of the $\delta_{2(i)}$ tracers,
\beq
\delta_{\bar{2}}(\q_L) = \frac{1}{\sum_i N_{2(i)}^{-1}} \, \sum_i N_{2(i)}^{-1} \, \delta_{2(i)}(\q_L).
\eeq
This averaging is optimal with respect to information on $\fnl$ in the Gaussian covariance approximation.

The quantity $\delta_{\bar{2}}$ can be useful because it is a single tracer as opposed to the set of multiple tracers $\{ \delta_{2(i)}\}$ (one for each bin of short modes) one needs to consider in the general case. It is characterized by parameters (obtained by averaging Eq.~(\ref{eq:bias dP})),
\beq
\label{eq:d2bar}
b_{\bar{2}} = \frac{47}{21} - \frac{1}{3} \, \langle n^{\delta}_s(k_S) \rangle, \quad b_{\bar{2}}'(q_L) = 4 \M^{-1}(q_L),
\quad N_{\bar{2}} = \frac{2 (2\pi)^3}{V_{\k,S}},
\eeq
where $\langle n^{\delta}_s(k_S) \rangle$ is the spectral index of the matter power spectrum averaged over short modes,
\beq
\langle n^{\delta}_s(k_S) \rangle \equiv \int_S \frac{4 \pi \, k_S^2 \, d k_S}{V_{\k, S}} \frac{d \ln P}{d \ln k}\left( k_S \right).
\eeq
In Figure \ref{fig:CV vs SN}, we approximate $\langle n^{\delta}_s(k_S) \rangle = -2.5$ (the value of the index at $k = 0.2 h/$Mpc).

\vskip 7pt

Analogously to the above treatment
of the position-dependent matter power spectrum, we also define the position-dependent halo power spectrum averaged over short modes,
\beq
\delta_{\bar{2h}}(\q_L) \equiv \frac{1}{\sum_i N_{2h(i)}^{-1}} \, \sum_i N_{2h(i)}^{-1} \, \delta_{2h(i)}(\q_L).
\eeq
The properties of this tracer are obtained by averaging the quantities in Eq.~(\ref{eq:d2h}),
\beq
b_{\bar{2h}} = \frac{47}{21} - \frac{1}{3} \, \langle n^{\delta}_s(k_S) \rangle + 4 \frac{b_{20}^{(h)}}{b_{10}^{(h)}}, \,
b_{\bar{2h}}'(q_L) = 4 \M^{-1}(q_L) \, \left( 1 + \frac{1}{2} \, \frac{b_{11}^{(h)}}{b_{10}^{(h)}}\right), \,
N_{\bar{2h}} = \left[ \int_S \frac{d^3 \k_S}{2(2\pi)^3} \, \left(\frac{\bar{n}_h \, b_{10}^{(h) \, 2} \, P(k_S)}{1 + \bar{n}_h \, b_{10}^{(h) \, 2} \, P(k_S)} \right)^2 \right]^{-1}.
\eeq
Note that the average spectral index $\langle n^{\delta}_s(k_S) \rangle$ in this case contains as an additional weight factor the quantity in the integrand of $N_{\bar{2h}}$. In Figure \ref{fig:biases2} (right panel) and Figure \ref{fig:BKhcontributions}, we use the approximation $\langle n^{\delta}_s(k_S) \rangle = -2.5$.

\section{Relations between halo bias parameters}
\label{app:biasparams}

In this Appendix, we will briefly review the relations between the halo bias parameters used in the main text.
We refer to e.g.~\cite{DJSreview16,mowhite96,giannporc10,baldaufetal11a,Baldauf:2012hs} and references therein for more details.
We will express
the non-Gaussian bias parameters $b_{01}^{(h)}$, $b_{11}^{(h)}$ and tidal-tensor bias $b_{s^2}^{(h)}$,
in terms of $b_{10}^{(h)}$ and $b_{20}^{(h)}$ (dropping the $(h)$ superscripts from here on).
We do this by considering relation between Lagrangian bias parameters, which are then converted to Eulerian bias parameters.

\subsection{Conversion between Lagrangian and Eulerian biases}

We assume a Lagrangian bias model of the form (ignoring stochastic terms),
\beq
\delta_h^L(\q) = b_{10}^L \, \delta_0(\q) + \fnl \, b_{01}^L \, \phi(\q) +  b_{20}^L \, \delta_0^2(\q) + \fnl \, b_{11}^L \, \phi(\q) \, \delta_0(\q) + \fnl^2 \, b_{02}^L \, \phi^2(\q),
\eeq
where $\delta_0(\q)$ is the linearly extrapolated primordial matter overdensity as a function of Lagrangian position $\q$ ({\it not} a wave vector here).
We have expanded the halo density to second order and we have again explicitly taken out factors of $\fnl$ from the non-Gaussian bias parameters.
We have here already assumed that there is zero tidal-tensor bias in the initial conditions.
It is this {\it assumption} that will allow us to express the Eulerian tidal-tensor bias in terms of the linear bias $b_{10}$.

The relation between Eulerian (no superscript for Eulerian quantities) and Lagrangian halo density is given by,
\beq
1 + \delta_h(\x) = \left( 1 + \delta_h^L(\q) \right) \, \left( 1 + \delta(\x) \right).
\eeq
Using perturbation theory, we can match the Lagrangian matter overdensity to the final Eulerian overdensity by,
\beq
\delta_0(\q) = \delta(\x) + a_2 \, \delta^2(\x) + a_{s^2} \, s^2(\x) + \mathcal{O}(\delta^3),
\eeq
with,
\beq
a_s = -\frac{17}{21} \quad \quad \text{and} \quad \quad a_{s^2} = -\frac{2}{7}.
\eeq

From the above relations it is then straightforward to obtain the Eulerian bias parameters (cf.~Eq.~(\ref{eq:halos 2nd})),
\bea
b_{10} &=& 1 + b_{10}^L, \quad b_{01} =  b_{01}^L  \nonumber \\
b_{20} &=& b_{20}^L + (1 + a_2) \, b_{10}^L, \quad b_{s^2} =  a_{s^2} \, b_{10}^L \nonumber \\
b_{11} &=& b_{11}^L + b_{01}^L, \quad  b_{02} = b_{02}^L
\eea

Thus, in particular, from the absence of  Lagrangian tidal-tensor bias, we find for the Eulerian tidal-tensor bias,
\beq
\label{eq:bs2}
b_{s^2} = -\frac{2}{7} \, b_{10}^L = -\frac{2}{7} \, \left( b_{10} - 1 \right).
\eeq

\subsection{Neglecting the convection term}

We have in the main text neglected the convection term that is generated by the transformation of the $b_{01}^L$ contribution from Lagrangian coordinates, $\q$, to Eulerian coordinates, $\x = \q + {\bf \Psi}$,
\beq
\delta_h(\x) \supset \fnl \, b_{01}^L \, \phi(\q) = \fnl \, b_{01}^L \, \phi(\x) - \fnl \, b_{01}^L \, {\bf \Psi}(\x) \cdot {\bf \nabla} \phi(\x).
\eeq
We have checked that the
resulting convection term, $- \fnl \, b_{01}^L \, {\bf \Psi}(\x) \cdot {\bf \nabla} \phi(\x)$, leads to a suppressed contribution to the halo mode-coupling signal.
Concretely, it describes a mode-coupling of the form \cite{tellarinietal15},
\beq
\delta_h(\k) \supset \int \frac{d^3 \q}{(2 \pi)^3} \, \tilde{\mathcal{N}}_2(\q, \k - \q) \, \tilde{\delta}(\q) \, \tilde{\delta}(\k - \q),
\eeq
with the symmetrized kernel,
\beq
\tilde{\mathcal{N}}_2(\k_1, \k_2) = \frac{\k_1 \cdot \k_2}{2 k_1^2} \, \M^{-1}(k_2) + \frac{\k_1 \cdot \k_2}{2 k_2^2} \, \M^{-1}(k_1).
\eeq
Thus, according to Eqs (\ref{eq:def d2hi}) and (\ref{eq:modulationfromkernel}),
the convection term produces a contribution to the modulation of the position-dependent power spectrum given by (expressed in terms of the  Eulerian bias parameters),
\beq
b_{2h(i)}'(\q_L) \supset  \frac{b_{01}}{b_{10}} \, \M^{-1}(q_L) \, \frac{\M(q_L)}{\M(k)} \,
\left( 1 + \mu^2 \, n_s^{\phi \delta}(k)  \right)
+ \frac{b_{01}}{b_{10}} \, \M^{-1}(q_L) \, \left(\frac{q_L}{k}\right)^2
\, \left[ 1 - \left(2 - n_s^\delta(k) \right) \, \mu^2 \right],
\eeq
up to terms suppressed by powers of $q_L/k$.
Here, $n_s^\delta(k)$ and $n_s^{\delta \phi}(k)$ are the spectral indices of the matter power spectrum and the matter-$\phi$ cross-spectrum, respectively.
Since the standard contributions to $b_{2h(i)}'(q_L)$ are of order $\M^{-1}(q_L)$, the signal from the convection term is strongly suppressed by factors $\M(q_L)/\M(k)$ and $(q_L/k)^2$.

\subsection{Non-Gaussian bias parameters from universal halo mass function}

Assuming a universal mass function, schematically \cite{DJSreview16},
\beq
\bar{n}_h \propto f\left(\frac{\delta_c}{\sigma}\right),
\eeq
where $\delta_c$ is the critical overdensity for collapse, and $\sigma$ the standard deviation of matter perturbations on the halo scale of interest,
one can derive the following relations between the Lagrangian bias parameters,
\beq
b_{01}^L = 2 \delta_c \, b_{10}^L \quad \quad \text{and} \quad \quad b_{11}^L = 4 \delta_c \, b_{20}^L - 2 b_{10}^L
\eeq
(we do not make use of $b_{02} = b_{02}^L$ in this paper).
Converting this to Eulerian biases according to the previous subsection finally gives,
\beq
\label{eq:b01 b11}
b_{01} = 2 \delta_c \, \left( b_{10}  - 1 \right), \quad
b_{11} = 4 \delta_c \, b_{20} - 2 \left(1 + \delta_c \, \left(1 + 2 a_2 \right) \right) \, \left( b_{10} - 1\right).
\eeq

\section{The halo bispectrum and stochastic noise}
\label{app:stochnoise}

We here discuss how the halo bispectrum in the position-dependent power spectrum approach relates to the leading-order (tree level) result in perturbation theory. This question has been worked out in great detail in \cite{chiang17}, but here we extend that work to include the leading order halo shot noise contributions. This requires the addition of second order stochastic noise terms to the halo bias expansion. While for simplicity we have neglected such terms in the main text, we here show how they reproduce the standard shot noise terms in the bispectrum.

In the main text, we have considered the bias model of Eq.~(\ref{eq:halos 2nd}) (we drop the term due to $b_{02}^{(h)}$ below),
\beq
\delta_h = b^{(h)}_{10} \, \delta + b^{(h)}_{01} \, \phi + \epsilon_h
+ b^{(h)}_{20} \, \delta^2 + \fnl \, b^{(h)}_{11} \, \delta \, \phi + b^{(h)}_{s^2} \, s^2.
\eeq
For the long mode halo perturbation, this led to
\beq
\delta_{h,L} = \left( b_h + \fnl \, b_h' \right) \, \delta_L + \epsilon_{h,L}
= b_{10}^{(h)} \, \delta_L
+ \fnl \, b_{01}^{(h)} \, \phi_L + \epsilon_h,
\eeq
with stochastic noise power spectrum,
\beq
N_h = \frac{1}{\bar{n}_h}.
\eeq
For the position-dependent halo power spectrum, the bias expansion led to,
\beq
\delta_{2h(i),L} = \left(b_{2h(i)} + \fnl \, b_{2h(i)}' \right) \, \delta_L + \epsilon_{2h(i)}
= 4 \left( \bar{F}_2(\mu, k_i) + \frac{b_{20}^{(h)}}{b_{10}^{(h)}} + \frac{b^{(h)}_{s^2}}{b_{10}^{(h)}} \, \left(  \mu^2 - \tfrac{1}{3} \right) \right) \, \delta_L
+ \fnl \, \left(4 + 2 \frac{b_{11}^{(h)}}{b_{10}^{(h)}} \right) \, \phi_L + \epsilon_{2h(i)}, \nonumber
\eeq
with stochastic noise (due to the long mode-independent variance in the short modes),
\beq
{\bf N}_{2h(i),2h(j)} = \frac{2 (2 \pi)^3}{V_{\k,i}} \, \left( \frac{1 + \bar{n}_h \, \left(b_{10}^{(h)}\right)^2 \, P(k_i)}{\bar{n}_h \, \left(b_{10}^{(h)}\right)^2 \, P(k_i)} \right)^2 \, \delta^{(K)}_{ij},  \quad \quad \text{with} \quad \quad \mu \equiv \hat{k}_i \cdot \hat{q}_L
\eeq

Ignoring any correlation between the two different types of stochastic noise, i.e.~setting,
\beq
{\bf N}_{h,2h(i)} = 0,
\eeq
this leads to the halo bispectrum expectation value,
\bea
&B^{(h)}&(\q_L, \k_i, -\k_i - \q_L)
= b_{10}^{(h) \, 2} \, P(k_i) \, P_{h,2h(i)}(q_L)
= b_{10}^{(h) \, 2} \, P(k_i) \, \left( b_h + \fnl \, b_h' \right) \, \left(b_{2h(i)} + \fnl \, b_{2h(i)}' \right) \, P(q_L) \nonumber \\
&=& b_{10}^{(h) \, 2} \, P(k_i) \, b_{10}^{(h)} \, 4 \left( \bar{F}_2(\mu, k_i) + \frac{b_{20}^{(h)}}{b_{10}^{(h)}} + \frac{b^{(h)}_{s^2}}{b_{10}^{(h)}} \, \left(  \mu^2 - \tfrac{1}{3} \right) \right) \,  \, P_{\delta \delta}(q_L) \nonumber \\
&+& \fnl \, b_{10}^{(h) \, 2} \, P(k_i) \, \left( b_{10}^{(h)} \,  \left(4 + 2 \frac{b_{11}^{(h)}}{b_{10}^{(h)}} \right) +   b_{01}^{(h)} \, 4 \left( \bar{F}_2(\mu, k_i) + \frac{b_{20}^{(h)}}{b_{10}^{(h)}} + \frac{b^{(h)}_{s^2}}{b_{10}^{(h)}} \, \left(  \mu^2 - \tfrac{1}{3} \right) \right) \right) \, P_{\delta \phi}(q_L) \nonumber \\
&+& \fnl^2 \, b_{10}^{(h) \, 2} \, P(k_i) \, b_{01}^{(h)} \, \left(4 + 2 \frac{b_{11}^{(h)}}{b_{10}^{(h)}} \right) \, P_{\phi \phi}(q_L),
\eea
where in the last equality, we have used the notation $P_{\delta \delta}(q_L) = P(q_L)$ for the matter power spectrum, $P_{\delta \phi}(q_L) = \M^{-1}(q_L) \, P(q_L)$ for the cross-spectrum, and $P_{\phi \phi}(q_L) \ \M^{-2}(q_L) \, P(q_L)$ for the primordial potential power spectrum.
The above is exactly the bispectrum given in Eqs (9 - 11) of \cite{chiang17}, except for three differences: (1) We here ignore all terms involving $\phi_S = \M^{-1}(k_S) \, \delta_S$, which are subdominant (because suppressed relative to the other signal terms by factors $\M^{-1}(k_S)/\M^{-1}(q_L)$). This corresponds to setting all terms with $P_{\delta \phi}(k)$ and $P_{\phi \phi}(k)$ in \cite{chiang17}, where $k$ is the short mode, to zero. (2) Our expression applies to the full bispectrum without averaging over angles, as reflected by the explicit $\mu$ dependence. (3) We have included tidal-tensor bias $b_{s^2}^{(h)}$ (although this contribution would average to zero if we considered only the monopole of the bispectrum).
Moreover, just like in \cite{chiang17}, our approach ignores non-linear contributions to
the long-mode halo overdensity, e.g.~the effect of terms of the type,
\beq
\delta_h(\q_L) \supset b_{20}^{(h)} \, \int \frac{d^3 \k_S}{(2 \pi)^3} \, \delta(\k_S) \, \delta(\q_L - \k_S).
\eeq

Within the approximations that we have spelled out, we have shown above that the position-dependent power spectrum approach properly reproduces the ``signal'' part of the tree level halo bispectrum.
However,
the standard tree level halo bispectrum also contains shot noise contributions.
Assuming Poisson statistics for the shot noise \cite{peebles80,scocc00},
\beq
B^{(h)}_{sn}(\k_1,\k_2, \k_3) = \frac{1}{\bar{n}_h} \, \left[P_{hh}(k_1) + P_{hh}(k_2) + P_{hh}(k_3) \right] + \frac{1}{\bar{n}_h^2},
\eeq
which in the squeezed limit becomes,
\beq
\label{eq:Bsn}
B^{(h)}_{sn}(\q_L, \k_i, -\k_i - \q_L) = \frac{1}{\bar{n}_h} \, P_{hh}(q_L) + \frac{2}{\bar{n}_h} \, P_{hh}(k_i) + \frac{1}{\bar{n}_h^2}
\eeq
(note that $P_{hh}(k) = b_{10}^{(h) \, 2} \, P(k)$ is the halo power spectrum without shot noise).
We will show below how these contributions can be included by taking into account the non-Gaussian nature of the halo stochastic noise $\epsilon_h$.
Note however that even in a conventional bispectrum forecast, we likely would not include the shot noise terms in  $B_{sn}$ as ``signal'' so neglecting the higher order stochastic terms does not throw away important information.
Moreover, the leading order halo shot noise contributions to the bispectrum {\it covariance} (and the covariance of the other statistics considered) {\it are} already included in this paper even though we neglect the mode-coupling involving the halo shot noise $\epsilon_h$.

Let us now add the mode-coupling due to the non-Gaussian nature of the halo stochastic noise.
For simplicity, from here one, we ignore terms related to primordial non-Gaussianity, i.e.~bias terms involving the primordial potential $\phi$.
Our aim is to capture the long-short mode-coupling due to the Poisson statistics of the halo shot noise
in terms of biasing parameters, choosing $\epsilon_h$ itself to contain no mode-coupling, i.e.~$\epsilon_{h,L}$ and $\epsilon_{h,S}$ are independent by construction. We then write,
\beq
\delta_{h,S} \supset b^{(h)}_{\delta_L \epsilon_S} \, \delta_L \, \epsilon_{h,S}
+ b^{(h)}_{\epsilon_L \delta_S} \, \epsilon_{h,L} \, \delta_S
+ 2 b_{\epsilon^2}^{(h)} \, \epsilon_{h,L} \, \epsilon_{h,S}.
\eeq
The position-dependent halo power spectrum modulation then becomes (again, ignoring the primordial non-Gaussianty terms),
\bea
\label{eq:d2h 2ndorderstoch}
\delta_{2h(i),L} &=& 4 \left( \bar{F}_2(\mu, k_i) + \frac{b_{20}^{(h)}}{b_{10}^{(h)}} + \frac{b^{(h)}_{s^2}}{b_{10}^{(h)}} \, \left(  \mu^2 - \tfrac{1}{3} \right) \right) \, \delta_L + \epsilon_{2h(i),L} \nonumber \\
&+& 2 b^{(h)}_{\delta_L \epsilon_S} \, \frac{1}{\bar{n}_h \, b^{(h) \, 2}_{10} \, P(k_i)} \, \delta_L \nonumber \\
&+& \left( 2 \frac{b^{(h)}_{\epsilon_L \delta_S}}{b^{(h)}_{10}} + 4 b^{(h)}_{\epsilon^2} \, \frac{1}{\bar{n}_h \, b^{(h) \, 2}_{10} \, P(k_i)}\right) \, \epsilon_{h,L},
\eea
where the terms due to deviations from Gaussianity in the stochastic noise are on the second and third lines.
We have thus effectively added a term to $b_{2h(i)}$ (second line) and have introduced stochastic contributions in addition to $\epsilon_{2h(i),L}$ (third line).

The parameters $b^{(h)}_{\delta_L \epsilon_S}$, $b^{(h)}_{\epsilon_L \delta_S}$ and $b_{\epsilon_L \epsilon_S}^{(h)}$ for Poisson noise
are obtained by considering the response of small-scale fluctuations to a change in the background matter density and halo number density.
First, the shot noise in the  power spectrum of $\delta_h(\x) = (n_h(\x) - \bar{n}_h)/\bar{n}_h$ depends on the mean value of $n_h$ in the local volume $V_S$, which is modulated by the long mode halo perturbation, giving
\beq
P_{sn,L} = \frac{\bar{n}_h \, \left( b_{10}^{(h)} \, \delta_L + \epsilon_{h,L} \right)}{\bar{n}^2_h}
= \frac{1}{\bar{n}_h} \, \left( b_{10}^{(h)} \, \delta_L + \epsilon_{h,L} \right).
\eeq
This tells us the modulation of $\epsilon_{h,S}$ by $\delta_L$ and $\epsilon_{h,L}$. In particular,
\beq
b^{(h)}_{\delta_L \epsilon_S} = \frac{1}{2} \, b_{10}^{(h)}, \quad \quad
b_{\epsilon_L \epsilon_S}^{(h)} = \frac{1}{4}.
\eeq

Next consider the effect on $\delta_{h,S}$ of
a short-wavelength matter density perturbation, $\delta_S$, in the presence of
a long-mode shot noise perturbation, $\epsilon_{h,L}$.
Locally, the halo number density perturbation will equal $b_{10}^{(h)} \, \delta_S$ times the local mean number density, $\bar{n}_h \, \left(1 + \epsilon_{h,L}\right)$, giving,
\beq
\delta_{h,S} = \frac{b_{10}^{(h)} \, \delta_S \, \bar{n}_h \, \left(1 + \epsilon_{h,L}\right)}{\bar{n}_h}
= \left(1 + \epsilon_{h,L}\right) \, b_{10}^{(h)} \, \delta_S.
\eeq
Thus, we can read off,
\beq
b^{(h)}_{\epsilon_L \delta_S} = b_{10}^{(h)}.
\eeq

The new terms lead to the following shot noise terms in the bispectrum,
\bea
B_{sn}(\q_L, \k_i, -\k_i - \q_L) &=& 2 b^{(h)}_{\delta_L \epsilon_S} \, b^{(h)}_{10} \, \frac{1}{\bar{n}_h} \, P(q_L)
+ 2 b^{(h)}_{\epsilon_L \delta_S} \, b^{(h)}_{10} \, \frac{1}{\bar{n}_h} \, P(k_i)
+ 4 b^{(h)}_{\epsilon_L \epsilon_S} \, \frac{1}{\bar{n}_h^2} \nonumber \\
&=& \frac{1}{\bar{n}_h} \, b_{10}^{(h) \, 2} \, P(q_L)
+ \frac{2}{\bar{n}_h} \, b_{10}^{(h) \, 2} \, P(k_i) +
\frac{1}{\bar{n}_h^2}.
\eea
This reproduces the desired result, Eq.~(\ref{eq:Bsn}).

\bibliographystyle{utphys}
\bibliography{refs}

\end{document}